\documentclass[a4paper,11pt]{article}
\pdfoutput=1 
\usepackage{jheppub}  
\usepackage{graphicx,color}
\usepackage{amsmath}
\usepackage{autobreak}
\usepackage{times}
\usepackage{epsfig,graphicx}
\usepackage[caption=false]{subfig}
\usepackage[english]{babel}
\usepackage[utf8]{inputenc} 
\usepackage{amssymb} 
\usepackage{amsmath} 
\usepackage{xspace} 
\usepackage{url} 
\usepackage{slashed} 

\usepackage[usenames,dvipsnames,svgnames,table]{xcolor}  
\usepackage{marginnote} 
\usepackage{pdfpages} 
\allowdisplaybreaks
\newcommand{\ben}{\begin{enumerate}}
\newcommand{\een}{\end{enumerate}}
\newcommand{\beq}{\begin{equation}}
\newcommand{\eeq}{\end{equation}}
\newcommand{\bal}{\begin{align}}
\newcommand{\eal}{\end{align}}

\newcommand{\bea}{\begin{eqnarray}}
\newcommand{\eea}{\end{eqnarray}}
\newcommand{\nn}{\nonumber}

\def\z#1{{\zeta_{#1}}}
\def\D#1{{{\cal D}_{#1}}}
\newcommand{\Ca}{C_A}
\newcommand{\Cf}{C_F}
\newcommand{\nf}{n_F}

\newcommand{\Lqf}{L_{qf}}

\newcommand{\F}{\hat{\cal F}}
\newcommand{\ph}{{\cal \varPhi}}
\newcommand{\Z}{{\cal Z}}
\newcommand{\K}{{\cal K}}
\newcommand{\G}{{\cal G}}
\newcommand{\eps}{\epsilon}
\newcommand{\ash}{\hat{a}_s}
\newcommand{\as}{a_s}
\newcommand{\mur}{\mu_r}
\newcommand{\muf}{\mu_f}

\newcommand{\KB}{\bar{{\cal K}}}
\newcommand{\GB}{\bar{{\cal G}}}
\newcommand{\gb}{\bar{{g}}}

\newcommand{\omg}{\omega}

\def\Dm1{{{\delta(1-z)}}}

\def\nf{{n^{}_{\! f}}}

\def\CIg#1{{{\Delta_{\rm gg}^{(#1),G}}}}
\def\CIq#1{{{\Delta_{\rm q\bar{q}}^{(#1),G}}}}

\def\A#1{{A_{#1}}}
\newcommand{\Lqrfrt}{L_{qr}L_{fr}^{2}}
\newcommand{\Lqrfr}{L_{qr}L_{fr}}
\newcommand{\Lqrtfr}{L_{qr}^{2}L_{fr}}

\def\g0#1DY{{g_{0#1}^{DY}}}
\def\gGRGG#1{{g_{0#1}^{gg}}}
\def\gGRQQ#1{{g_{0#1}^{q\bar{q}}}}

\def\gNB#1{{g_{#1}}}
\newcommand{\Lqr}{L_{qr}}
\newcommand{\Lfr}{L_{fr}}

\newcommand{\iW}{\omega^{-1}}
\def\LogmW1{{{\ln (1-\omega)}}}
\def\w{{\omega}}
\def\WbimW{{\frac{\omega}{(1-\omega)}}}
\def\LogomWtIMW{{\frac{\ln(1-\omega)}{(1-\omega)}}}
\def\LogomWtIMWt{{\frac{\ln (1-\omega)}{(1-\omega)^2}}}
\def\LogtmWtIMWt{{\frac{\ln (1-\omega)^2}{(1-\omega)^2}}}
\def\LogttmWtIMWt{{\frac{\ln (1-\omega)^3}{(1-\omega)^2}}}

\def\LogtmWtIMW{{\frac{\ln (1-\omega)^2}{(1-\omega)}}}
\def\WttmWtimWt{{\frac{\omega(2-\omega)}{(1-\omega)^2}}}
\def\WtbimWt{{\frac{\omega^2}{(1-\omega)^2}}}

\def\btzAIII{{\frac{A_3}{\beta_0^2}}}
\def\btzAII{{\frac{A_2}{\beta_0}}}
\def\btzAI{{A_1}}

\def\btzDII{{\frac{D_2}{\beta_0}}}
\def\btzDI{{D_1}}

\def\btztAIV{{\frac{A_4}{\beta_0^2}}}
\def\btztAIII{{\frac{A_3}{\beta_0}}}
\def\btztAII{{A_2}}
\def\btztAI{{\beta_0 A_1}}

\def\btztDIII{{\frac{D_3}{\beta_0}}}
\def\btztDII{{D_2}}
\def\btztDI{{\beta_0 D_1}}
\def\bobt{{\frac{\beta_1 \beta_2}{\beta_0^5}}}

\def\AAo{{\frac{A_1}{\beta_{0}}}}
\def\AAt{{\frac{A_2}{\beta_{0}^2}}}
\def\DDo{{\frac{D_1}{\beta_{0}}}}

\newcommand{\dRAoNR }{\frac{d_R^{abcd}d_A^{abcd}}{N_R}}
\newcommand{\dRFoNR }{\frac{d_R^{abcd}d_F^{abcd}}{N_R}}

\newcommand{\btoo}{{\bigg(\frac{\beta_{2}}{\beta_0^{3}}\bigg)}}
\newcommand{\bthr}{{\bigg(\frac{\beta_{3}}{\beta_0^{4}}\bigg)}}

\title{Resummed inclusive cross-section in ADD model at N$^3$LL+NNLO}
\author[a,b]{Goutam Das,}
\author[c]{M. C. Kumar }
\author[c]{ and Kajal Samanta }
\affiliation[a]{Theory Group, Deutsches Elektronen-Synchrotron (DESY), Notkestrasse 85, D-22607 Hamburg, Germany}
\affiliation[b]{Theoretische Physik 1, Naturwissenschaftlich-Technische Fakult{\"a}t, Universit{\"a}t Siegen,
Walter-Flex-Strasse 3, 57068 Siegen, Germany}
\affiliation[c]{Department of Physics, Indian Institute of Technology Guwahati, Guwahati-781039, India}
\emailAdd{goutam.das@uni-siegen.de}
\emailAdd{mckumar@iitg.ac.in}
\emailAdd{kajal.samanta@iitg.ac.in}
\abstract{We present three loop soft-plus-virtual (SV) corrections to the spin-2 production at the Large Hadron Collider (LHC). For this calculation, we make use of the recently computed quark and gluon three loop form factors for the spin-2 production, the universal soft-collinear coefficients as well as the mass factorization kernels. The SV coefficients are presented up to next-to-next-to-next-to leading order (N$^3$LO).  We also use these coefficients at three loops to compute the resummed prediction for inclusive cross-section to next-to-next-to-next-to leading logarithmic accuracy (N$^3$LL) matched to next-to-next-to leading order (NNLO). We use the standard technique to derive the Mellin N-dependent coefficients and also the N-independent coefficients to achieve the resummation using the minimal prescription matching procedure. Considering the spin-2 propagator in the large extra dimensional (ADD) model, we also study the numerical impact of these three-loop SV corrections as well as the resummed predictions on the di-lepton invariant mass distribution at the 13 TeV LHC. We find that the conventional scale uncertainties in the NNLO+N$^3$LL resummed results substantially get reduced to as low as 2\% in the high invariant mass region. We also estimate the PDF uncertainties in our predictions that will be useful in the experimental searches for large extra dimensions.}

\begin{document} 
\preprint{DESY 19-171}
\keywords{Resummation, Phenomenology of Large extra dimensions}
\maketitle

\section{Introduction} \label{intro}
The Standard Model (SM) of particle physics has been successful so far in describing the dynamics of the fundamental particles. The discovery of the Higgs boson \cite{Aad:2012tfa,Chatrchyan:2012xdj} at the CERN Large Hadron Collider (LHC) has been a milestone in establishing the theory. Presently, the SM is being tested to unprecedented accuracy at the LHC in order to measure the Higgs couplings to the fermions and gauge bosons of SM, to improve the accuracy of the universal parton distribution functions (PDFs) as well as to search for any possible deviations from the SM results that can hint a sign of beyond SM (BSM) physics. To achieve this, one needs a robust and highly precise theory predictions, thanks to the recent developments both in the electroweak and QCD precision studies. At the LHC, the initial state partons being colored, the QCD corrections are very dominant and higher and higher terms in the perturbative expansion are often needed to have a reliable prediction which can be compared with experimental outcome.

Precise theory predictions are already available for many SM and BSM processes at the LHC.  Particularly the Higgs and pseudo-scalar Higgs (Spin-0) boson productions in gluon fusion  \cite{Harlander:2002wh,Harlander:2002vv,Anastasiou:2002yz,Anastasiou:2002wq,Ravindran:2003um} as well as in bottom quark annihilation  \cite{Harlander:2003ai} are already available at NNLO accuracy. For the SM Higgs, even the complete N$^3$LO results  \cite{Anastasiou:2015ema,Mistlberger:2018etf,Duhr:2019kwi} are also available very recently and the corrections are found to be well within $3-5\%$. The exclusive observable are also being calculated at the same accuracy. For example the NNLO corrections to  rapidity distributions for Higgs are also available in the context of LHC \cite{Anastasiou:2004xq,Anastasiou:2011qx,Buehler:2012cu}. The SM Higgs rapidity is also known to N$^3$LO accuracy in gluon fusion \cite{Dulat:2018bfe,Cieri:2018oms}. Recently there is renewed interest in the di-Higgs production and associated production in the context of Higgs properties measurement (see for example \cite{deFlorian:2013jea,Borowka:2016ehy,Banerjee:2018lfq,deFlorian:2016uhr,H:2018hqz,Kumar:2014uwa,Brein:2011vx,Ferrera:2011bk,Ferrera:2014lca,Ahmed:2019udm,Bhattacharya:2019oun}).

The di-lepton production through the decay of spin-1 gauge bosons on the other hand provides one of the cleanest channels to be measured at the collider experiments. Consequently, the di-lepton production process has been of interest both in theory as well as in experiments and also looked in the context of BSM searches. The phenomenological implications of this process are exposed in the deviations from the SM predictions either in the form of contact interactions or heavy resonances. For the case of spin-1 particle (W/Z) production (Drell-Yan process) at the LHC, NNLO corrections are available for decades.  DY inclusive  cross-section is known at NNLO \cite{Hamberg:1990np,Harlander:2002wh}. Drell-Yan rapidity is also known at the same accuracy \cite{Anastasiou:2003yy,Anastasiou:2003ds,Catani:2009sm,Melnikov:2006kv,Gavin:2012sy}.
 
Spin-2 particle production on the other hand is available very recently at NNLO accuracy for di-lepton production for both generic universal \cite{Ahmed:2016qhu} and non-universal \cite{Banerjee:2017ewt} couplings. Spin-2 production in the context of large extra dimensional models like ADD \cite{ArkaniHamed:1998rs} or RS \cite{Randall:1999ee} has got much attention in the context of BSM searches. In ADD and RS models, gravitons couple to the energy momentum tensor and consequently they couple with equal strength to all the SM particles (universal couplings). For such universal coupling scenario, since graviton couples to quarks as well as gluons, their production cross sections at the LHC are very important and have been studied well phenomenologically. Consequently, searches for extra dimensions at the LHC in di-lepton \cite{Sirunyan:2018ipj,Khachatryan:2016zqb}, di-photon missing energy \cite{Khachatryan:2014tva} signals have been carried out yielding stringent bounds on the model parameters. On the theory side, there is extensive  study on spin-2 production. Di-photon, di-lepton and di-gauge boson rates are provided in ADD  \cite{Frederix:2012dp,Frederix:2013lga,Kumar:2007af,Kumar:2008dn,Kumar:2008pk,Kumar:2009nn,Agarwal:2009xr,Agarwal:2010sp} and in RS models  \cite{Das:2014tva,Kumar:2011yta,Agarwal:2009zg,Agarwal:2010sn,Mathews:2005bw} at NLO as well as the tri-gauge bosons production \cite{Kumar:2011jq,Das:2015bda}. The generic universal and non-universal spin-2 processes have been automated \cite{Das:2016pbk} in M{\scriptsize AD}G{\scriptsize RAPH}5\_{\scriptsize A}MC@NLO \cite{Alwall:2014hca} framework recently.

Out of several extra dimensional models, ADD model provides a very simple solution to the hierarchy problem and  has been looked for extensively at the LHC. In the ADD model all the SM particles are confined to four dimensional brane whereas gravity can propagate through the $4+n$ dimensional bulk. These extra dimensions are compactified with periodic boundary conditions which leads to a tower of Kaluza-Klein (KK) modes. These KK modes lead to non-resonant excess in high invariant mass of di-lepton pairs which results from the decay of virtual gravitons. The search for non-resonant enhancement from models like ADD has been searched at the LHC from time to time. It is observed in the NLO QCD computation \cite{Mathews:2004xp} that the K-factors in the di-lepton production case are potentially large and range up to 60\%. This is because the graviton couples to quarks like the gauge bosons do in the SM, but also to gluons and hence mimic large K-factors of the Higgs boson production case at the LHC. This leads to the computation of the NNLO QCD corrections to the di-lepton production process in extra dimension models \cite{Ahmed:2016qhu}. The NNLO QCD corrections are found to contribute to the total cross sections another 10\% of the LO predictions. The NNLO K-factors are thus quite different from those of the SM.

To minimize the theory uncertainties it is imperative to go beyond the NNLO in QCD. First step towards higher orders beyond NNLO is to get the SV predictions by calculating the most singular terms at the higher order. SV calculation has been successfully performed in case of SM inclusive Higgs production \cite{Anastasiou:2014vaa,Moch:2005ky,Laenen:2005uz,Ravindran:2005vv,Ravindran:2006cg,Idilbi:2005ni}, associated production \cite{Kumar:2014uwa}, bottom quark annihilation  \cite{Ahmed:2014cha}, DY production \cite{Ravindran:2006bu,Ahmed:2014cla}, pseudo-scalar Higgs production  \cite{Ahmed:2015qda}   at N$^3$LO level as well as for rapidity \cite{Ravindran:2006bu,Ravindran:2007sv,Ahmed:2014uya,Ahmed:2014era} and has been shown that it constitutes a significant contribution to the cross-section.  Another way to improve the accuracy of the inclusive cross-section over NNLO is to resum threshold enhanced logarithms to all order. These logarithms play an important contribution when partonic threshold variable $z$ takes the limit 1. The resummation is well understood in the Mellin-N space and is possible due to complete factorization of the soft function amplitude as well as phase space in Mellin-N space. The threshold resummation has been successfully applied to many SM process for example Higgs production \cite{Catani:2003zt,Moch:2005ky,Catani:2014uta,Bonvini:2014joa,Bonvini:2016frm,H:2019dcl} (see  also \cite{Ahmed:2015sna} for renormalisation group improved prediction.), DY production \cite{Moch:2005ky,Catani:2014uta} as well as pseudo-scalar production  \cite{Ahmed:2016otz} (see \cite{Schmidt:2015cea,deFlorian:2007sr} for earlier works).

The resummation is very important for the differential observable. The threshold enhanced resummation has been performed consistently in double Mellin space for rapidity and for $x_F$ distributions \cite{Catani:1989ne,Westmark:2017uig,Banerjee:2017cfc,Banerjee:2018vvb} (see also \cite{Lustermans:2019cau} for SCET based factorization and resummation). Resummation is essential for observable which are very sensitive to infrared physics for example transverse momentum distribution where logarithms of the type $\ln (Q^2/p_T^2)$ can be very large in the infrared region thus spoiling the fixed order (FO) prediction.  Resummation is thus very important to correctly describe the low $p_T$ region and results are available up to N$^3$LL accuracy for many important SM processes.  The Higgs $p_T$ spectrum is known to NNLO+N$^3$LL accuracy \cite{Chen:2018pzu,Bizon:2018foh} and the uncertainty is found to be reduced by $60\%$ compared to NLO+NNLL in the low $p_T$ region. For the pseudo-scalar production, the $p_T$ spectrum is known to NNLO$_A$+NNLL \cite{Agarwal:2018vus} and the scale variation is found to be improved to $20\%$ in the low-$p_T$ region. Drell-Yan $p_T$ spectrum is also known to same accuracy \cite{Bozzi:2010xn,Chen:2018pzu}.

In this article we improve the inclusive cross-section for spin-2 production in di-lepton channel within ADD model beyond NNLO accuracy. First, we calculate the compete SV results at N$^3$LO using the form-factor at three loops and the universal soft function at the same. Second, we apply the standard threshold resummation technique and extract the process-dependent constant pieces required up to N$^3$LL level. The paper is organized as follows: in Sec.\ \ref{sec:theory}, we study the theoretical formalism where we have collected all the formulas for the DY production in ADD. We also present the formalism to calculate the SV contributions as well as the resummed coefficients required for the N$^3$LL accuracy. In Sec.\ \ref{sec:numerical}, we present the numerical results in the context of 13 TeV LHC and then summarize our results.

\section{Theoretical Framework} \label{sec:theory}
The hadronic cross-section for standard DY production at the hadron collider is given by,
\begin{align}\label{eq:hadronic-xsect}
{d \sigma^{P_1 P_2} \over d Q}\left(\tau,Q^2\right)
&=\frac{Q}{S}\sum_{ab={q,\overline q,g}} \int_0^1 dx_1
\int_0^1 dx_2~ f_a^{P_1}(x_1,\mu_f^2) ~
f_b^{P_2}(x_2,\mu_f^2)
\nonumber\\[2ex] &
\times 
\sum_{I\in \{\gamma,Z,G\}}\int_0^1 dz \,\, 
\Delta^I(z,Q^2,\mu_f^2)
\delta(\tau-z x_1 x_2)\,.
\end{align}
Here $S$ and $\hat{s}$ denote the center-of-mass energy in the hadronic and partonic frame respectively. The hadronic and partonic threshold variables $\tau$ and $z$ are defined as
\begin{align}
\tau=\frac{Q^2}{S}, \qquad z= \frac{Q^2}{\hat{s}} \,.
\end{align}
They are thus related by $\tau = x_1 x_2 z$. The partonic cross-section gets contribution from virtual photon and $Z$ boson as in the standard DY process in SM, in addition, it also gets contribution from spin-2 mediator ($G$) decaying to leptons. The SM in here is treated as background for the signal defined by diagrams with spin-2 production. Notice that the signal and background completely get separated from each other in the cross-section after performing the phase-space integration for invariant mass distribution. This gives opportunity to calculate the SM and ADD contributions completely separately and there is no interference term between them. Whereas in the SM case, there is only quark annihilation channel at the born level, in the ADD case, both quark annihilation as well as gluon fusion channels are present already at the born level. 

The partonic cross-section in the above Eq.\ (\ref{eq:hadronic-xsect}) can have two separate kind of contributions, one which is more singular when $z\to1$ known as the soft-virtual contribution  and the other is regular contribution which is finite in the limit $z\to1$. Thus the decomposition of the partonic cross-section has the following form,
\begin{align}\label{eq:sv-decompose}
\Delta^I(z,Q^2,\mu_f^2)  &= {\cal F}^{(0)}_I \Big(\delta_{ab}\Delta^{{\rm(sv)},I}_{ab} + \Delta^{{\rm(reg)},I}_{ab} \Big),
\end{align}
where ${\cal F}^{(0)}_I$ is the pre-factor which depends on the specific model in consideration. In case of ADD model, the pre-factor has the following form,
\begin{align}
{\cal F}^{(0)}_{\rm ADD} = \frac{\kappa^4Q^6}{320\pi^2} |{\cal D}(Q^2)|^2 \,,
\end{align}
with 
\begin{align}
{\cal D}(Q^2) = 16 \pi \bigg( \frac{Q^{d-2}}{ \kappa^2 M_S^{d+2}}  \bigg) {\cal I}\bigg( \frac{M_S}{Q}\bigg) \,.
\end{align}
The summation over the non-resonant KK modes depends on the number of extra dimensions present in the model and yields
\begin{align}
{\cal I}(\omega)&=- \sum_{k=1}^{d/2-1} {1 \over 2 k} \omega^{2 k}
 -{1 \over 2} \log(\omega^2-1)\,, \qquad \qquad \qquad  d={\rm even}\,,
\label{eq16}
\end{align}
\begin{align}
{\cal I}(\omega)&=
-\sum_{k=1}^{(d-1)/2} {1\over 2 k-1} \omega^{2 k-1}
 +{1 \over 2} \log \left({\omega+1}\over{\omega-1}\right) \,,
\quad \quad  d={\rm odd}\,.
\end{align}
The pre-factor for DY case has the following expression,
\begin{align}
\begin{autobreak}
{\cal F}_{\rm DY}^{(0)} =
{4 \alpha^2 \over 3 Q^2} \Bigg[Q_q^2 
- {2 Q^2 (Q^2-M_Z^2) \over  \left((Q^2-M_Z^2)^2
+ M_Z^2 \Gamma_Z^2\right) c_w^2 s_w^2} Q_q g_e^V g_q^V 
+ {Q^4 \over  \left((Q^2-M_Z^2)^2+M_Z^2 \Gamma_Z^2\right) c_w^4 s_w^4}\Big((g_e^V)^2
+ (g_e^A)^2\Big)\Big((g_q^V)^2+(g_q^A)^2\Big) \Bigg]\,,
\end{autobreak}
\end{align}
where $M_Z$ and $\Gamma_Z$ are the mass and the decay width of the $Z$-boson, $\alpha$ is the fine structure constant, $c_w,s_w$ are sine and cosine of Weinberg angle respectively. 
\begin{align}
 g_a^A = -\frac{1}{2} T_a^3 \,, \qquad g_a^V = \frac{1}{2} T_a^3  - s_w^2 Q_a \,,
\end{align}
$Q_a$ being electric charge and $T_a^3$ is the weak isospin of the electron or quarks.

The SV cross-section for spin-2 production is known to two loops in \cite{deFlorian:2013wpa} after the subsequent calculation of the two loop form factors \cite{deFlorian:2013sza}. Recently the complete NNLO correction has been performed after calculating the regular piece at second order in strong coupling \cite{Ahmed:2016qhu} with reverse unitarity method \cite{Anastasiou:2002yz}. In the next two sections we improve this accuracy by first calculating the SV cross-section at the three loops and then in the next section we resum large threshold logarithms up to N$^3$LL accuracy and matched with existing NNLO fixed order result.

\subsection{Soft-virtual cross-section}
The SV cross-section constitutes a significant contribution to the partonic cross-section and can be computed order by order in strong coupling,
\begin{align}
\Delta^{{\rm(sv)},I}_{ab}  = \sum_{i=0}^{\infty} a_s^i \Delta_{\rm ab}^{(i),I} \,.
\end{align}
For SV cross-section, only $q\bar{q}$ and $gg$ channels contribute which appear in the born process for spin-2 production. The threshold enhanced partonic soft-virtual cross-section can be written \cite{Ravindran:2005vv,Ravindran:2006cg} as 
\begin{align}
  \label{eq:sigma}
  \Delta^{{\rm{(sv)}},I} (z, Q^2, \mu_r^{2}, \mu_f^2) = 
  {\cal C} \exp \Big( \Psi_{I} \left(z, Q^2, \mu_r^2, \mu_f^2,
  \epsilon \right)  \Big)  \Big|_{\epsilon = 0} \qquad \text{with } \qquad I = {q,g} \,.
 \end{align}
Here $\Psi_{I}$ is a finite distribution in the limit $\epsilon \to 0$. The symbol $ {\cal C}$ denotes the Mellin convolution (denoted below as $ \otimes$) which in the above expression should be treated as
\begin{align}
  \label{eq:conv}
  {\cal C} \exp \Big( {f(z)} \Big) = \delta(1-z) + \frac{1}{1!} f(z) + \frac{1}{2!} f(z) \otimes f(z) + \cdots \,,
\end{align}
with $f(z)$ being a function containing only $\delta(1-z)$ and plus distributions. The finite exponent in the above gets contribution from the form factor \big(${\F}_{I} (\hat{a}_s, q^2 = -Q^2, \mu^2, \epsilon)$\big), soft-collinear function \big($\ph_{I} (\hat{a}_s, z, Q^2, \mu^2, \epsilon)$\big) as well as mass factorization kernels  \big($\Gamma_{I} (\hat{a}_s, z, \mu_f^2, \mu^2, \epsilon)$\big) and can be written as 
\begin{align}\label{sv:psi}
  \Psi_{I} \Big(z, Q^2, \mu_r^2, \mu_f^2, \epsilon \Big)  
  = &\bigg( \ln \Big[ \Z_{I} (\hat{a}_s, \mu_r^2, \mu^2, \epsilon) \Big]^2 
      + \ln \Big|  {\F}_{I} (\hat{a}_s, q^2, \mu^2, \epsilon)   \Big|^2 \bigg) \delta(1-z) 
      \nonumber\\
    & + 2 \ph_{I} (\hat{a}_s, z, Q^2, \mu^2, \epsilon) 
      - 2 {\cal C} \ln \Gamma_{I} (\hat{a}_s, z, \mu_f^2, \mu^2, \epsilon) \, .
\end{align}
Here $\mu$ has been introduced to define the strong coupling ($\ash$) dimensionless in the $d=4+\eps$ dimensions. $\Z_{I} (\hat{a}_s, \mu_r^2, \mu^2, \epsilon)$ denotes the overall UV renormalization constant which for the ADD model is unity since gravity couples to the standard model universally leading to conserved tensorial current. Notice that both quark and gluon subprocesses are present at the born level for the gravity production in contrary to the standard DY production. Therefore one needs to know the both quark and gluon form factor for gravity production. This has been achieved sometime ago in \cite{Ahmed:2015qia} up to 3-loop. 

The bare quark and gluon form factors satisfy the Sudakov K+G equation which follows as a consequence of the gauge invariance as well as renormalisation group invariance and can be given as,
\begin{align}
\frac{d \ln \F_I}{d\ln q^2} =   \frac{1}{2} \bigg[ {\K}_I(\ash, \frac{\mur^2}{\mu^2},\eps) + {\G}_I(\ash, \frac{q^2}{\mur^2},\frac{\mur^2}{\mu^2},\eps)\bigg] \,.
\end{align}
The function $\K$ contains all the infrared poles in $\eps$ whereas the function $\G$ is finite in the limit $\eps \to 0$. The renormalisation group invariance leads to the following solutions of these functions in terms of cusp anomalous dimensions ($A_I$):
\begin{align}
\frac{d\K_I}{d\ln\mur^2} = - \frac{d\G_I}{d\ln\mur^2}  = \sum_{i=1}^{\infty} \as(\mur) A_I^{(i)} \,.
\end{align}
The cusp anomalous dimensions are known to fourth order \cite{Moch:2004pa,Lee:2016ixa,Moch:2017uml,Grozin:2018vdn,Henn:2019rmi,Davies:2016jie,Lee:2017mip,Gracey:1994nn,Beneke:1995pq,Moch:2017uml,Moch:2018wjh,Henn:2019rmi,Lee:2019zop} (estimate at five loops can be found in \cite{Herzog:2018kwj}) and are collected in App. \ref{app:B2}. The $\mur$ independent piece of the $\G_I$ can be written in perturbative series as
\begin{align}
\G_I(\as(q),\eps) = \sum_{j=1}^{\infty} \as(q) \G_I^{(j)}(\eps)\,,
\end{align} 
where the coefficients $\G_I^{(j)}(\eps)$  can be decomposed as 
\begin{align}
 \G_I^{(i)}(\eps) = 2 \Big(B_I^{(i)} - \gamma_I^{(i)} \Big) + f_I^{(i)} + C_I^{(i)}  + \sum_{k=1}^{\infty} \epsilon^k g_I^{(i,k)}    \,,
\end{align}
where 
\begin{align}
C_I^{(1)}  &=   0  \nn\\
C_I^{(2)}  &= - 2 \beta_0 g_I^{(1,1)}  \nn\\
C_I^{(3)}  &= - 2 \beta_1 g_I^{(1,1)} - 2 \beta_0 \Big( g_I^{(2,1)} + 2 \beta_0  g_I^{(1,2)} \Big) \,.
\end{align}
The coefficients $g_I^{(i,k)}$ can be found from explicit calculation of quark and gluon form factors. These have been calculated at the three loops and are collected in Eq.\ (5.16-5.17) in \cite{Ahmed:2015qia}.

The UV anomalous dimensions $\gamma_I^{(i)} $ are identically zero due to the conservation of QCD energy-momentum tensor as mentioned earlier. Similar to the cusp anomalous dimension, the coefficients $f_I^{(i)}$ have been found to be maximally non-abelian i.e. they satisfy 
\begin{align}
f_{I,g}^{(i)} = \frac{C_F}{C_A} f_{I,q}^{(i)} \,.
\end{align}
In addition they are found to be same as those appear in the quark and gluon form factor up to three loops. 
 
The initial state collinear singularities are removed using the Altarelli-Parisi (AP) splitting kernels $\Gamma_{I} (\hat{a}_s, \mu_f^2, \mu^2, z, \epsilon)$. They satisfy the well-known DGLAP evolution given as,
 \begin{align}
 \frac{d\Gamma_{I} (  z, \muf^2, \eps)}{d\ln \muf^2} = \frac{1}{2} P(z,\muf^2) \otimes \Gamma_{I} (  z, \muf^2, \eps)\,,
 \end{align}
 where $P(z,\muf^2) $ is the AP splitting functions. The perturbative expansion for these splitting functions has the the following form:
 \begin{align}
 P(z,\muf^2) = \sum_{i=0}^{\infty} \as^{i+1}(\muf) P^{(i)}(z) \,.
 \end{align}
 As already discussed, only the $q\bar{q}$ and $gg$ channels contribute to the SV cross-section and thus we find that, only the diagonal terms of the splitting functions contribute to the SV cross-section. The diagonal part of the splitting functions is known to contain the $\delta(1-z)$ and distributions and can be written as,
\begin{align}
P_{II}^{(i)} = 2 \Big[B^{(i+1)}_I \delta(1-z) + A_I^{(i+1)} {\cal D}_0 \Big] + P_{II}^{(reg,i)}(z) \,.
\end{align}
The splitting functions are known exactly to three loops \cite{Vogt:2004mw,Moch:2004pa,Moch:2014sna} and partial results are available for four-loop as well \cite{Davies:2016jie,Moch:2017uml,Moch:2018wjh}. Recently the complete four-loop result is also available completely analytically \cite{Henn:2019swt}.

The finiteness of the soft-virtual cross-section demands that the soft-collinear function $\ph$ will also satisfy similar Sudakov type equation like the form factor i.e. one can write
\begin{align}
\frac{d\ph_I}{d \ln Q^2} =   \frac{1}{2} \bigg[ \KB_I(\ash, z, \frac{\mur^2}{\mu^2},\eps) + \GB_I(\ash, z, \frac{Q^2}{\mur^2},\frac{\mur^2}{\mu^2},\eps)\bigg] \,,
\end{align}
where $\KB_I(\ash, z, \frac{\mur^2}{\mu^2},\eps)$ contains all the poles and $\GB_I(\ash, z, \frac{Q^2}{\mur^2},\frac{\mur^2}{\mu^2},\eps)$ is finite in the dimensional regularization such that $\Psi$ becomes finite as $\eps \to 0$. The solution to the above equation has been found \cite{Ravindran:2005vv,Ravindran:2006cg} to be 
\begin{align}\label{eq:phi-sol}
\ph_I =  \sum_{j=1}^{\infty} \ash^j       \frac{j\eps}{1-z}  \bigg(  \frac{Q^2(1-z)^2}{\mu^2}  \bigg)^{j\eps/2}   {\cal S}_{\eps}^j ~\hat{\ph}_I^{(j)}(\eps) \,.
\end{align}
$\hat{\ph}_I^{(j)}$ can be found from the solution of the form factor by the replacement  as $A_I \to -A_I, \G_I(\eps) \to \GB_I(\eps) $. Notice that $ \GB_I(\eps) $ are now new finite $z$-independent coefficients coming from the soft function whereas the $z$ dependence has been taken out in Eq.\ (\ref{eq:phi-sol}). This can be found by comparing the poles and non-poles terms in $\hat{\ph}^{(j)}$ with those coming from the form factors, overall renormalisation constants, splitting kernel and the lower order SV terms.

Using the following expansion
\begin{align}
\frac{1}{(1-z)} \Big[ (1-z)^2 \Big]^{j\eps/2} = \frac{1}{j\eps} \delta(1-z) + \sum_{k=0}^{\infty} \frac{(j\eps)^k}{k!} {\cal D}_k   \,,
\end{align}
one can finally find the finite soft function $\GB$ as,
\begin{align}
\GB_I^{(i)} = - f_I^{(i)} \delta(1-z) + 2 A_I^{(i)} {\cal D}_0 +    \bar{C}_I^{(i)}    + \sum_{k=1}^{\infty} \eps^k \gb_I^{(i,k)} \,,
\end{align}
where 
\begin{align}
\bar{C}_I^{(1)} &= 0, \nn\\
\bar{C}_I^{(2)} &= -2 \beta_0 \gb_I^{(1,1)}(z), \nn\\
\bar{C}_I^{(3)} &= -2 \beta_1 \gb_I^{(1,1)}(z) - 2 \beta_0 \Big( \gb_I^{(2,1)}(z) + 2 \beta_0 \gb_I^{(1,2)}(z)    \Big) \,. 
\end{align}
It is worth noting that $\GB$ as well as the complete soft function $\ph_I$ satisfy the maximally non-abelian property up to three loops. Moreover $\ph_I$ is also universal in the sense that it only depends on the initial legs and is completely unaware of the color neutral final state. Up to three loops all the coefficients are known for quark and gluon initiated processes \cite{Anastasiou:2014vaa,Ahmed:2014cla}.

Finally plugging all these functions and coefficients into the Eq.\ (\ref{sv:psi}) and expanding in the powers of $\as(\mur)$, we obtain the soft-virtual cross-section up to third order.  The born level results are trivial and presented below,
\begin{align}
\Delta_{\rm q\bar{q}}^{(0),DY} &= \frac{2\pi}{n_c} \delta(1-z)\,, \nn\\
\Delta_{\rm q\bar{q}}^{(0),G}&=  \frac{\pi}{8 n_c} \delta(1-z)\,, \nn\\
\Delta_{\rm gg}^{(0),G} &= \frac{\pi}{2 (n_c^2-1)} \delta(1-z) \,.
\end{align}
The results up to two loops are also available in \cite{deFlorian:2013wpa}. The new three-loop results are calculated here for the first time and collected in the App.\ \ref{app:A}.

\subsection{Resummation}
The inclusive cross-section can also be improved with the inclusion of threshold enhanced logarithms by resumming them to all orders. These threshold logarithms arise from soft and collinear emissions from virtual and real diagrams. The leading contribution arises from the most singular soft-virtual terms containing plus distributions which can be resummed to all orders in a systematic way. The resummation is conveniently performed in Melin-N space where the threshold limit $z\to1$ translates into large-N limit i.e. $N\to \infty$. In the Mellin space, the large-N behavior of the born normalized partonic cross-section at all orders can be organized \cite{Catani:1996yz,Catani:1989ne,Moch:2005ba} as,
\begin{align}
(d\hat{\sigma}_N/dQ)/(d\hat{\sigma}_{\rm LO}/dQ) = g_0^I \exp \Big( G_N^I \Big) \,,
\end{align}
$(d\hat{\sigma}_{\rm LO}/dQ)$ contains the born normalisation i.e. for the SM DY,
\begin{align}
(d\hat{\sigma}_{\rm LO}/dQ) = {\cal F}^{(0)}_{\rm DY} \bigg\{ \frac{2\pi}{n_c}\bigg\} \,,
\end{align}
whereas for ADD,
\begin{align}
(d\hat{\sigma}_{\rm LO}/dQ) = {\cal F}^{(0)}_{\rm ADD} \bigg\{ \frac{\pi}{8 n_c}, \frac{\pi}{2(n_c^2-1)}\bigg\} \qquad \text{for} ~~ q\bar{q}~ \text{and} ~gg~ \text{respectively.}
\end{align}
The exponent $G_N^I$ resums large-N terms at all orders and is given in terms of universal cusp anomalous dimensions $A$ and constants $D$  and has the following form,
\begin{align}
G_N^I = \int_0^1 dz \frac{z^{N-1} - 1}{1 - z} \bigg[ \int_{\muf^2}^{Q^2(1-z)^2} \frac{d\omg^2}{\omg^2} 2~A(a_s(\omg^2)) + D(a_s(q^2(1-z)^2))\bigg] \,.
\end{align}
%
$G_N^I$ can be also written in a resummed perturbative series. Recalling that in the context of resummation $a_s \ln \bar{N}  \sim {\cal O}(1)$, one can write,
\begin{align}\label{eq:resum-exponent}
G_N^I = \ln \bar{N} ~g_1^I + g_2^I + a_s ~g_3^I + a_s^2~ g_4^I \,,
\end{align}
where $\bar{N} = N\exp(\gamma_e)$ with $\gamma_e = 0.577216\dots$ is the Euler Gamma. Successive terms in the above expression determines the logarithmic accuracy. For example, the first coefficient ($g_1^I$) resums all leading logarithms (LL) at all orders, whereas the first two coefficients ($g_1^I + g_2^I$) also resums next to leading logarithms (NLL) and so on. Note that the universality of the resummed exponent is a direct consequence of the soft-gluon emission near the partonic threshold. The exponent is thus universal in the sense that it will only depend on the initial legs being gluons or quarks. The expressions for the resummed exponents $g^I_i$ can be found in \cite{Moch:2005ba,Catani:2003zt} up to N$^3$LL order, also see \cite{das:dis} for N$^4$LL order in DIS. For consistency, we have also  derived the same and collected in the App.\ \ref{app:B2}.

The process dependent coefficient $g_0^I$ on the other hand depends on the specific process under consideration. It gets contribution from the entire form factor as well as from the $\delta(1-z)$ coming from the soft part. It can be also written as a perturbative series as,
\begin{align}
g_0^I = 1+ a_s g_{01}^I + a_s^2 g_{02}^I + a_s^3 g_{03}^I + \dots \,.
\end{align}
For the quark initiated spin-2 production and gluon initiated spin-2 production we have extracted those from the soft-virtual results up to the third order in the strong coupling. These are collected in App.\ \ref{app:B2}. We again remind the reader that for NLL accuracy one needs coefficient $g_{01}$ in the above expansion, at NNLL one needs up to $g_{02}$ and so on. The resummed expression in the Mellin space has to be finally inverse Mellin transformed and matched with the fixed order result. We follow the standard Minimal prescription \cite{Catani:1996yz} to take care of the Landau pole issue in the Mellin inversion routine. The matched cross-section has the following form,
\begin{align}\label{Eq:matched}
\bigg[\frac{d\sigma}{dQ}\bigg]_{N^nLL+N^mLO} &= \frac{Q}{S}
\sum_{ab\in\{q,\bar{q},g\}}
 \frac{d\hat{\sigma}_{LO}}{dQ}  \int_{c-i\infty}^{c+i\infty} \frac{dN}{2\pi i} (\tau)^{-N} \delta_{ab}f_{a,N}(\muf^2) f_{b,N}(\muf^2) \nn\\
&\times \bigg( \bigg[ \frac{d\hat{\sigma}_N}{dQ} \bigg]_{N^nLL} - \bigg[ \frac{d\hat{\sigma}_N}{dQ} \bigg]_{tr}     \bigg)+ \bigg[\frac{d\sigma}{dQ}\bigg]_{N^mLO}  \,.
\end{align}
The second term inside the bracket has been introduced to remove double counting of singular terms which are already present in the FO result i.e. in the last term of the above expression. In particular, for N$^3$LL matching with NNLO, we need the resummed expression keeping up to ${\cal O}(a_s^2)$ terms in the resummed exponent Eq.\ (\ref{eq:resum-exponent}) and subtracting all the leading singular terms that are present in  the NNLO cross-section through subtracting the expanded resummed cross-section up to that order. The matched formula in Eq.\ (\ref{Eq:matched}) thus gives opportunity to match different orders in FO and resum series. In the next section we will improve the existing NNLO cross-section by resumming large threshold logarithms to N$^3$LL accuracy by matching the later to NNLO results. In the next section we study the phenomenological effect of SV cross-section and resummed prediction for ADD model.

\section{Numerical Results}\label{sec:numerical}
In this section we present our numerical results for three loop soft-virtual QCD correction to the di-lepton production in the ADD model at LHC. The LO, NLO and NNLO parton level cross sections are convoluted with the respective order by order parton distribution function (PDF) taken from {\tt lhapdf} \cite{Buckley:2014ana}. However, for N$^3$LO$_{\rm sv}$ corrections we convoluted the partonic coefficient functions with the NNLO PDFs due to the unavailability of N$^3$LO PDFs. The corresponding strong coupling constant $a_s(\mu_r^2) = \alpha_s(\mu_r^2)/(4\pi)$ is also provided by the {\tt lhapdf}. The fine structure constant is taken to be $\alpha_{\rm em} = 1/128$ and the weak mixing angle is $\sin^2\theta_w = 0.227$. Here the results are presented for n$_f = 5$ flavors  in the massless limit of quarks.  The default choice for the center of mass energy of LHC is 13 TeV and the choice for the PDF set is MMHT2014. Except for the scale variations, we have used the factorization ($\mu_f$) and  renormalisation ($\mu_r$) scales to be the invariant mass of the di-lepton, i.e. $\mu_f = \mu_r = Q$.  We also note that there have been several experimental searches at the LHC for extra dimensions in the past, yielding stringent bounds on the ADD model parameters, the cut-off scale $M_s$ and the number of extra dimensions $d$. Such analyses have already used the K-factors that have been computed in the extra dimension models. There are several experimental data available regarding the  lower bound of the model parameters $M_S$ and $d$. The lower limits on the scale $M_S$ obtained from both  ATLAS and CMS collaborations using $7$ TeV data \cite{Aad:2012bsa,Chatrchyan:2012kc} are $M_S = 2.4$ TeV  corresponding to $d = 3$ in HLZ formalism \cite{Han:1998sg}. After the availability of $8$ TeV data this lower bound further pushed to $M_S = 3.3$ TeV for $d = 3$ \cite{Aad:2014wca,Khachatryan:2014fba}. Now $13$ TeV data are also available and the bound in $M_S$ is given by ATLAS is $5.5$ TeV using di-photon channel \cite{Aaboud:2017yyg}. CMS collaboration also studied the same and the lower bounds are found to be $5.6$ TeV for di-lepton channel \cite{Sirunyan:2018ipj} and $5.7$ TeV for di-photon channel \cite{Sirunyan:2018wnk}. Here in our work, for our phenomenological study to assess the impact of QCD corrections,  we choose $M_S = 4$ TeV and $d = 3$. The computational details of the QCD corrections presented here are model independent,  a numerical estimate of the theory predictions for  any other choice of the model parameters is straight-forward. For completeness, we also study the dependence of the invariant mass distributions on the model parameters considering the recent bounds on $M_S$ for different extra dimensions.
\subsection{Threshold corrections up to N$^3$LO$_{\rm sv}$}
First, we will present in Table-\ref{tabled} the relative contributions from different logarithmic terms $D_i$ as well as the $\delta(1-z)$ term with respect to $\mathcal{D}_5$ to the invariant mass distribution of the di-lepton at $a_s^3$ level. 
%
\begin{table}[h!]
\begin{center}
{\scriptsize
\resizebox{15.2cm}{1.4cm}{%
	\begin{tabular}{ |p{1.0cm}|p{2.0cm}|p{1.0cm}|p{1.0cm}|p{1.0cm}|p{1.0cm}|p{1.0cm}|p{1.0cm}|p{1.0cm}|p{1.0cm}| }
\hline
	Q (GeV) & $(+)\mathcal{D}_{5}$ (pb/GeV) & $(+)\frac{\mathcal{D}_{0}}{\mathcal{D}_{5}}$  & $(-)\frac{\mathcal{D}_{1}}{\mathcal{D}_{5}}$ & $(-)\frac{\mathcal{D}_{2}}{\mathcal{D}_{5}}$ & $(-)\frac{\mathcal{D}_{3}}{\mathcal{D}_{5}}$ & 
$(+)\frac{\mathcal{D}_{4}}{\mathcal{D}_{5}}$ 
&  $(-)\frac{\sum\mathcal{D}_{i}}{\mathcal{D}_{5}}$ & $(+)
	\delta /\mathcal{D}_{5}$ & tot /$\mathcal{D}_{5}$
\\
\hline
\hline
	$100$ & $0.3560\times~10^{-8}$ & $0.0552$ & $0.1635$ & $0.5890$ & $0.6312$ & $0.2143$ & $0.1144$ & $0.1036$ & $-0.0108$\\
\hline
	$1000$ & $0.2002\times~10^{-5}$ & $0.0398$ & $0.1447$ & $0.5584$ & $0.6159$ & $0.2157$ & $0.0632$ & $0.0466$  & $-0.0166$\\
\hline
	$2000$ & $0.5106\times~10^{-5}$ & $0.0333$ & $0.1378$ & $0.5445$ & $0.6054$ & $0.2207$ & $0.0334$ & $0.0340$  & $+0.0006$\\  
\hline
	$3000$ & $0.6431\times~10^{-5}$ & $0.0284$ & $0.1328$ & $0.5332$ & $0.5951$ & $0.2273$ & $0.0053$ & $0.0269$  & $+0.0216$\\  
\hline
\end{tabular}
}
\caption{Contribution of large logarithms, the constant term $\delta(1-z)$ and the total SV correction to the di-lepton
invariant mass distribution at $3$-loop level in the ADD model for $13$TeV LHC.}
\label{tabled}
}
\end{center}
\end{table}

%
\noindent
The $\delta(1-z)$ terms are process dependent and need explicit computation while the $\mathcal{D}_i$ can be predicted from the universal nature of the infrared structures in QCD as well as the lower order process dependent contributions. We note that the sub-leading logarithms $\mathcal{D}_3$ and $\mathcal{D}_2$ are negative and are comparable in magnitude to the leading logarithmic $\mathcal{D}_5$ contribution. As a result, the contribution from the sum of logarithmic terms is negative but comparable in magnitude to that of $\delta(1-z)$ term. Consequently, the sign of total soft-plus-virtual (SV) correction at three-loop level i.e. $a_s^3 \Delta^{(3),\text{G}}_{ab}$ crucially depends on the relative weightage of these two kind of terms. It can be seen that SV contribution is negative at lower $Q(\sim 100 \text{GeV})$ but becomes positive for $Q(> 2000 \text{GeV})$.
\begin{figure}[h!]
  \centering

                   {\includegraphics[trim=0 0 0 0,clip,width=0.48\textwidth,height=0.40\textwidth]
                   {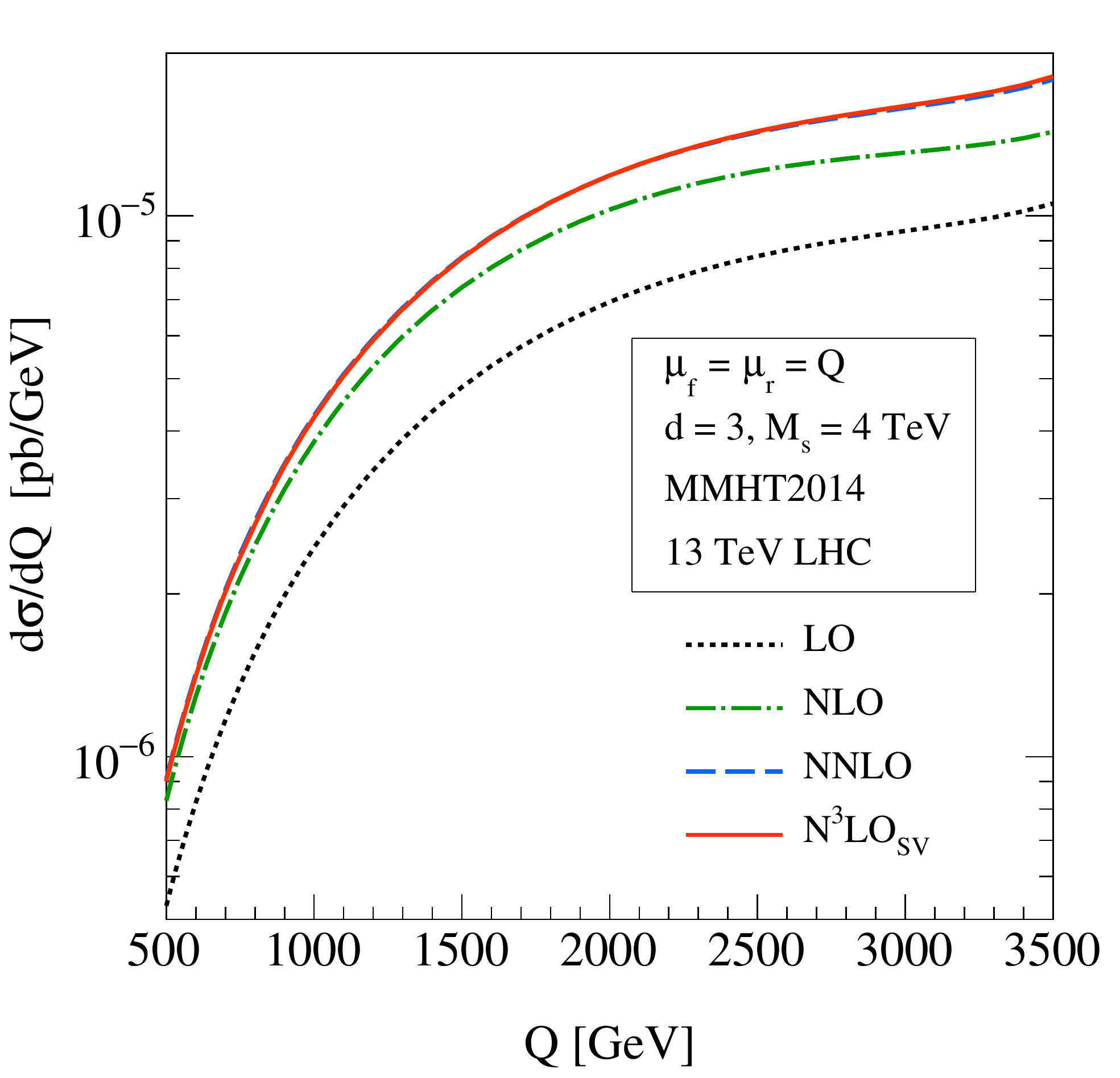}\label{fig:3}}
  \hskip0.05cm   
                   {\includegraphics[trim=0 0 0 0,clip,width=0.48\textwidth,height=0.40\textwidth]
                   {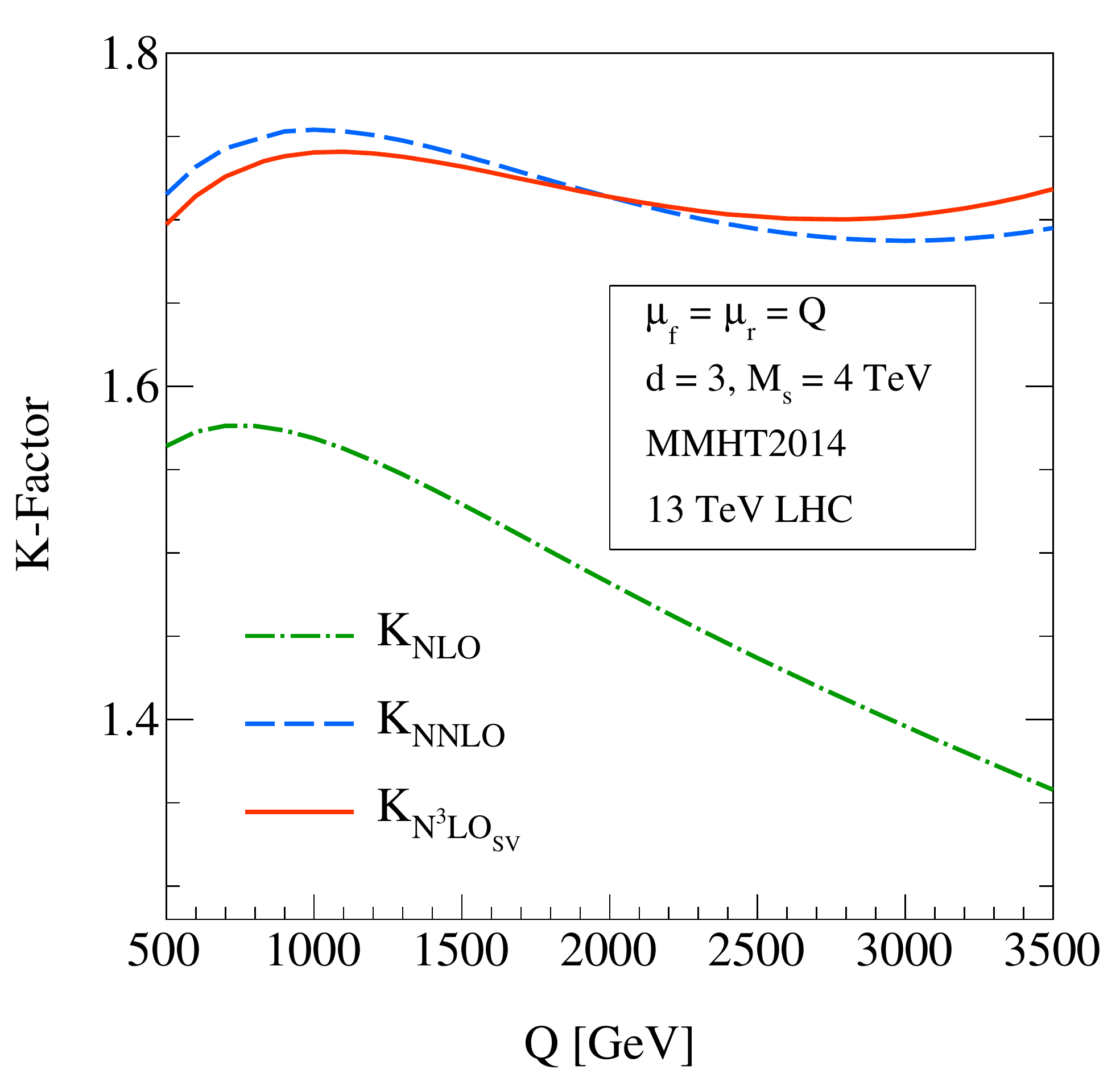}\label{fig:4}}\\

                   {\includegraphics[trim=0 0 0 0,clip,width=0.48\textwidth,height=0.40\textwidth]
                   {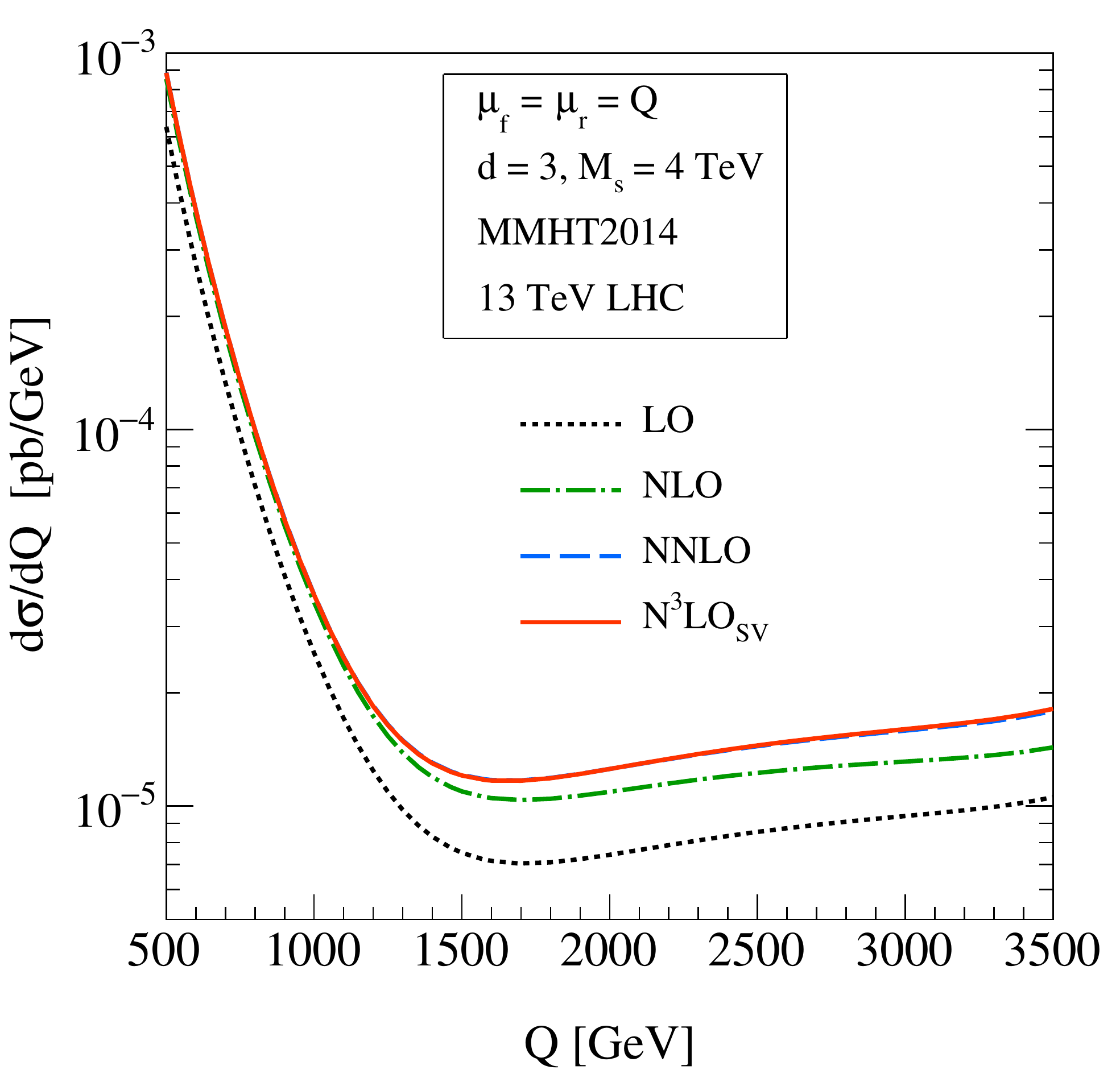}\label{fig:5}}
  \hskip0.05cm
                   {\includegraphics[trim=0 0 0 0,clip,width=0.48\textwidth,height=0.40\textwidth]
                   {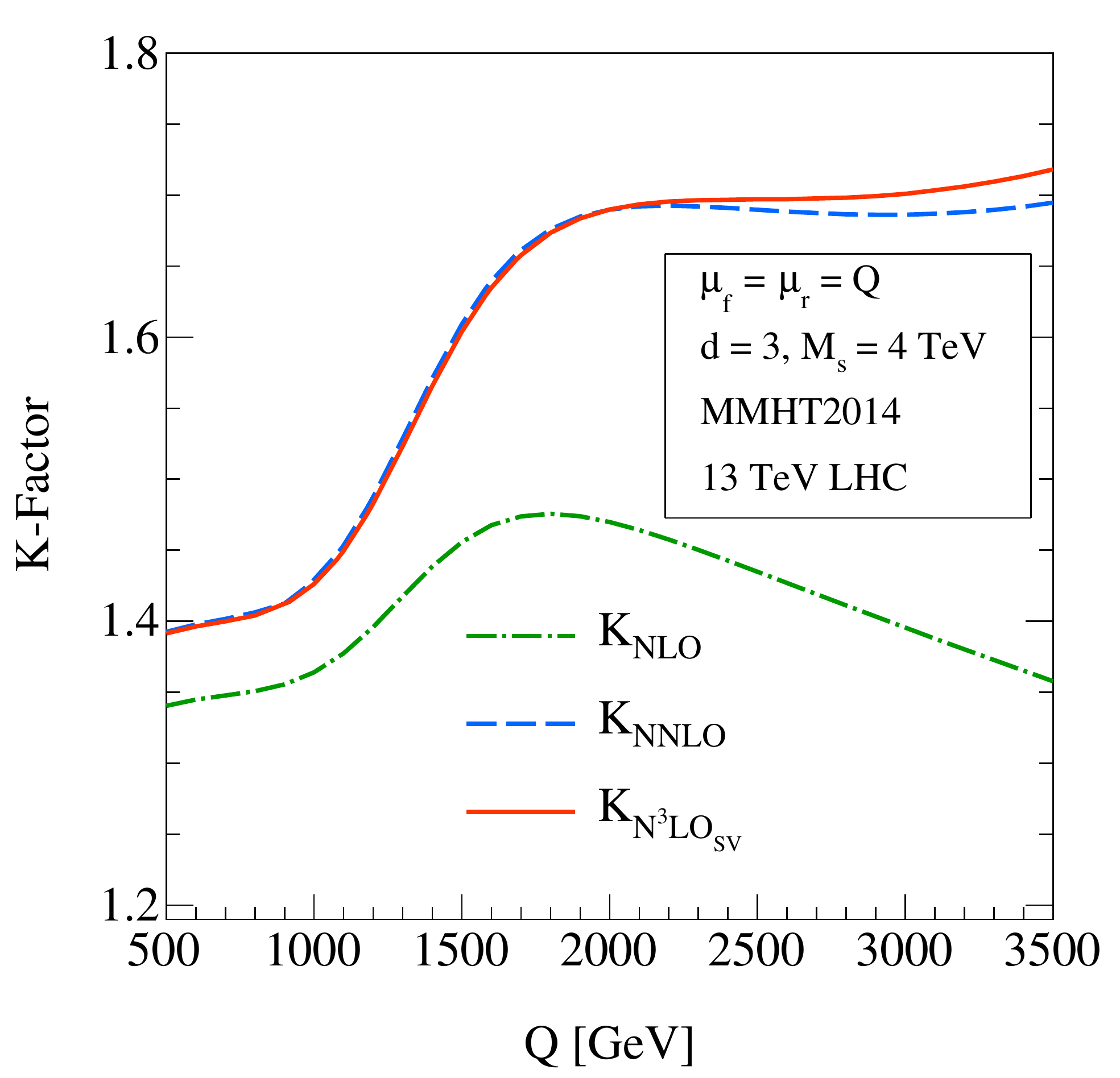}\label{fig:6}}		   
      
\caption{Invariant mass distribution of di-lepton at hadronic center of mass energy $13$ TeV for ADD model and signal (left panel from top to bottom) and their corresponding K factor on the right panel. (from top to bottom)}
\label{fig:order}
\end{figure}
Next, in Fig.(\ref{fig:order}) we present the di-lepton invariant mass distribution for the pure ADD model (GR) case and the signal (SM+GR), along with the corresponding K-factors to N$^3$LO$_{\text{sv}}$ in QCD. The NLO corrections in the high $Q$-region around $Q=2500$ GeV contribute by about 40\% of LO, while NNLO corrections add an additional $25\%$ of LO to the total invariant mass distribution. The NNLO corrections are too large enough to truncate the perturbation theory at this order and necessitates the computation of higher order corrections for the convergence of the perturbation series. The three-loop SV corrections that we have computed here are found to contribute an additional $(1-2)\%$ of LO to the invariant mass distribution,  demonstrating a very good convergence of the perturbation theory. We also note that the three-loop SV corrections are negative in the low $Q$-region while in the high $Q$-region they are positive because of threshold enhancement.
\begin{figure}[h !!!!]
  \centering
                   {\includegraphics[trim=0 0 0 0,clip,width=0.48\textwidth,height=0.40\textwidth]
                   {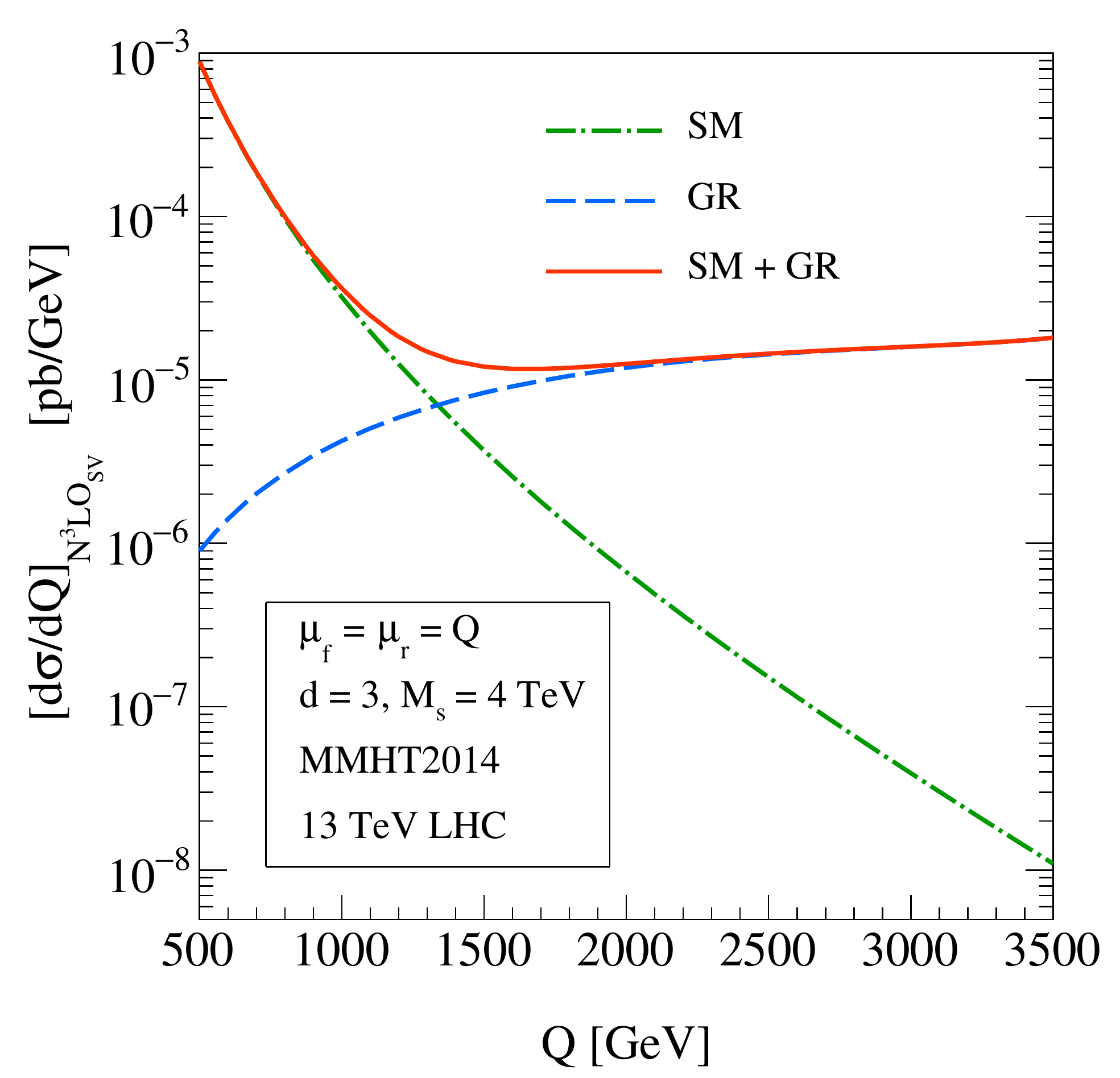}\label{fig:7}}
  \hskip0.05cm
                   {\includegraphics[trim=0 0 0 0,clip,width=0.48\textwidth,height=0.40\textwidth]
                   {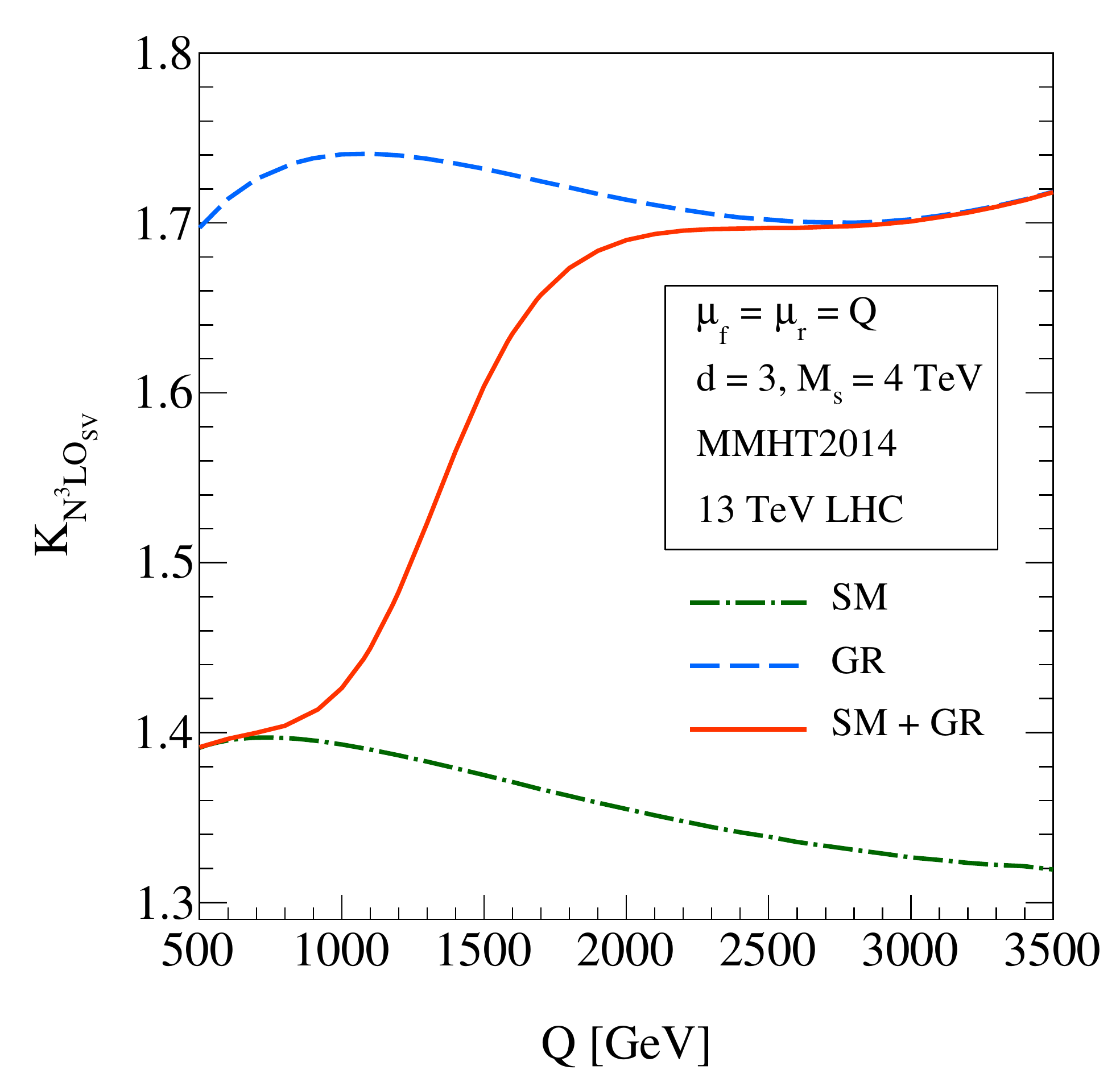}\label{fig:8}}
                  
 \caption{Invariant mass distribution (left panel) of di-lepton at hadronic center of mass energy $13$ TeV for SM, ADD,signal and the corresponding K factors (right panel) at N$^3$LO$_{\rm sv}$ level.}
\label{fig:model}
\end{figure}
In Fig.(\ref{fig:model}) we present invariant mass distributions (left panel) and the corresponding K-factors (right panel) for the SM background, GR and the signal up to N$^3$LO$_{\rm sv}$ in QCD. At low $Q$ values of less than 800 GeV most of the signal contribution is coming from SM and as we go to high $Q$ value the GR contribution starts to dominate as the number of accessible KK modes will increase with $Q$.  Therefore the signal K factor at high $Q$ value is completely dominated by ADD model which receives contributions from both quark-anti-quark annihilations as well as gluon fusion channel even at LO in contrast to the SM case where there is only quark-anti-quark annihilation at LO. This results in larger K-factors for the signal compared to those of the SM background, 
\begin{equation}
	\begin{aligned}	
\text{K}_\text{NLO} = \frac{ d\sigma^{\rm NLO}/dQ} {d\sigma^{\rm LO}/dQ } \quad 
\text{K}_\text{NNLO} = \frac{ d\sigma^{\rm NNLO}/dQ} {d\sigma^{\rm LO}/dQ } \quad 
		\text{K}_\text{N$^3$LO$_{\rm sv}$} = \frac{ d\sigma^{\text{N$^3$LO$_{\text{sv}}$}}/dQ} {d\sigma^{\rm LO}/dQ }.
	\end{aligned}
\label{eqk}
\end{equation}
In Eq.(\ref{eqk}) we define K-factors for the signal at different orders in QCD. In Table-\ref{tablek} we present these K-factors as a function of the invariant mass of the di-lepton. As the three-loop SV corrections change sign for higher $Q$ values as mentioned above, the signal K-factors at N$^3$LO level ($\text{K}_\text{N$^3$LO$_{\rm sv}$}$)  are smaller (larger) than $\text{K}_\text{NNLO}$ for about $Q<2000$ ($Q>2000$) GeV.
%
\begin{table}
\begin{center}
{\scriptsize
\resizebox{5.5cm}{3.5cm}{%
\begin{tabular}{ |p{0.8cm}|p{0.8cm}|p{0.8cm}|p{0.8cm}| }
\hline
	Q(GeV) & K$_{\text{NLO}}$  & K$_{\text{NNLO}}$ & K$_{\text{N$^3$LO$_{\rm sv}$}}$  \\
\hline
\hline
$200$  &   $1.298$  &   $1.340$  &   $1.341$   \\
\hline
$400$  &   $1.333$ &    $1.384$  &   $1.383$   \\
\hline
$600$  &   $1.345$  &   $1.398$  &   $1.396$   \\
\hline
$800$  &   $1.351$  &   $1.406$  &   $1.404$   \\
\hline
$1000$  &   $1.364$  &   $1.429$  &   $1.426$  \\
\hline
$1200$  &   $1.396$  &   $1.488$  &   $1.483$  \\
\hline
$1400$  &   $1.439$  &   $1.571$  &   $1.566$  \\
\hline
$1600$  &   $1.468$  &   $1.640$  &   $1.635$  \\
\hline
$1800$  &   $1.476$  &   $1.676$  &   $1.674$  \\
\hline
$2000$  &   $1.470$  &   $1.690$  &   $1.690$  \\
\hline
$2200$  &   $1.458$  &   $1.693$  &   $1.696$  \\
\hline
$2400$  &   $1.443$  &   $1.691$  &   $1.697$  \\
\hline
$2600$  &   $1.427$  &   $1.688$  &   $1.697$  \\
\hline
$2800$  &   $1.411$  &   $1.687$  &   $1.698$  \\
\hline
$3000$  &   $1.396$  &   $1.686$  &   $1.701$  \\
\hline
\end{tabular}
}}
\caption{Fixed order K-factors for the signal of the di-lepton invariant mass distribution at the LHC up to N$^3$LO$_{\rm sv}$.}
\label{tablek}
\end{center}
\end{table}

\begin{figure}[h !!!!]
  \centering
                   {\includegraphics[trim=0 0 0 0,clip,width=0.48\textwidth,height=0.40\textwidth]
                   {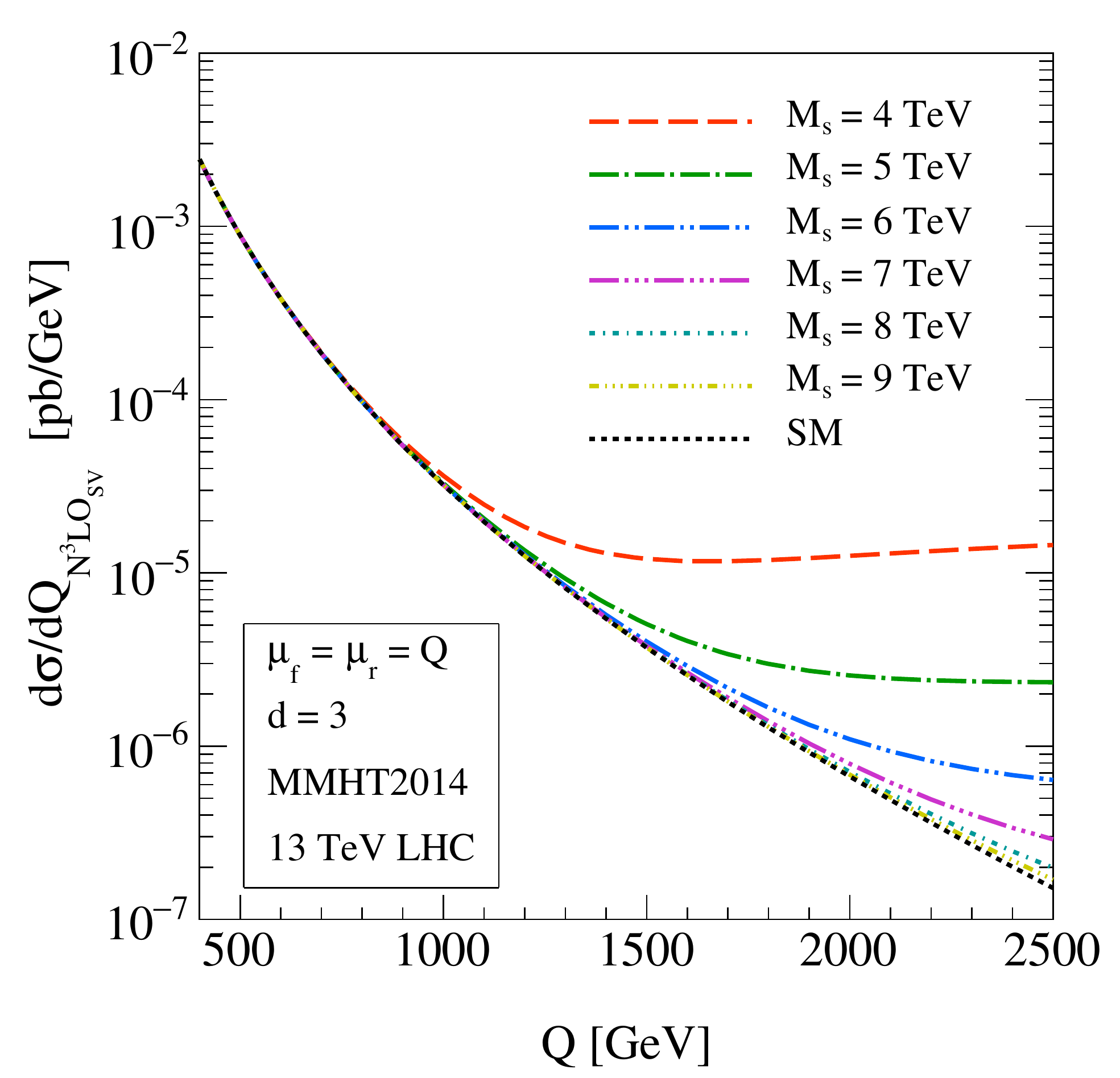}\label{fig:9}}
  \hskip0.05cm
                   {\includegraphics[trim=0 0 0 0,clip,width=0.48\textwidth,height=0.40\textwidth]
                   {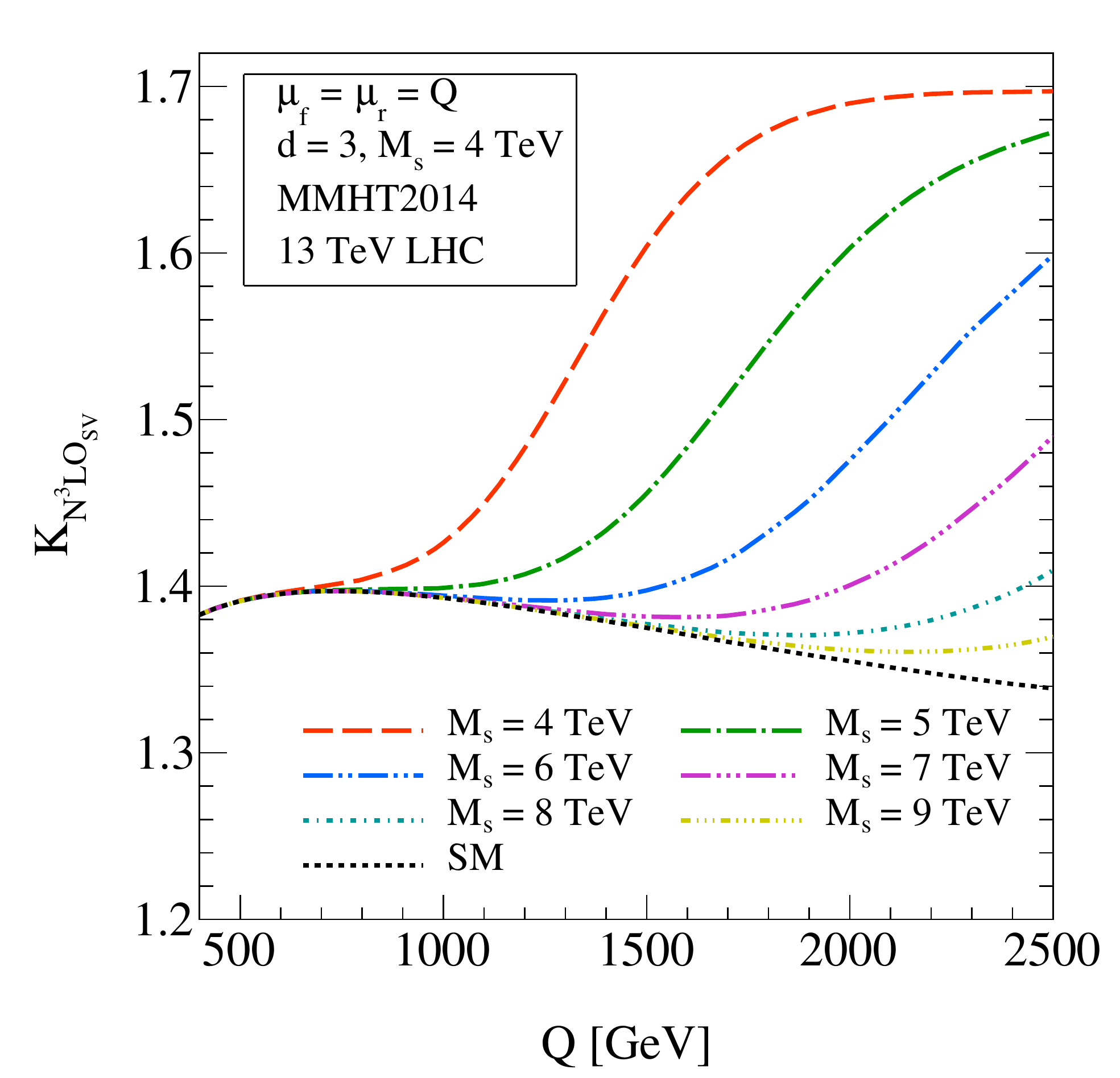}\label{fig:10}}
                  
 \caption{Invariant mass distribution of di-lepton at hadronic center of mass energy $13$ TeV for signal with $d = 3$ and different $M_S$ value and their corresponding K-factors on the right panel at N$^3$LO$_{\rm sv}$ level.}                   
\label{fig:ms-var}
\end{figure}

\begin{figure}[h !!!!]
  \centering
                   {\includegraphics[trim=0 0 0 0,clip,width=0.48\textwidth,height=0.40\textwidth]
                   {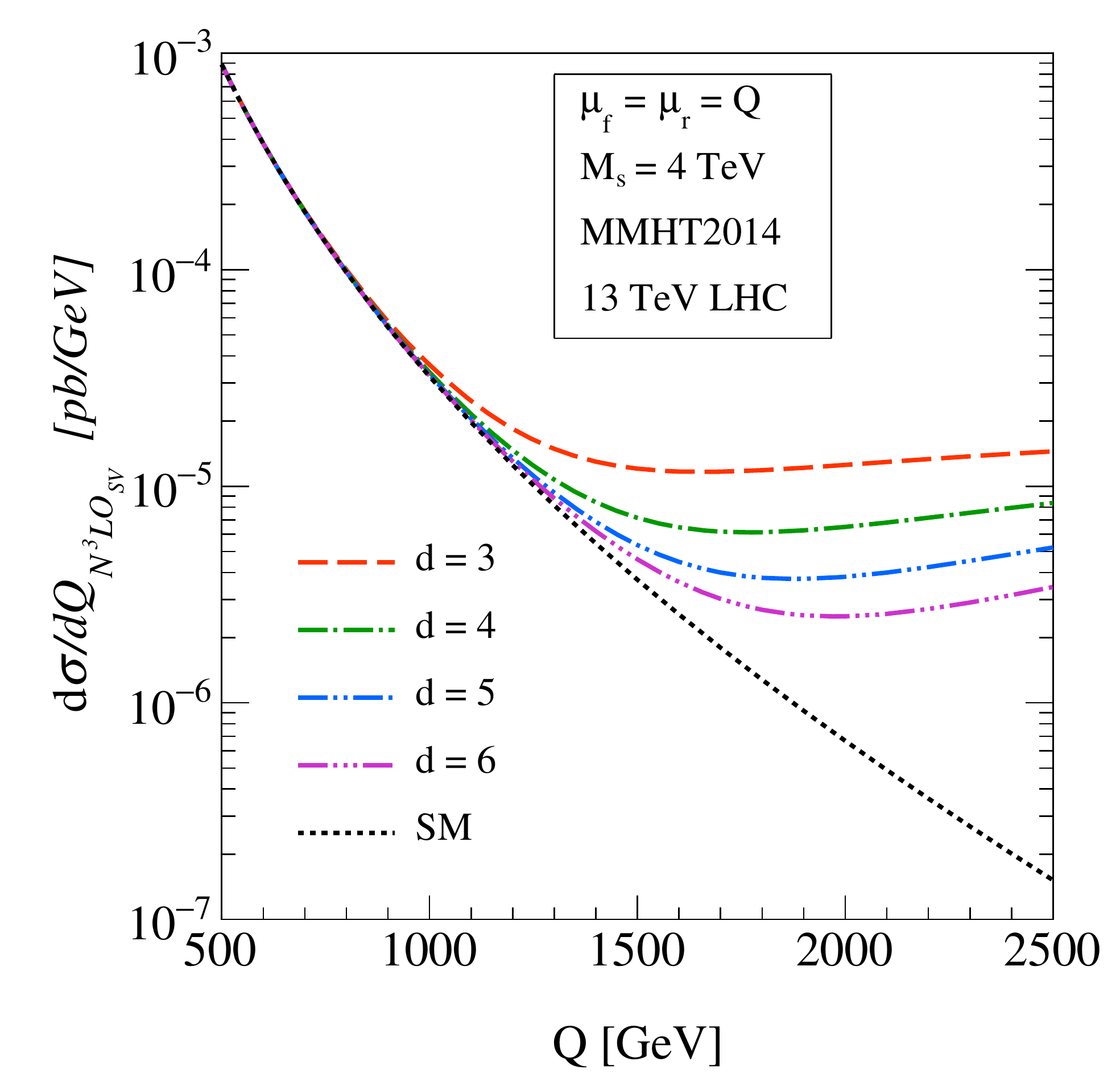}\label{fig:11}}
  \hskip0.05cm
                   {\includegraphics[trim=0 0 0 0,clip,width=0.48\textwidth,height=0.40\textwidth]
                   {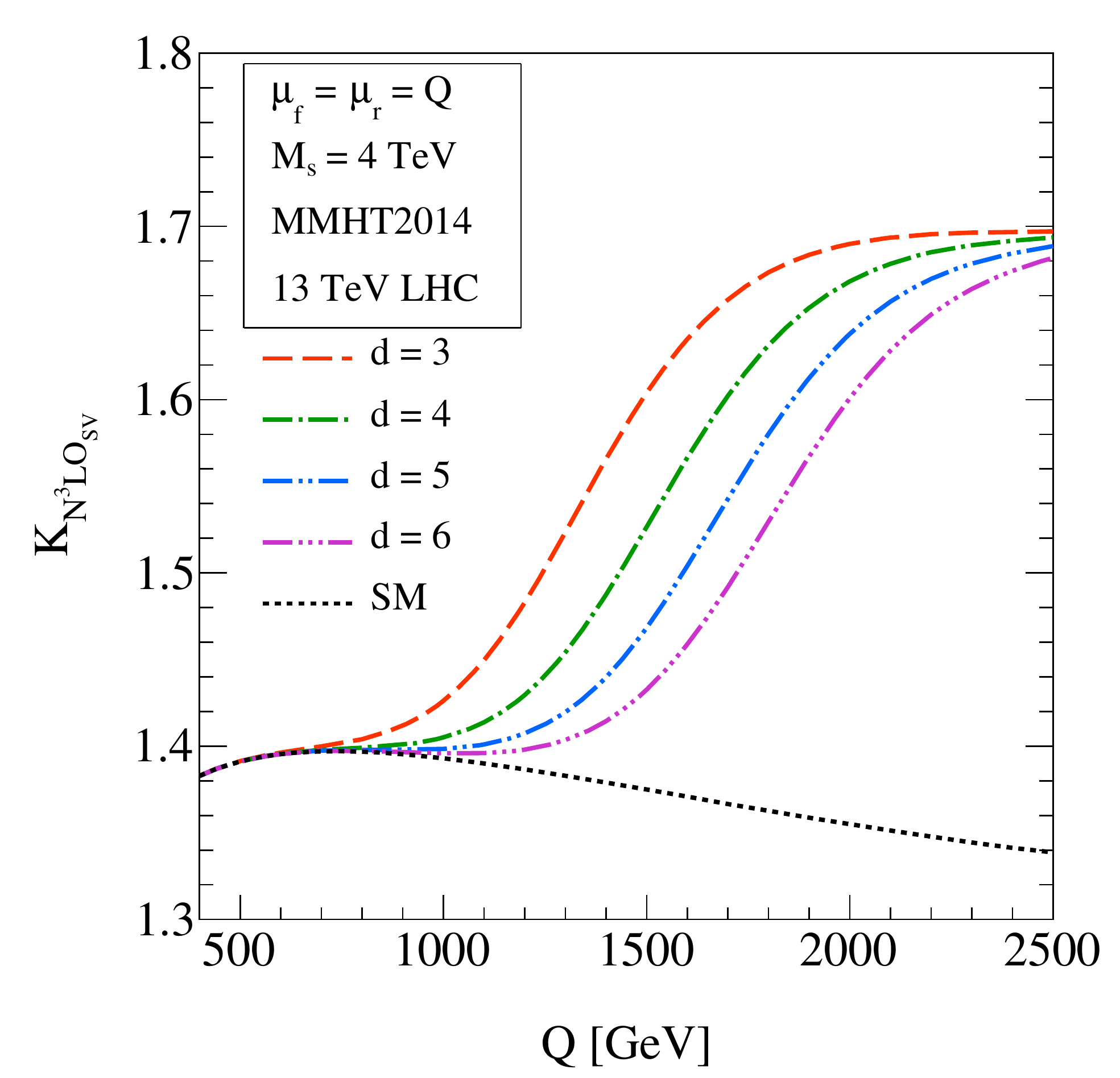}\label{fig:12}}
                  
 \caption{Invariant mass distribution of di-lepton at hadronic center of mass energy $13$ TeV for signal with $M_S = 4$ and different $d$ value and their corresponding K factor on the right panel at N$^3$LO$_{\rm sv}$ level.}                   
\label{fig:d-var}
\end{figure}

We also study the dependence of our results on the ADD model parameters namely the scale $M_S$ and the number of extra dimensions $d$. In Fig.(\ref{fig:ms-var}) we present the invariant mass distribution (left) and the corresponding K-factors (right) for different values of $M_S$ keeping $d=3$ fixed. From the figure, we can see that the invariant mass distribution decreases with increase in $M_S$ for any given value of $Q$ and $d$ simply because of the scale $M_S$ suppression in the gravity propagator.
Similarly, we present in Fig.(\ref{fig:d-var}) the invariant mass distribution (left) and the relevant K-factors (right) for different values of $d$ keeping $M_S=4$ TeV fixed. From the Fig.(\ref{fig:d-var}) we can see that the cross section decreases with the number of extra dimensions $d$ because of the fact that the mass of the graviton mode increases with increasing $d$ resulting in the less number of accessible graviton modes. 

\begin{figure}[h !!!!]
  \centering
                   {\includegraphics[trim=0 0 0 0,clip,width=0.70\textwidth,height=0.4\textwidth]
                   {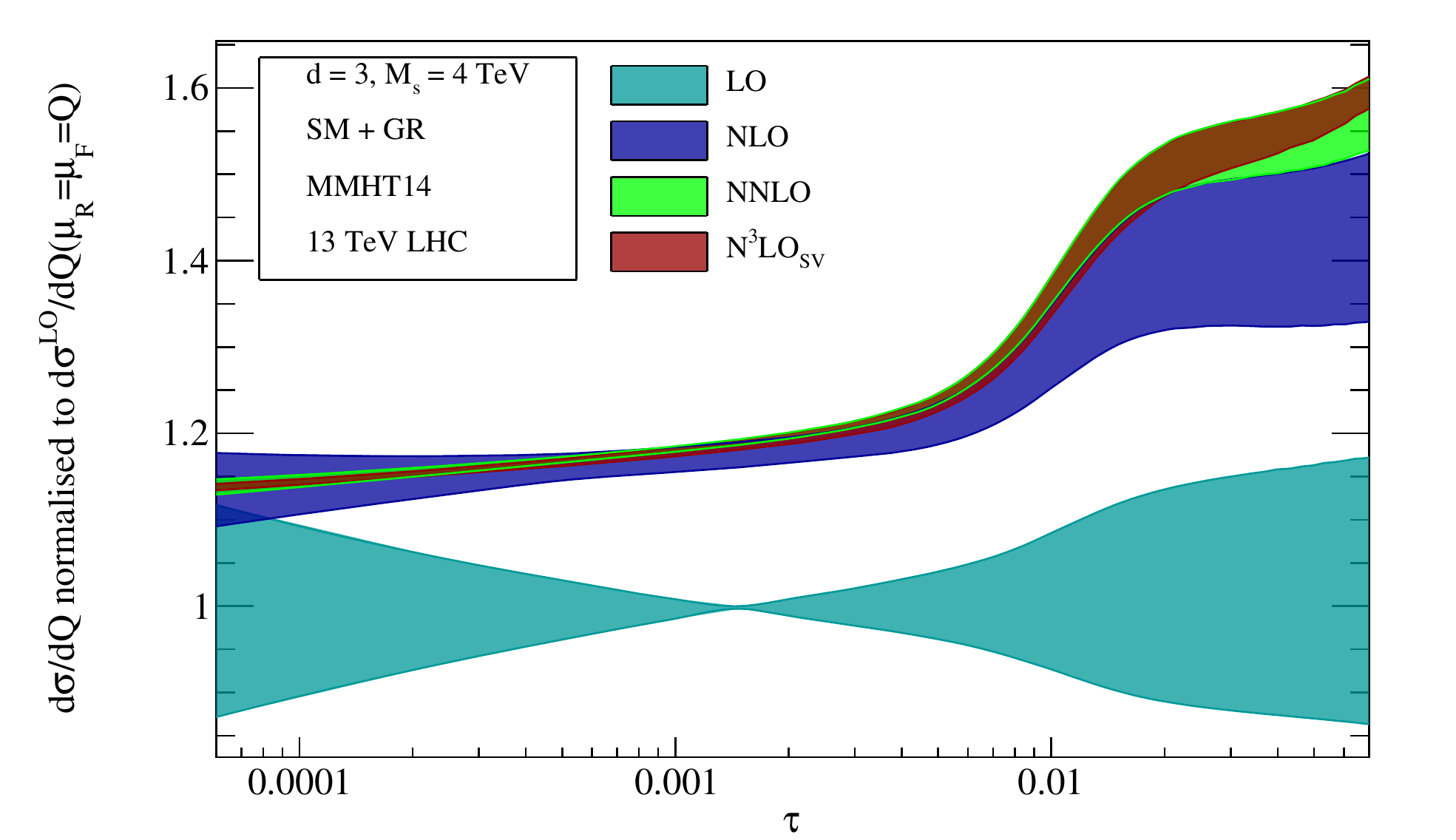}\label{fig:13}}
                   
 \caption{Seven point scale variation of invariant mass distribution of di-lepton at hadronic center of mass energy $13$ TeV for signal with $M_S = 4$ and $d = 3$.}                   
\label{fig:scale_fo}
\end{figure}

We have considered different sources of theoretical uncertainties in our analysis. Firstly we considered the uncertainties due to the presence of two unphysical scales $\mu_r$ and $\mu_f$ in the theory and secondly those coming from the non-perturbative  parton distribution function in the calculation. For the scale uncertainties we vary $\mu_r$ and $\mu_f$ simultaneously from $Q/2$ to $2Q$ by putting the constraint that the ratio of unphysical scales is less than $2$, as
\begin{align}
	\Big|\text{ln}\frac{\mu_r}{Q}\Big| \leq \text{ln }2, \quad \Big|\text{ln}\frac{\mu_f}{Q}\Big| \leq \text{ln }2,	
	\quad \Big|\text{ln}\frac{\mu_r}{\mu_f}\Big| \leq \text{ln }2.
	\label{eqscale}	
\end{align}	
The last condition in Eq.\ (\ref {eqscale}) ensures that no unusual choice of the scales is considered. This results in $7$ different combinations of the scale {\it viz.} $\Big(\mu_r/Q, \mu_f/Q \Big) = (1/2, 1/2), (1/2, 1), (1, 1/2)$, $(1, 1), (2, 1), (1, 2), (2, 2)$. With this choice, we estimate the $7$-point scale uncertainties in our predictions to N$^3$LO$_{\rm sv}$ and the results are depicted in Fig.(\ref{fig:scale_fo}). The upper and lower band of a particular order respectively corresponds to the maximum and minimum values of the invariant mass distributions normalized by LO computed with the default choice of scales. These normalized distributions are obtained by taking the order by order PDFs for both the numerator and the denominator. The scale uncertainties are found to get reduced significantly from LO to N$^3$LO$_{\rm sv}$. For example at $Q=2500$ GeV, the scale uncertainties at LO are 28\%, at NLO they are 18\%, at NNLO 7\% and at N$^{3}$LO$_{\rm sv}$ they are 5\%. For $Q=3000$ GeV, the scale uncertainties reduce from 30\% at LO to  about 4\% at N$^{3}$LO$_{\rm sv}$. It is expected that the scale uncertainties get significantly reduced with the inclusion of missing process dependent regular terms at $a_s^3$ level, as well as the convolution with the N$^3$LO level PDFs that are yet to be available.

We also estimate the uncertainties coming from the non-perturbative PDFs. For this we calculate the uncertainty in two different ways, (i) the uncertainty due to the intrinsic error in the PDFs that result from various experimental errors from the global fits, (ii) the uncertainty due to the choice of PDFs provided by different groups. In both the cases we use the PDF sets MMHT2014, CT14, NNPDF31, AMMP16 and PDF4LHC15 provided from the lhapdf.
For the case-(i) we calculate the intrinsic PDF uncertainties using $51$ sets for MMHT2014, $57$ sets for CT14, $101$ sets for NNPDF31, $30$ sets for ABMP16 and $31$ sets for PDF4LHC15. To this end we use all PDF sets extracted at NNLO level. In Table-\ref{tpdf2} we present these uncertainties for the di-lepton invariant mass distribution to N$^3$LO$_{\rm sv}$.
%
\begin{table}[h!]
\begin{center}
\scriptsize{	
\begin{tabular}{| *{7}{c|} }
       \hline	
	{} & \multicolumn{2}{c|}{\% of Uncertainty at $Q = 100$GeV} &\multicolumn{2}{c|}{\% of Uncertainty at $Q = 1000$GeV} & 
	\multicolumn{2}{c|}{\% of Uncertainty at $Q = 2500$GeV} \\

    \cline{2-3} \cline{4-5} \cline{6-7}
	PDF Name & N$^3$LO$_{\rm sv}$   &  NNLO+N$^3$LL &  N$^3$LO$_{\rm sv}$   &  NNLO+N$^3$LL & N$^3$LO$_{\rm sv}$   &  NNLO+N$^3$LL \\
    \hline
MMHT$2014$     & $3$ & $3$ & $5$ & $5$ & $12$ & $14$   \\
    \hline
CT$14$         & $7$ & $8$ & $10$ & $10$ & $32$ & $31$    \\
    \hline
ABMP$16$       & $2$ & $2$ & $3$ & $3$ & $12$ & $12$    \\
    \hline
NNPDF$31$      & $2$ &  $2$&  $5$ &  $5$ & $7$ & $7$    \\
    \hline
PDF$4$LHC$15$  & $4$ & $4$ & $5$ & $5$ & $16$ & $16$    \\
    \hline
\end{tabular}
	}
    \end{center}
\caption{Intrinsic PDF uncertainties for different PDF choices. These
uncertainties are given for both fixed order as well as the resummed cross sections
for a given value of $Q=2500$ GeV.}
\label{tpdf2}
\end{table}

\begin{figure}[h !!!!]
  \centering
                   {\includegraphics[trim=0 0 0 0,clip,width=0.48\textwidth,height=0.40\textwidth]
                   {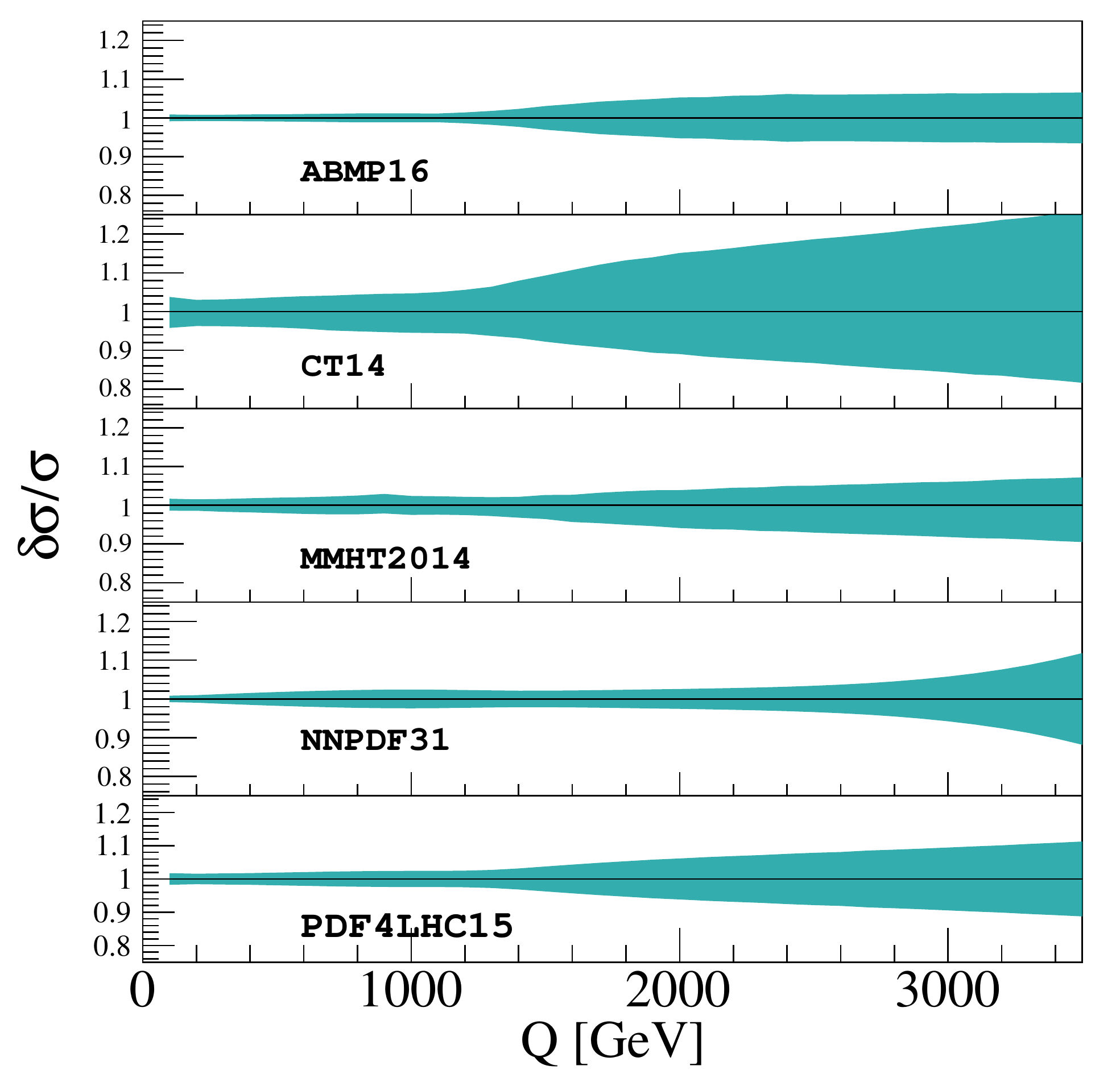}\label{fig:14}}
  \hskip0.05cm
                   {\includegraphics[trim=0 0 0 0,clip,width=0.48\textwidth,height=0.42\textwidth]
                   {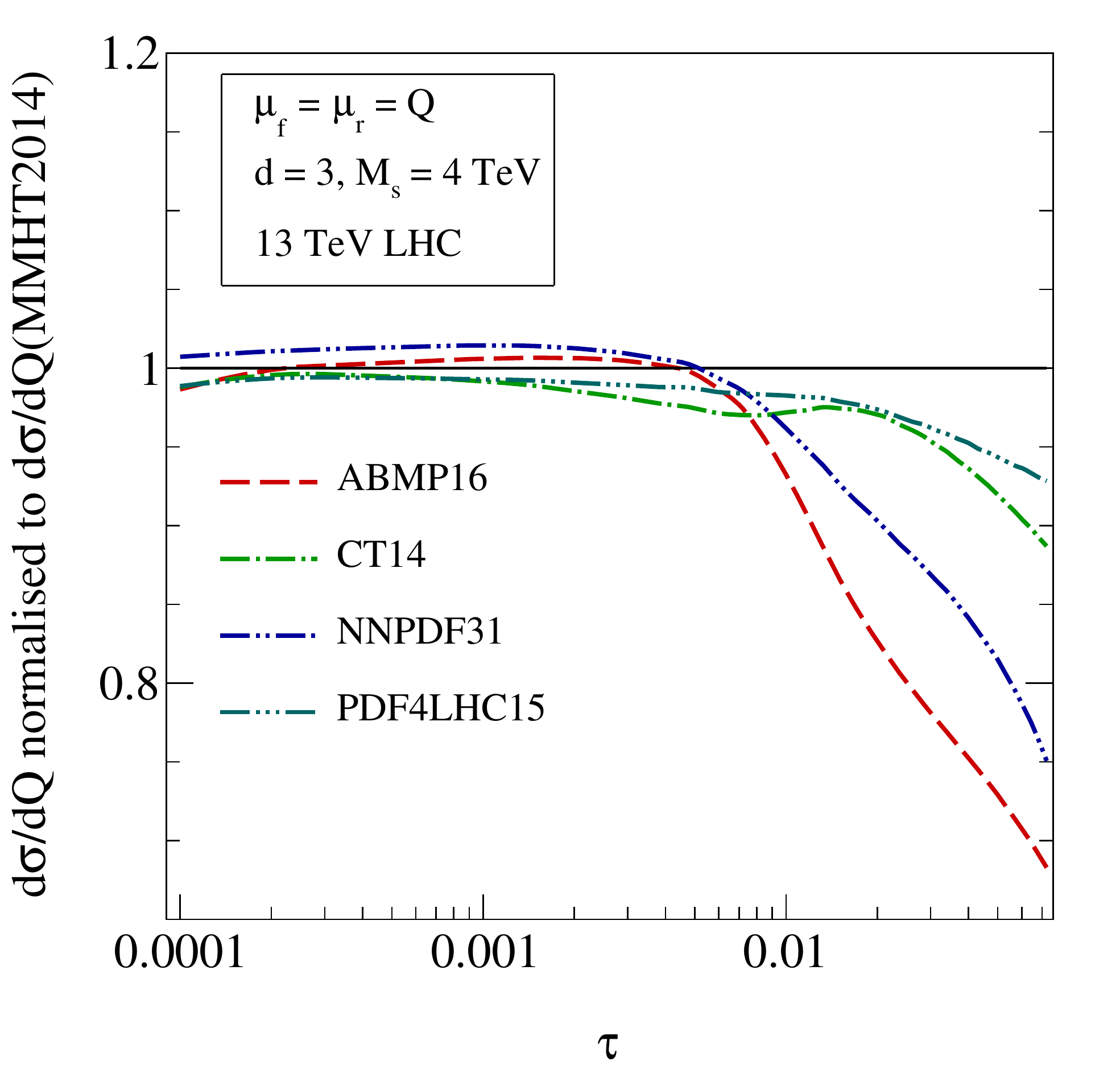}\label{fig:15}}
                  
 \caption{PDF uncertainty for different PDF (left panel) at NNLO order and the result for different PDF (right panel) normalized by our default PDF MMHT2014nnlo.}                   
\label{fig:pdf_fo}
\end{figure}

In Fig.(\ref{fig:pdf_fo}) we present intrinsic uncertainty (left panel) plot for different PDFs as a function of $Q$. At high $Q$ region ($\sim 1500$GeV) these uncertainties are high due to the availability of less number of experimental data. In the right panel of Fig.(\ref{fig:pdf_fo}) we present the relative contribution of different PDFs with respect to our default PDF choice.
%
\subsection{Resummed results up to NNLO+N$^3$LL}
\begin{figure}[h !!!!]
  \centering
                   {\includegraphics[trim=0 0 0 0,clip,width=0.48\textwidth,height=0.40\textwidth]
                   {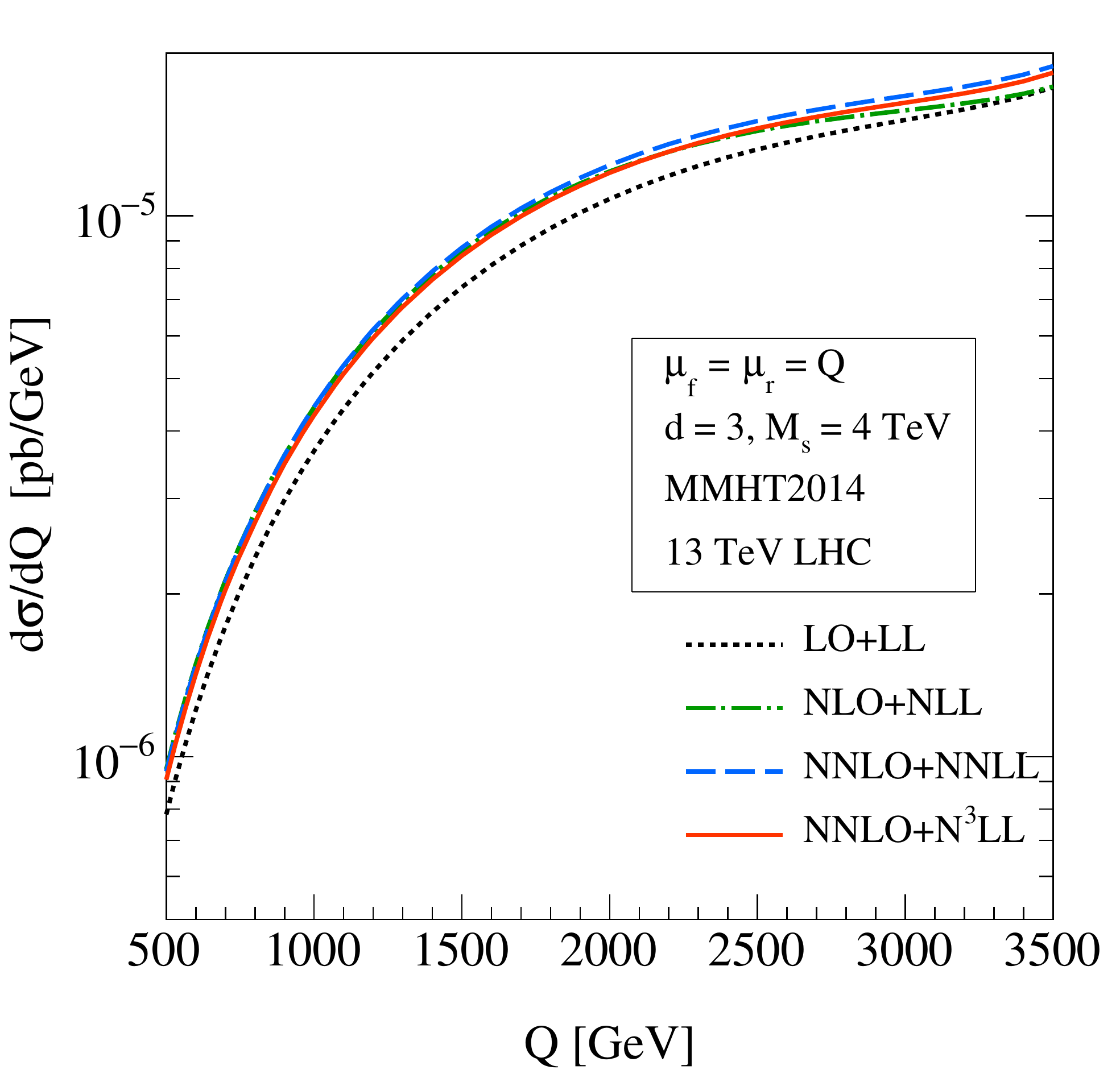}\label{fig:17}}
  \hskip0.05cm    
      
                   {\includegraphics[trim=0 0 0 0,clip,width=0.48\textwidth,height=0.40\textwidth]
                   {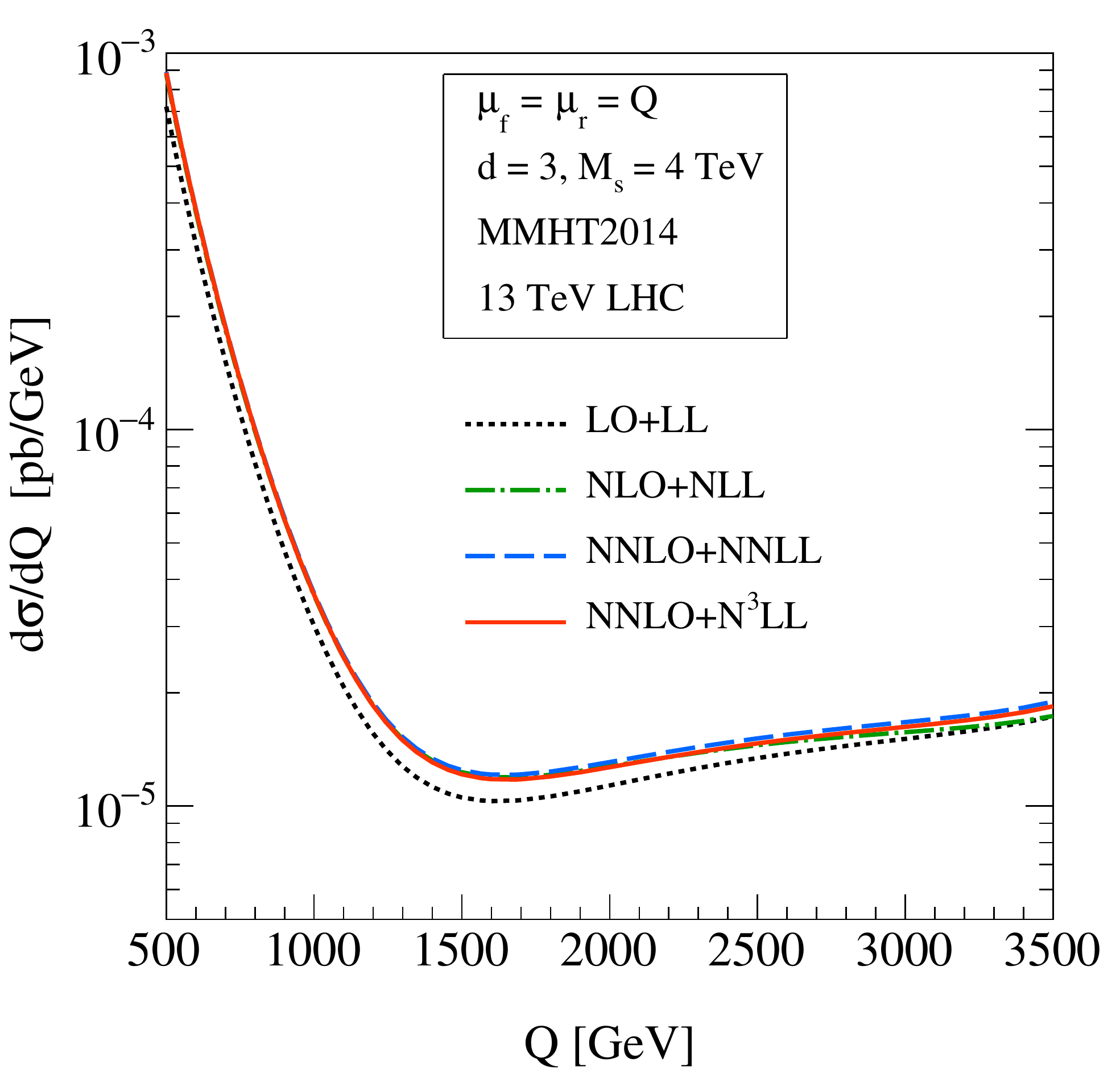}\label{fig:18}}
                   {\includegraphics[trim=0 0 0 0,clip,width=0.48\textwidth,height=0.40\textwidth]
                   {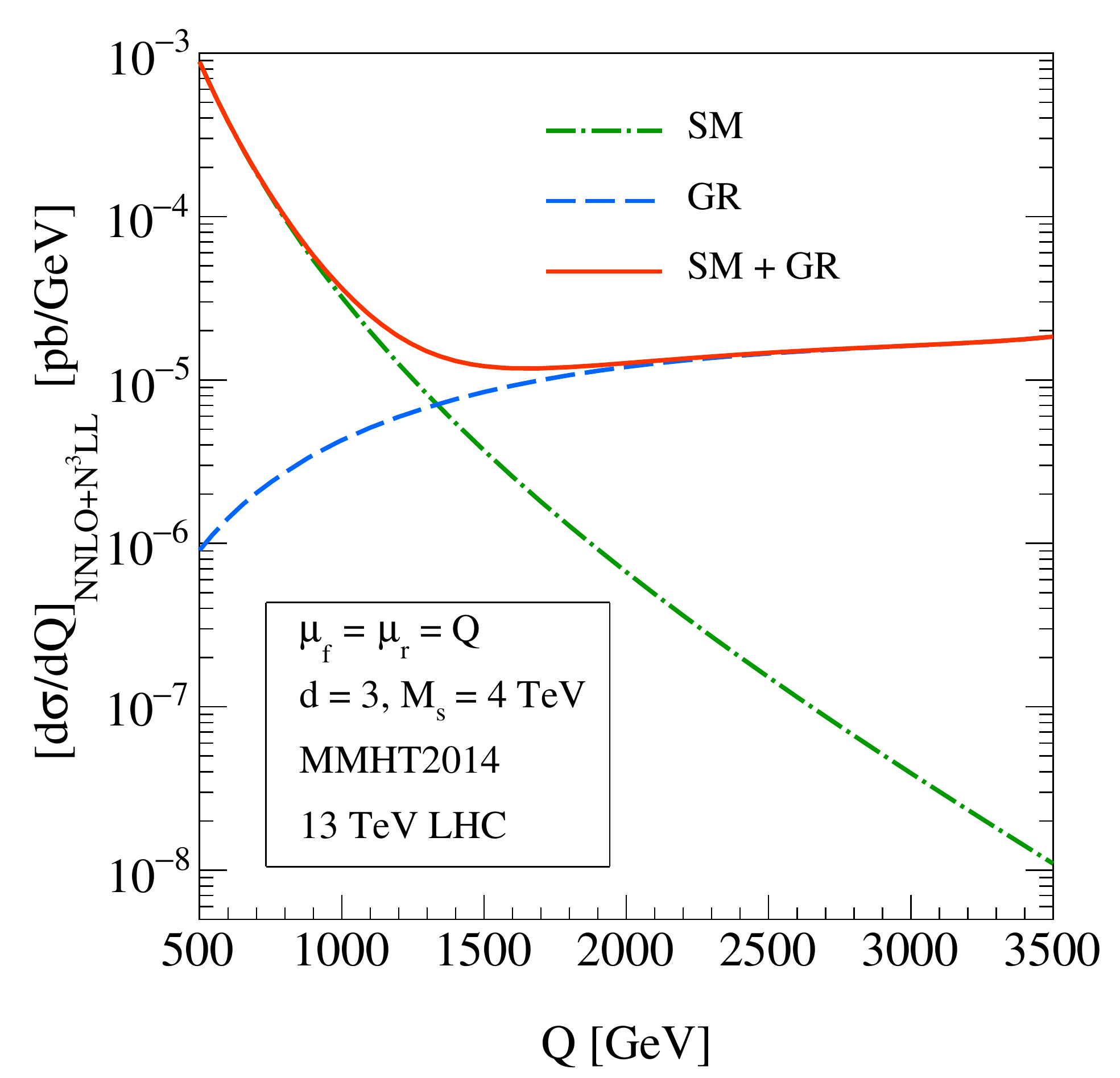}\label{fig:19}}
                   
\caption{Invariant mass distribution of di-lepton at hadronic center of mass energy $13$ TeV for ADD (top), signal (bottom left) and model(bottom right).}
\label{resum_inv}
\end{figure}

\begin{figure}[h !!!!]
  \centering
                   {\includegraphics[trim=0 0 0 0,clip,width=0.48\textwidth,height=0.40\textwidth]
                   {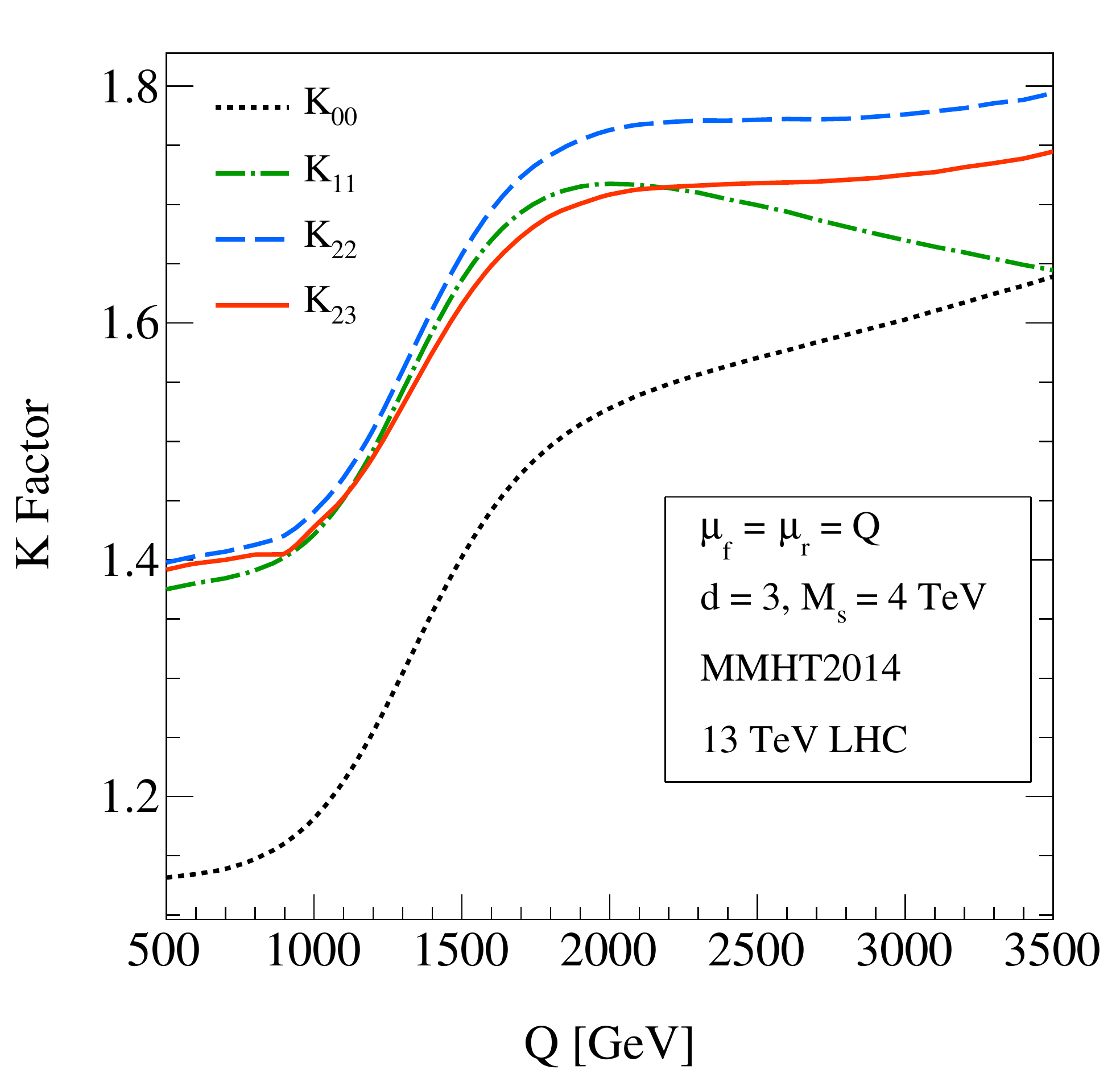}\label{fig:20}}
                  
 \caption{K-factor of di-lepton channel for different order at hadronic center of mass energy $13$ TeV, $M_s = 4$, $d = 3$ for signal}   
\label{resum_kfac}
\end{figure}

\begin{figure}[h !!!!]
  \centering
                   {\includegraphics[trim=0 0 0 0,clip,width=0.48\textwidth,height=0.40\textwidth]
                   {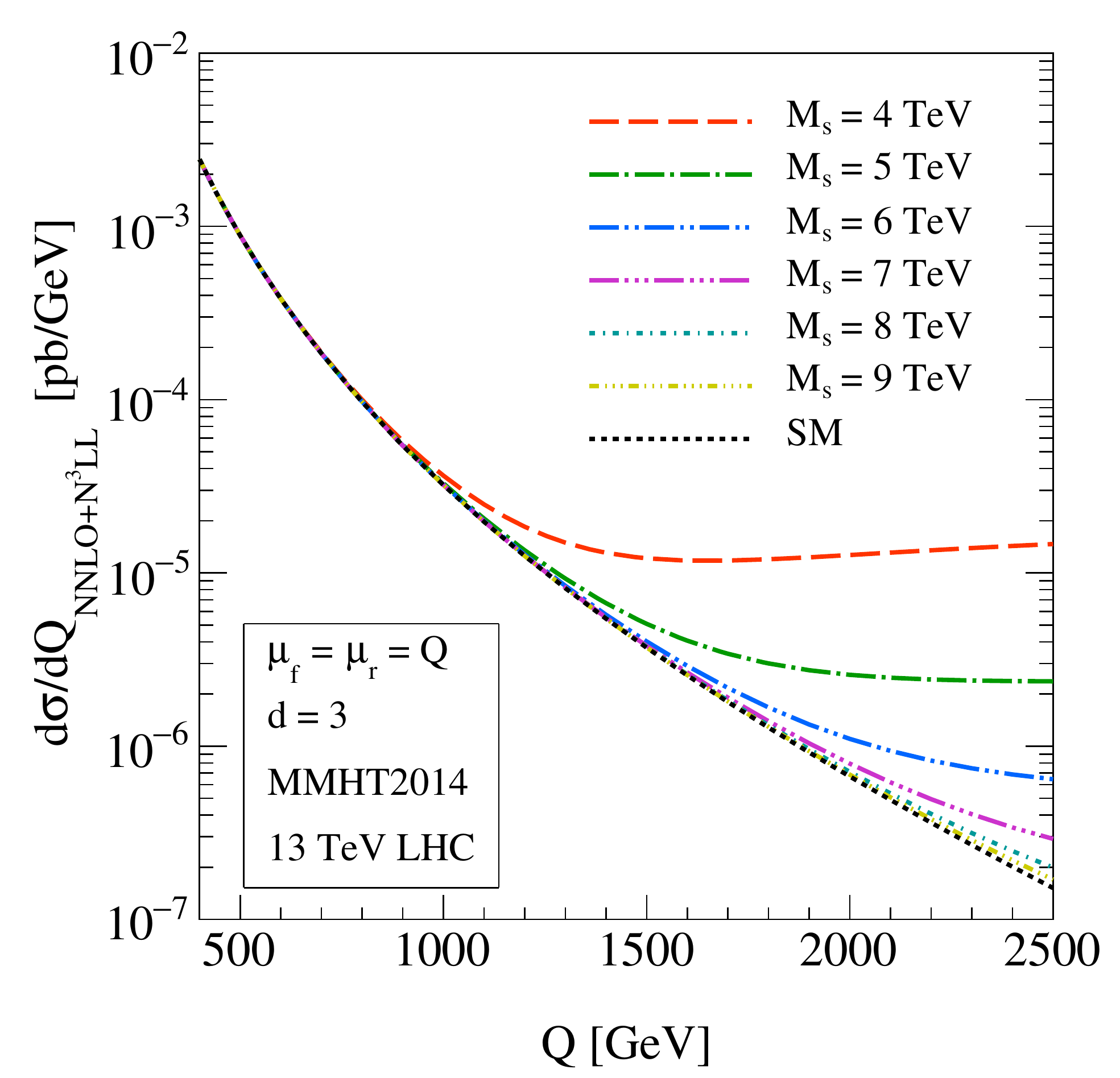}\label{fig:21}}
  \hskip0.05cm
                   {\includegraphics[trim=0 0 0 0,clip,width=0.48\textwidth,height=0.40\textwidth]
                   {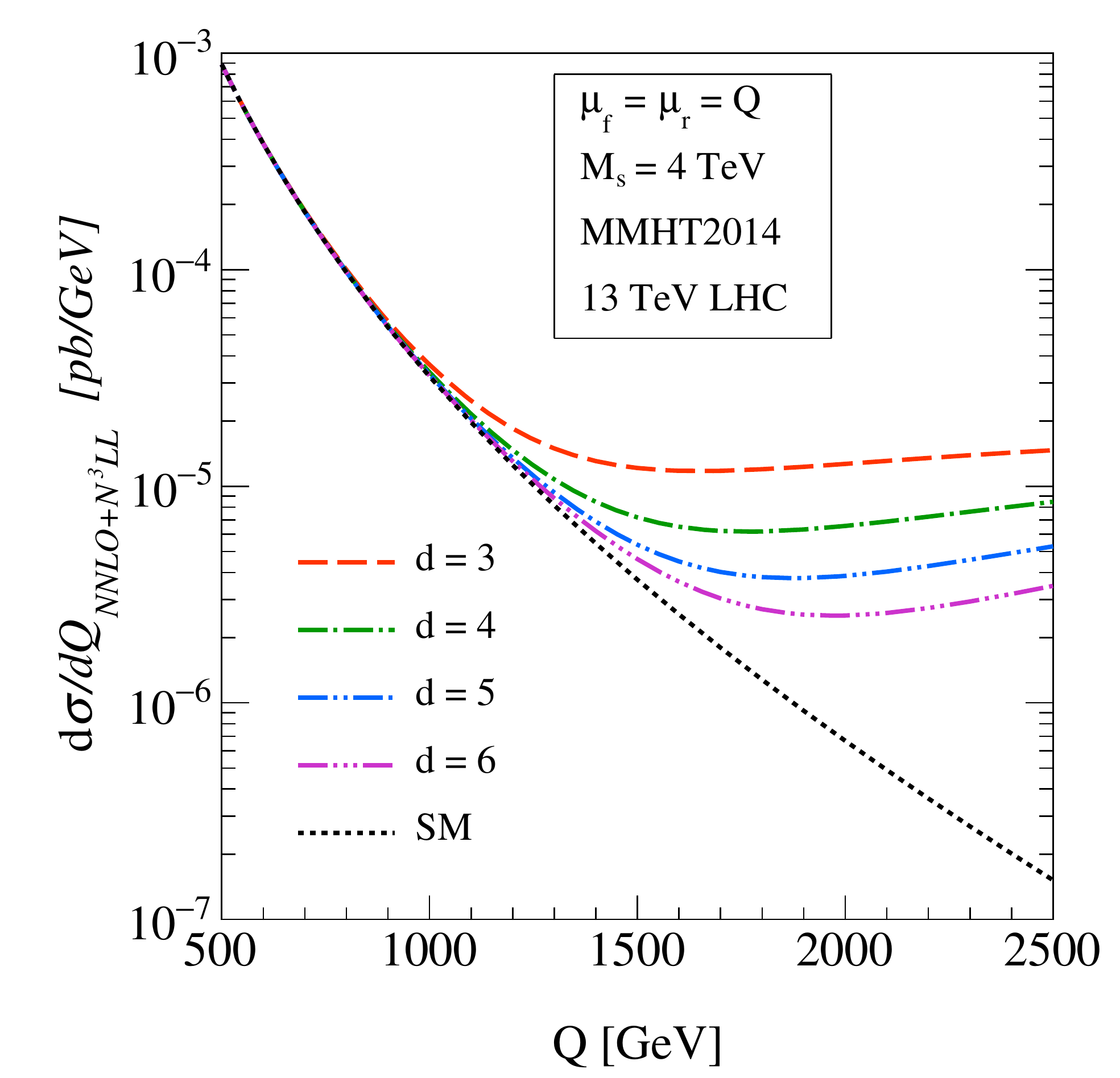}\label{fig:22}}
                
 \caption{Effect of model parameters for invariant mass distribution of di-lepton at hadronic center of mass energy $13$ TeV for signal at NNLO+N$^3$LL level. Left panel for $M_S$ variation for $d = 3$ and right panel for $d$ variation for $M_S = 4$.}
\label{model_parameters}
\end{figure}

In this section we present numerical results for di-lepton production through spin-2  propagator at the LHC to NNLO+N$^3$LL in QCD. In this numerical calculation we use the same choice of QCD parameters and the ADD model as in the computation of three-loop SV corrections (fixed order). For the inverse Mellin transformation Eq.\ (\ref{Eq:matched}), we use $c = 1.9$, $\phi = 3\pi/4$ and $N = 75$.  In Fig.\ (\ref{resum_inv}) we present the numerical result for invariant mass distribution of di-lepton for pure gravity, signal and the corresponding SM background. The behavior of these plots is similar to that of the fixed order results presented in Figs.(\ref{fig:order})\&(\ref{fig:model}). We notice significant enhancement of these resummed results over the fixed order ones, for example at $Q = 2400$ GeV, there is $26\%$  enhancement at NLO+NLL over NLO, $8\%$  at NNLO+NNLL over NNLO and $2\%$ at NNLO+N$^3$LL over N$^3$LO$_{\rm sv}$.  The corresponding mass dependent K-factors up to NNLO+N$^3$LL are shown in Fig.\ (\ref{resum_kfac}). For phenomenological purpose, we defined the resummed K-factors Eq.\ (\ref{kr}) as
%
\begin{table}[h!]
\begin{center}
{\scriptsize
\resizebox{6.5cm}{3.5cm}{%
\begin{tabular}{ |p{1.0cm}|p{0.8cm}|p{0.8cm}|p{0.8cm}|p{0.8cm}| }
\hline
	Q(GeV) & K$_{00}$  & K$_{11}$ & K$_{22}$ & K$_{23}$ \\
\hline
\hline
$200$  &   $1.130$  &   $1.333$  &   $1.346$  &   $1.341$ \\
\hline
$400$  &   $1.130$ &    $1.367$  &   $1.389$  &   $1.383$ \\
\hline
$600$  &   $1.135$  &   $1.380$  &   $1.403$  &   $1.397$ \\
\hline
$800$  &   $1.147$  &   $1.391$  &   $1.413$  &   $1.404$ \\
\hline
$1000$  &   $1.182$  &   $1.421$  &   $1.441$  &   $1.428$ \\
\hline
$1200$  &   $1.255$  &   $1.493$  &   $1.510$  &   $1.487$ \\
\hline
$1400$  &   $1.356$  &   $1.593$  &   $1.611$  &   $1.575$ \\
\hline
$1600$  &   $1.442$  &   $1.670$  &   $1.696$  &   $1.649$ \\
\hline
$1800$  &   $1.496$  &   $1.708$  &   $1.742$  &   $1.691$ \\
\hline
$2000$  &   $1.528$  &   $1.718$  &   $1.763$  &   $1.709$ \\
\hline
$2200$  &   $1.548$  &   $1.714$  &   $1.770$  &   $1.715$ \\
\hline
$2400$  &   $1.564$  &   $1.705$  &   $1.771$  &   $1.717$ \\
\hline
$2600$  &   $1.577$  &   $1.694$  &   $1.772$  &   $1.719$ \\
\hline
$2800$  &   $1.590$  &   $1.681$  &   $1.773$  &   $1.721$ \\
\hline
$3000$  &   $1.603$  &   $1.670$  &   $1.776$  &   $1.725$ \\
\hline
\end{tabular}
}}
\caption{Resummed K-factors, defined in Eq.\ref{kr}, for di-lepton invariant mass distribution at the LHC to various logarithmic accuracy in QCD.}
\label{tablekr}
\end{center}
\end{table}

\begin{equation}
\begin{aligned}	
	&\text{K}_{00} = \frac{ d\sigma^{\rm LO+LL}/dQ} {d\sigma^{\rm LO}/dQ } \quad \quad
         \text{K}_{11} = \frac{ d\sigma^{\rm NLO+NLL}/dQ} {d\sigma^{\rm LO}/dQ } \\
	&\text{K}_{22} = \frac{ d\sigma^{\text{N$^2$LO+N$^2$LL}}/dQ} {d\sigma^{\rm LO}/dQ } \quad \quad
         \text{K}_{23} = \frac{ d\sigma^{\text{N$^2$LO+N$^3$LL}}/dQ} {d\sigma^{\rm LO}/dQ }  \,.
\end{aligned}
\label{kr}
\end{equation}	
As can be seen from the figure, the K-factor $K_{22}$ could be as large as 1.8 for $Q>2000 GeV$, while the resummation of the logarithms beyond NNLL decrease the cross sections by about 5\% resulting in $K_{23}$ to be around $1.75$. These precise QCD predictions are expected to augment experimental searches for large extra dimensions at the LHC. To this end, in Table-(\ref{tablekr}), we give numerical values for the mass dependent resummed K-factors up to NNLO+N$^3$LL accuracy. For completeness, we also study the dependence on the ADD model parameters $M_S$ and $d$, and the corresponding results are depicted in Fig.(\ref{model_parameters}).

\begin{figure}[h !!!!]
  \centering
                   {\includegraphics[trim=0 0 0 0,clip,width=0.80\textwidth,height=0.50\textwidth]
                   {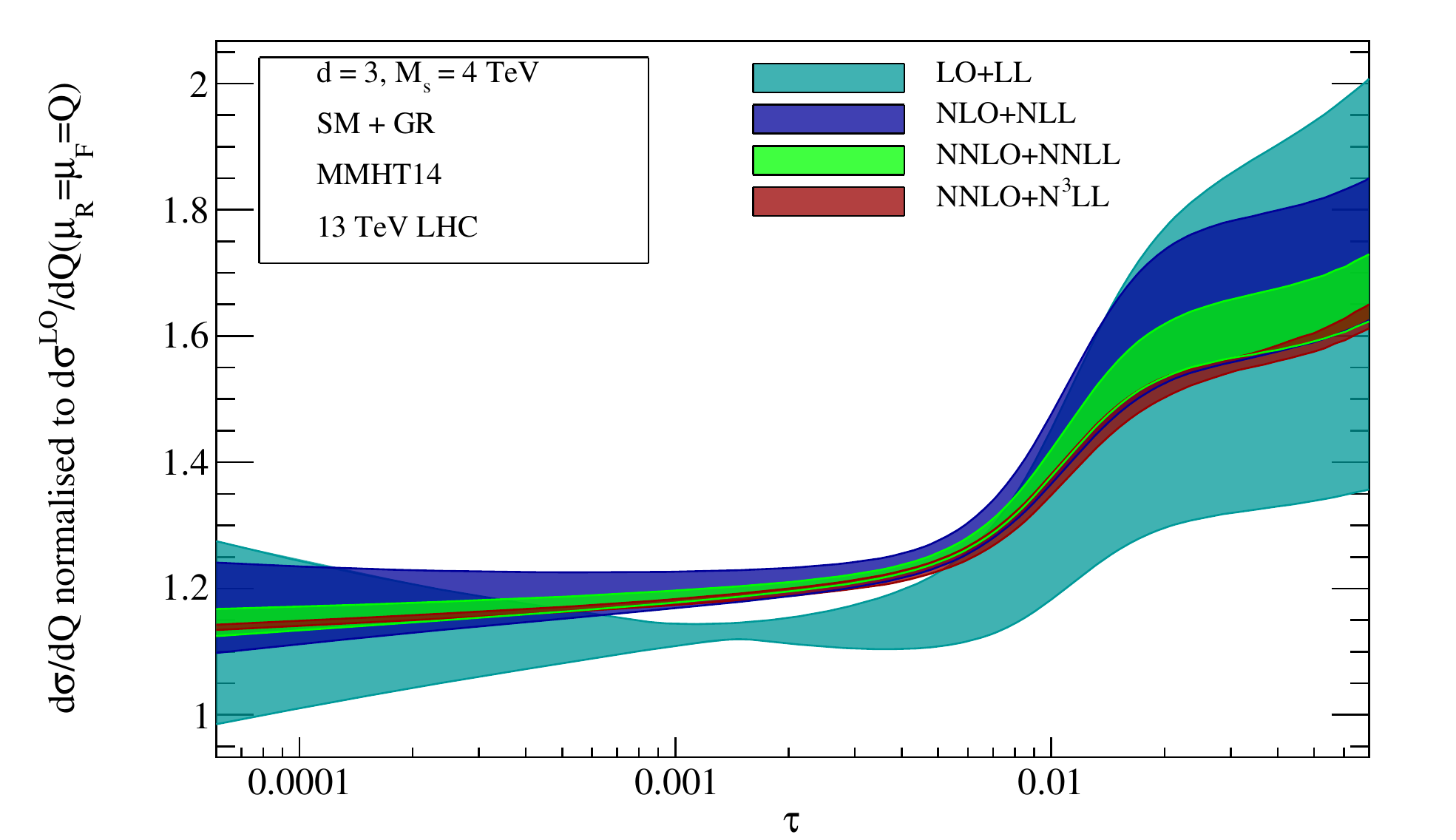}\label{fig:23}}

 \caption{Seven point scale variation of invariant mass distribution of di-lepton at hadronic center of mass energy $13$ TeV for signal with $M_S = 4$ and $d = 3$.}
\label{resum_scale}
\end{figure}

Finally we estimate the uncertainties in our resummed results due to the unphysical scales $\mu_r$ and $\mu_f$, and those due to the parton densities that are non-perturbative in nature. For scale uncertainties we follow the same procedure as in fixed order case by taking the $7$-point scale variation and the results are shown in Fig.(\ref{resum_scale}) as a function of the di-lepton invariant mass $Q$. The scale uncertainties are found to get reduced significantly from LO+LL to NNLO+N$^3$LL. For example  at $Q = 2500$ GeV, the scale uncertainties are $56\%$ at LO+LL, $22\%$ at NLO+NLL, $10\%$ at NNLO+NNLL and are as small as $2\%$ at NNLO+N$^3$LL.  We observe that the scale uncertainty bands  at higher orders lie inside the ones at lower orders. The scale uncertainties are conventionally used for estimating the contribution from the missing higher order contributions. In that sense, these resummed results have better theory predictions over the fixed order ones.

\begin{figure}[h !!!!]
  \centering
                   {\includegraphics[trim=0 0 0 0,clip,width=0.48\textwidth,height=0.40\textwidth]
                   {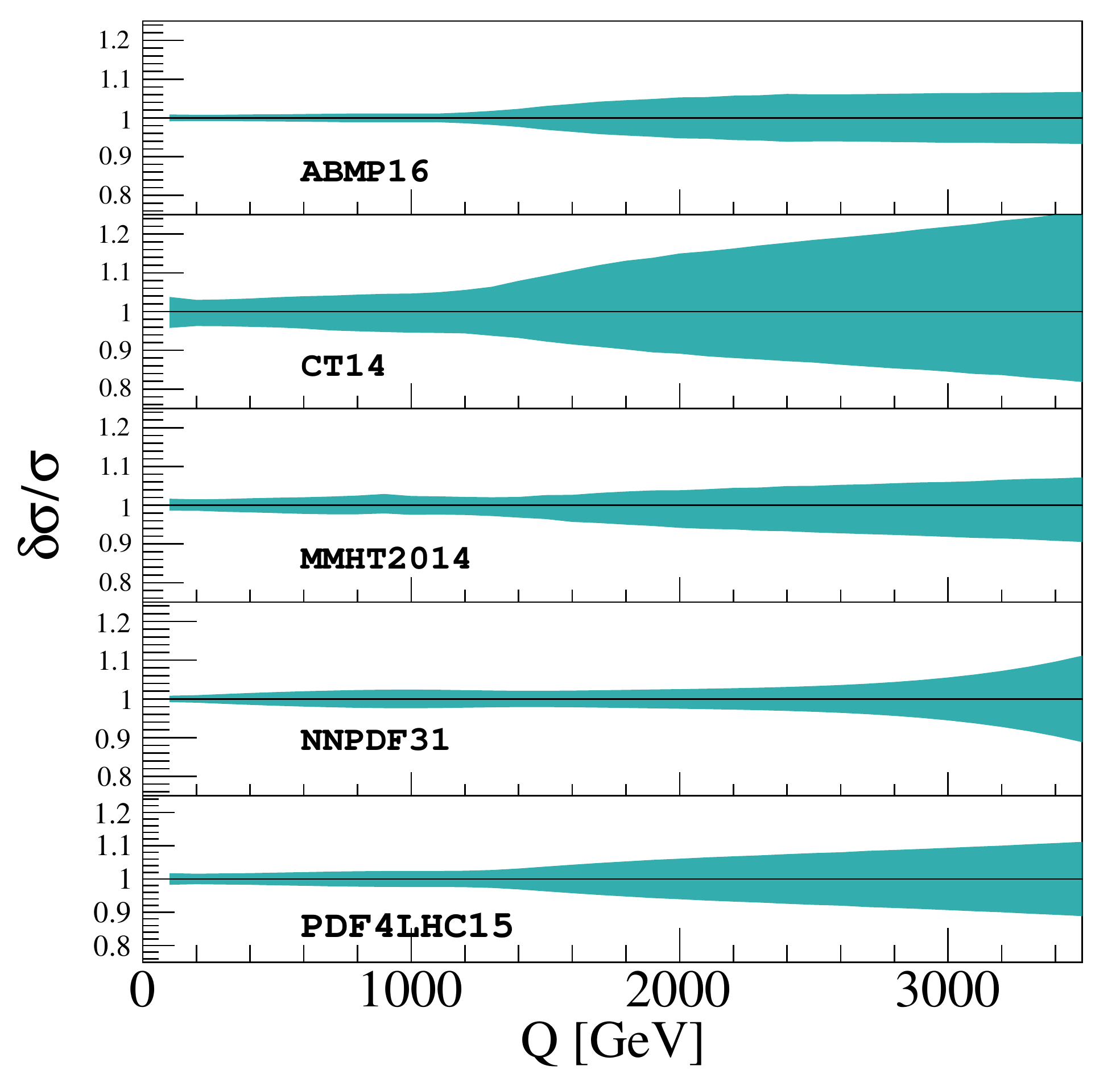}\label{fig:24}}
  \hskip0.05cm
                   {\includegraphics[trim=0 0 0 0,clip,width=0.48\textwidth,height=0.42\textwidth]
                   {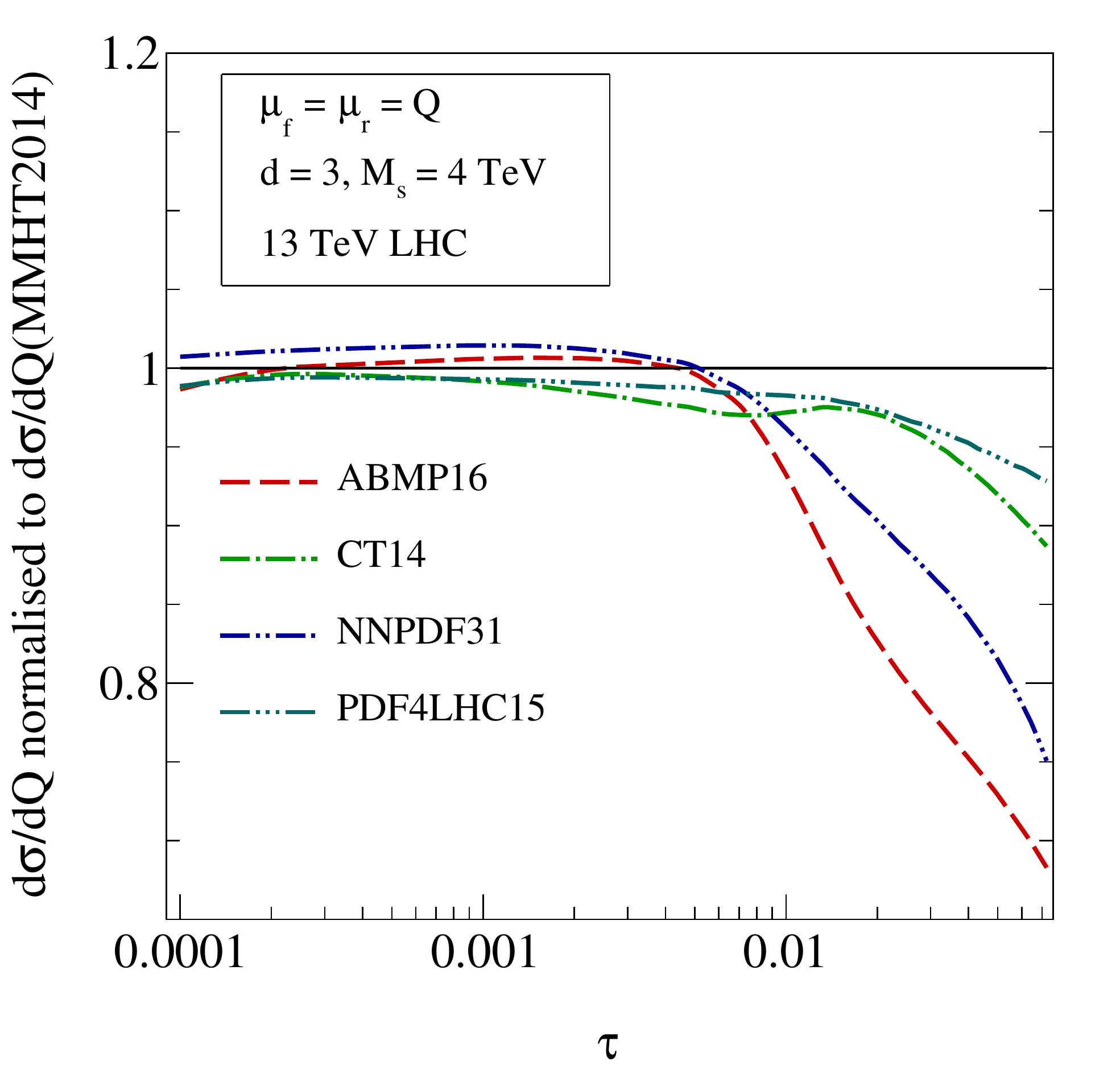}\label{fig:25}}
                  
 \caption{The intrinsic PDF uncertainties for different PDF groups are shown in the left panel as a function of the di-lepton invariant mass $Q$. In the right panel, the invariant mass distributions for different PDF groups (computed with central set) normalized with that obtained from the default choice MMHT2014nnlo PDF set.}
 \label{resum_pdf}
\end{figure}

The intrinsic uncertainties in a given PDF set as well as those from the choice of the PDF group itself are estimated as in the fixed order case. We present these results in Fig.(\ref{resum_pdf}). We observe that the intrinsic PDF uncertainties are very much similar to those of the fixed order case as can be seen from the Table.\ref{tpdf2}. This is simply because the results for resummation of the threshold logarithms still use the parton densities extracted at NNLO accuracy. Moreover, we also present the uncertainties due to the choice of the PDFs group in terms of the distributions normalized with respect to those obtained from MMHT2014 group.
%

\section{Conclusions}\label{conc}
In this article we have computed the higher order QCD corrections beyond NNLO for the spin-2 production at hadron colliders. Specifically, we have calculated  three-loop SV corrections to the spin-2 production, thanks to the recent computation of the quark and gluon form factors at three loop level. We have performed a detailed phenomenological study at N$^3$LO$_{\rm sv}$ in QCD and presented our numerical results for the di-lepton invariant mass distribution in the ADD model for 13TeV LHC. The three-loop SV corrections are about 2\% over the existing NNLO result. The conventional 7-point scale uncertainties of about 8\% at NNLO in the high invariant mass region get reduced to about 5\% at three-loop level. In addition we have also extracted the process-dependent coefficients coming from the form factor and the soft-collinear function to third order. Using these coefficients we perform resummation of large threshold logarithms up to N$^3$LL accuracy. We also study the numerical impact of these resummed result after matching it to NNLO fixed order result. While the quantitative enhancement of these resummed results is approximately 2\% over the known fixed order NNLO results, the resummed predictions reduce the scale uncertainties signficiantly to as low as 2\%. For completeness, we also estimate the PDF uncertainties in our predictions using the parton densities available at NNLO level from various groups. The uncertainties from these non-perturbative inputs are estimated be minimum of about 10\%.  Finally, we conclude that the perturbation theory predictions in QCD for massive spin-2 production are now very precise and are at par comparable to the accuracy that is achieved for the well known weak bosons (Z/W) and the most sought Higgs boson in the SM.

\section*{Acknowledgements}
We thank S. Alekhin, P. Mathews, S. Moch, V. Ravindran and A. Vogt for useful discussions.
The algebraic computations have been done with the latest version of the symbolic manipulation system {\sc Form}~\cite{Vermaseren:2000nd,Ruijl:2017dtg}. The research of G.D. is supported partially by DESY postdoctoral fellowship and by the \textit{Deutsche Forschungsgemeinschaft} (DFG) within the Collaborative Research Center TRR 257 (“\textit{Particle Physics Phenomenology after the Higgs Discovery}”).
\appendix

\section{Soft-Virtual coefficients}\label{app:A}

Here we present the SV coefficient for spin-2 production up to three loops for both quark and gluon initiated channels. The third loop results are new.
\begin{align} 
\begin{autobreak}
\CIq1 =
\frac{\pi}{8 n_c}\Bigg[ 
\Dm1    \bigg\{ \bigg(
- 20
+ 8  \z2
+ \bigg(6\bigg)  \Lqf\bigg)  \Cf \bigg\}      
+ \D1    \bigg\{ \bigg(16\bigg)  \Cf \bigg\}      
+ \D0    \bigg\{ \bigg(\bigg(8\bigg)  \Lqf\bigg)  \Cf \bigg\}
\Bigg]\end{autobreak} 
\\ 
\begin{autobreak} 
\CIq2 =
\frac{\pi}{8 n_c}\Bigg[ 
\Dm1    \bigg\{ \bigg(
- \frac{5941}{36}
+ 92  \z3
+ \frac{328}{9}  \z2
- \frac{12}{5}  \z2^2
+ \bigg(79
- 24  \z3\bigg)  \Lqf
+ \bigg(-11\bigg)  \Lqf^2\bigg)  \Ca  \Cf
+ \bigg(\frac{461}{18}
+ 8  \z3
- \frac{64}{9}  \z2
+ \bigg(-14\bigg)   \Lqf
+ \bigg(2\bigg)  \Lqf^2\bigg)  \Cf  \nf
+ \bigg(\frac{2293}{12}
- 124  \z3
- 70  \z2
+ \frac{8}{5}  \z2^2
+ \bigg(
- 117
+ 176  \z3
+ 24  \z2\bigg)  \Lqf
+ \bigg(18
- 32  \z2\bigg)  \Lqf^2\bigg)  \Cf^2 \bigg\}      
+ \D3    \bigg\{ \bigg(128\bigg)  \Cf^2 \bigg\}      
+ \D2    \bigg\{ \bigg(
- \frac{176}{3}\bigg)  \Ca  \Cf
+ \bigg(\frac{32}{3}\bigg)  \Cf  \nf
+ \bigg(\bigg(192\bigg)  \Lqf\bigg)  \Cf^2 \bigg\}      
+ \D1    \bigg\{ \bigg(
- 320
- 128  \z2
+ \bigg(64\bigg)  \Lqf^2
+ \bigg(96\bigg)  \Lqf\bigg)  \Cf^2
+ \bigg(
- \frac{160}{9}
+ \bigg( \frac{32}{3}\bigg)  \Lqf\bigg)  \Cf  \nf
+ \bigg(\frac{1072}{9}
- 32  \z2
+ \bigg(
- \frac{176}{3}\bigg)  \Lqf\bigg)  \Ca  \Cf \bigg\}      
+ \D0    \bigg\{ \bigg(
- \frac{1616}{27}
+ 56  \z3
+ \frac{176}{3}  \z2
+ \bigg(
- \frac{44}{3}\bigg)  \Lqf^2
+ \bigg(\frac{536}{9}
- 16   \z2\bigg)  \Lqf\bigg)  \Ca  \Cf
+ \bigg(\frac{224}{27}
- \frac{32}{3}  \z2
+ \bigg(
- \frac{80}{9}\bigg)  \Lqf
+ \bigg(\frac{8}{3}\bigg)  \Lqf^2\bigg)  \Cf   \nf
+ \bigg(256  \z3
+ \bigg(
- 160
- 64  \z2\bigg)  \Lqf
+ \bigg(48\bigg)  \Lqf^2\bigg)  \Cf^2 \bigg\}
\Bigg]\end{autobreak} 
\\ 
\begin{autobreak} 
\CIq3 =
\frac{\pi}{8 n_c}\Bigg[ 
\Dm1    \bigg\{ \bigg(
- \frac{1970041}{972}
- \frac{4292}{3}  \z5
+ \frac{282365}{81}  \z3
- \frac{400}{3}   \z3^2
+ \frac{9685}{27}  \z2
- \frac{1028}{3}  \z2  \z3
+ \frac{5459}{27}  \z2^2
+ \frac{13264}{315}  \z2^3
+ \bigg(
- \frac{971}{3}
+ 88  \z3\bigg)  \Lqf^2
+ \bigg(\frac{242}{9}\bigg)  \Lqf^3
+ \bigg(\frac{36310}{27}
+ 80  \z5
- \frac{9176}{9}   \z3
- \frac{224}{9}  \z2
+ \frac{68}{5}  \z2^2\bigg)  \Lqf\bigg)  \Ca^2  \Cf
+ \bigg(
- \frac{2807}{2}
+ \frac{3344}{3}  \z5
+ \frac{4300}{3}  \z3
+ \frac{10336}{3}  \z3^2
- \frac{1082}{3}  \z2
- 432  \z2  \z3
+ \frac{892}{5}  \z2^2
- \frac{184736}{315}  \z2^3
+ \bigg(
- 342
+ 1056  \z3
+ 640  \z2
- \frac{1792}{5}  \z2^2\bigg)  \Lqf^2
+ \bigg( 36
+ \frac{512}{3}  \z3
- 192  \z2\bigg)  \Lqf^3
+ \bigg(\frac{2231}{2}
+ 5664  \z5
- 4128  \z3
+ 120  \z2
- 2752  \z2  \z3
- \frac{336}{5}  \z2^2\bigg)  \Lqf\bigg)  \Cf^3
+ \bigg(
- \frac{697}{3}
+ \frac{5536}{9}  \z5
+ \frac{6200}{9}  \z3
- \frac{1256}{27}  \z2
- \frac{5504}{9}  \z2  \z3
+ \frac{3536}{135}  \z2^2
+ \bigg(
- 116
+ 160  \z3
+ \frac{640}{9}  \z2\bigg)  \Lqf^2
+ \bigg(12
- \frac{64}{3}  \z2\bigg)  \Lqf^3
+ \bigg(\frac{3022}{9}
- \frac{4432}{9}  \z3
- \frac{1936}{27}  \z2
+ \frac{112}{15}  \z2^2\bigg)  \Lqf\bigg)  \Cf^2  \nf
+ \bigg(
- \frac{7345}{243}
- \frac{2416}{81}  \z3
+ \frac{1744}{81}  \z2
+ \frac{128}{27}  \z2^2
+ \bigg(
- \frac{28}{3}\bigg)  \Lqf^2
+ \bigg(\frac{8}{9}\bigg)  \Lqf^3
+ \bigg(\frac{820}{27}
+ \frac{64}{9}  \z3
- \frac{32}{9}  \z2\bigg)  \Lqf\bigg)  \Cf  \nf^2
+ \bigg(\frac{130871}{243}
+ \frac{136}{3}  \z5
- \frac{29488}{81}   \z3
- \frac{16348}{81}  \z2
+ \frac{224}{3}  \z2  \z3
- \frac{7988}{135}  \z2^2
+ \bigg(
- \frac{11732}{27}
+ \frac{976}{9}   \z3
+ \frac{256}{9}  \z2
- \frac{8}{5}  \z2^2\bigg)  \Lqf
+ \bigg(
- \frac{88}{9}\bigg)  \Lqf^3
+ \bigg(114
- 16  \z3\bigg)   \Lqf^2\bigg)  \Ca  \Cf  \nf
+ \bigg(\frac{328511}{108}
- \frac{26824}{9}  \z5
- \frac{57388}{9}  \z3
+ \frac{3280}{3}  \z3^2
+ \frac{8894}{27}  \z2
+ \frac{33728}{9}  \z2  \z3
- \frac{26288}{135}  \z2^2
- \frac{20816}{315}  \z2^3
+ \bigg(
- \frac{43727}{18}
+ 240  \z5
+ \frac{39688}{9}  \z3
+ \frac{10696}{27}  \z2
- 1120  \z2  \z3
- \frac{256}{15}   \z2^2\bigg)  \Lqf
+ \bigg(
- 66
+ \frac{352}{3}  \z2\bigg)  \Lqf^3
+ \bigg(683
- 1024  \z3
- \frac{4288}{9}  \z2
+ 128  \z2^2\bigg)  \Lqf^2\bigg)  \Ca  \Cf^2 \bigg\}      
+ \D5    \bigg\{ \bigg(512\bigg)  \Cf^3 \bigg\}      
+ \D4    \bigg\{ \bigg(
- \frac{7040}{9}\bigg)  \Ca  \Cf^2
+ \bigg(\frac{1280}{9}\bigg)  \Cf^2  \nf
+ \bigg(\bigg(1280\bigg)  \Lqf\bigg)  \Cf^3 \bigg\}      
+ \D3    \bigg\{ \bigg(
- 2560
- 3072  \z2
+ \bigg(768\bigg)  \Lqf
+ \bigg(1024\bigg)  \Lqf^2\bigg)  \Cf^3
+ \bigg(
- \frac{2560}{9}
+ \bigg(\frac{2560}{9}\bigg)  \Lqf\bigg)  \Cf^2  \nf
+ \bigg(
- \frac{2816}{27}\bigg)  \Ca  \Cf  \nf
+ \bigg(\frac{256}{27}\bigg)  \Cf  \nf^2
+ \bigg(\frac{7744}{27}\bigg)  \Ca^2  \Cf
+ \bigg(\frac{17152}{9}
- 512  \z2
+ \bigg(
- \frac{14080}{9}\bigg)  \Lqf\bigg)  \Ca  \Cf^2 \bigg\}      
+ \D2    \bigg\{ \bigg(
- \frac{28480}{27}
+ \frac{704}{3}  \z2
+ \bigg(\frac{3872}{9}\bigg)  \Lqf\bigg)  \Ca^2  \Cf
+ \bigg(
- \frac{2368}{9}
+ 1344  \z3
+ \frac{11264}{3}  \z2
+ \bigg(\frac{7520}{3}
- 768  \z2\bigg)  \Lqf
+ \bigg(-1056\bigg)  \Lqf^2\bigg)  \Ca   \Cf^2
+ \bigg(
- \frac{640}{27}
+ \bigg(\frac{128}{9}\bigg)  \Lqf\bigg)  \Cf  \nf^2
+ \bigg(\frac{160}{9}
- \frac{2048}{3}  \z2
+ \bigg(
- \frac{1088}{3}\bigg)  \Lqf
+ \bigg(192\bigg)  \Lqf^2\bigg)  \Cf^2  \nf
+ \bigg(\frac{9248}{27}
- \frac{128}{3}  \z2
+ \bigg(
- \frac{1408}{9} \bigg)  \Lqf\bigg)  \Ca  \Cf  \nf
+ \bigg(10240  \z3
+ \bigg(
- 3840
- 4608  \z2\bigg)  \Lqf
+ \bigg(256\bigg)  \Lqf^3
+ \bigg(1152\bigg)  \Lqf^2\bigg)  \Cf^3 \bigg\}      
+ \D1    \bigg\{ \bigg(
- \frac{15068}{3}
- 4160  \z3
- \frac{14720}{9}  \z2
+ \frac{3648}{5}  \z2^2
+ \bigg(
- \frac{704}{3}\bigg)   \Lqf^3
+ \bigg(\frac{3824}{9}
- 256  \z2\bigg)  \Lqf^2
+ \bigg(\frac{59248}{27}
+ 512  \z3
+ \frac{9280}{3}  \z2\bigg)  \Lqf \bigg)  \Ca  \Cf^2
+ \bigg(
- \frac{32816}{81}
+ 384  \z2
+ \bigg(
- \frac{704}{9}\bigg)  \Lqf^2
+ \bigg(\frac{9248}{27}
- \frac{128}{3}  \z2\bigg)  \Lqf\bigg)  \Ca  \Cf  \nf
+ \bigg(\frac{1600}{81}
- \frac{256}{9}  \z2
+ \bigg(
- \frac{640}{27}\bigg)  \Lqf
+ \bigg(\frac{64}{9}\bigg)   \Lqf^2\bigg)  \Cf  \nf^2
+ \bigg(\frac{1856}{3}
+ 1280  \z3
+ \frac{2816}{9}  \z2
+ \bigg(
- \frac{10240}{27}
- \frac{1792}{3}  \z2\bigg)  \Lqf
+ \bigg(
- \frac{416}{9}\bigg)  \Lqf^2
+ \bigg(\frac{128}{3}\bigg)  \Lqf^3\bigg)  \Cf^2  \nf
+ \bigg(\frac{124024}{81}
- 704  \z3
- \frac{12032}{9}  \z2
+ \frac{704}{5}  \z2^2
+ \bigg(
- \frac{28480}{27}
+ \frac{704}{3}  \z2\bigg)  \Lqf
+ \bigg( \frac{1936}{9}\bigg)  \Lqf^2\bigg)  \Ca^2  \Cf
+ \bigg(\frac{9172}{3}
- 1984  \z3
+ 4000  \z2
- \frac{14208}{5}  \z2^2
+ \bigg(
- 1872
+ 11008  \z3
- 1152  \z2\bigg)  \Lqf
+ \bigg(
- 992
- 2048  \z2\bigg)  \Lqf^2
+ \bigg( 384\bigg)  \Lqf^3\bigg)  \Cf^3 \bigg\}      
+ \D0    \bigg\{ \bigg(
- \frac{594058}{729}
- 384  \z5
+ \frac{40144}{27}  \z3
+ \frac{98224}{81}  \z2
- \frac{352}{3}  \z2   \z3
- \frac{2992}{15}  \z2^2
+ \bigg(
- \frac{7120}{27}
+ \frac{176}{3}  \z2\bigg)  \Lqf^2
+ \bigg(\frac{968}{27}\bigg)  \Lqf^3
+ \bigg(\frac{62012}{81}
- 352  \z3
- \frac{6016}{9}  \z2
+ \frac{352}{5}  \z2^2\bigg)  \Lqf\bigg)  \Ca^2  \Cf
+ \bigg(
- \frac{1058}{27}
- \frac{5728}{9}  \z3
+ \frac{3104}{27}  \z2
- \frac{1472}{15}  \z2^2
+ \bigg(
- \frac{632}{3}
- \frac{320}{3}   \z2\bigg)  \Lqf^2
+ \bigg(\frac{3232}{9}
+ 640  \z3
+ \frac{832}{9}  \z2\bigg)  \Lqf
+ \bigg(32\bigg)  \Lqf^3\bigg)  \Cf^2  \nf
+ \bigg(
- \frac{3712}{729}
+ \frac{320}{27}  \z3
+ \frac{640}{27}  \z2
+ \bigg(
- \frac{160}{27}\bigg)  \Lqf^2
+ \bigg(\frac{32}{27}\bigg)   \Lqf^3
+ \bigg(\frac{800}{81}
- \frac{128}{9}  \z2\bigg)  \Lqf\bigg)  \Cf  \nf^2
+ \bigg(\frac{125252}{729}
- \frac{2480}{9}  \z3
- \frac{29392}{81}  \z2
+ \frac{736}{15}  \z2^2
+ \bigg(
- \frac{16408}{81}
+ 192  \z2\bigg)  \Lqf
+ \bigg(
- \frac{352}{27}\bigg)   \Lqf^3
+ \bigg(\frac{2312}{27}
- \frac{32}{3}  \z2\bigg)  \Lqf^2\bigg)  \Ca  \Cf  \nf
+ \bigg(\frac{32320}{27}
+ \frac{24224}{9}  \z3
- \frac{18752}{27}  \z2
- 1472  \z2  \z3
+ \frac{1408}{3}  \z2^2
+ \bigg(
- \frac{25834}{9}
- 1744  \z3
- \frac{4192}{9}  \z2
+ \frac{1824}{5}  \z2^2\bigg)  \Lqf
+ \bigg(\frac{3848}{3}
- 192  \z3
+ \frac{1472}{3}  \z2\bigg)  \Lqf^2
+ \bigg(-176\bigg)  \Lqf^3\bigg)  \Ca  \Cf^2
+ \bigg(12288  \z5
- 5120  \z3
- 6144  \z2  \z3
+ \bigg(
- 936
+ 2432  \z3
- 576  \z2\bigg)  \Lqf^2
+ \bigg(144
- 256  \z2\bigg)  \Lqf^3
+ \bigg(\frac{4586}{3}
+ 544  \z3
+ 2000  \z2
- \frac{7104}{5}  \z2^2\bigg)  \Lqf\bigg)  \Cf^3 \bigg\}
\Bigg]
\end{autobreak} 
\end{align}

\begin{align} 
\begin{autobreak}
\CIg1 =
\frac{\pi}{2 (n_c^2-1)}\Bigg[ 
\Dm1    \bigg\{ \bigg(
- \frac{203}{9}
+ 8  \z2
+ \bigg(\frac{22}{3}\bigg)  \Lqf\bigg)  \Ca
+ \bigg(\frac{35}{9}
+ \bigg(
- \frac{4}{3}\bigg)   \Lqf\bigg)  \nf \bigg\}      
+ \D1    \bigg\{ \bigg(16\bigg)  \Ca \bigg\}      
+ \D0    \bigg\{ \bigg(\bigg(8\bigg)  \Lqf\bigg)  \Ca \bigg\}
\Bigg]\end{autobreak} 
\\ 
\begin{autobreak} 
\CIg2 =
\frac{\pi}{2 (n_c^2-1)}\Bigg[ 
\Dm1    \bigg\{ \bigg(
- \frac{2983}{162}
+ \frac{64}{3}  \z3
- \frac{94}{9}  \z2
+ \bigg(
- \frac{44}{9}\bigg)  \Lqf^2
+ \bigg(\frac{647}{27}
- \frac{16}{3}  \z2\bigg)  \Lqf\bigg)  \Ca  \nf
+ \bigg(\frac{1225}{324}
+ \frac{8}{3}  \z2
+ \bigg(
- \frac{70}{27}\bigg)   \Lqf
+ \bigg(\frac{4}{9}\bigg)  \Lqf^2\bigg)  \nf^2
+ \bigg(\frac{61}{3}
- 16  \z3
+ \bigg(-4\bigg)  \Lqf\bigg)  \Cf  \nf
+ \bigg(\frac{7801}{324}
- \frac{88}{3}  \z3
- \frac{224}{9}  \z2
- \frac{4}{5}  \z2^2
+ \bigg(
- \frac{1657}{27}
+ 152  \z3
+ \frac{88}{3}  \z2\bigg)   \Lqf
+ \bigg(\frac{121}{9}
- 32  \z2\bigg)  \Lqf^2\bigg)  \Ca^2 \bigg\}      
+ \D3    \bigg\{ \bigg(128\bigg)  \Ca^2 \bigg\}      
+ \D2    \bigg\{ \bigg(
- \frac{176}{3}
+ \bigg(192\bigg)  \Lqf\bigg)  \Ca^2
+ \bigg(\frac{32}{3}\bigg)  \Ca  \nf \bigg\}      
+ \D1    \bigg\{ \bigg(
- \frac{2176}{9}
- 160  \z2
+ \bigg(\frac{176}{3}\bigg)  \Lqf
+ \bigg(64\bigg)  \Lqf^2\bigg)  \Ca^2
+ \bigg(\frac{400}{9}
+ \bigg(
- \frac{32}{3}\bigg)  \Lqf\bigg)  \Ca  \nf \bigg\}      
+ \D0    \bigg\{ \bigg(
- \frac{1616}{27}
+ 312  \z3
+ \frac{176}{3}  \z2
+ \bigg(
- \frac{1088}{9}
- 80  \z2\bigg)  \Lqf
+ \bigg( 44\bigg)  \Lqf^2\bigg)  \Ca^2
+ \bigg(\frac{224}{27}
- \frac{32}{3}  \z2
+ \bigg(\frac{200}{9}\bigg)  \Lqf
+ \bigg(-8\bigg)  \Lqf^2\bigg)  \Ca  \nf \bigg\}
\Bigg]\end{autobreak} 
\\ 
\begin{autobreak} 
\CIg3 =
\frac{\pi}{2 (n_c^2-1)}\Bigg[ 
\Dm1    \bigg\{ \bigg(
- \frac{303707}{810}
- \frac{28636}{9}  \z5
- \frac{186194}{135}  \z3
+ \frac{13216}{3}  \z3^2
+ \frac{923}{3}  \z2
+ \frac{8944}{3}  \z2  \z3
+ \frac{29416}{135}  \z2^2
- \frac{64096}{105}  \z2^3
+ \bigg(
- \frac{1319}{9}
+ 5984  \z5
- \frac{3496}{3}  \z3
+ \frac{15232}{27}  \z2
- 3872  \z2  \z3
- \frac{396}{5}   \z2^2\bigg)  \Lqf
+ \bigg(\frac{110}{3}
+ \frac{968}{3}  \z3
+ \frac{736}{3}  \z2
- \frac{1152}{5}  \z2^2\bigg)  \Lqf^2
+ \bigg(\frac{512}{3}  \z3
- \frac{352}{3}  \z2\bigg)  \Lqf^3\bigg)  \Ca^3
+ \bigg(
- \frac{241}{9}
+ 160  \z5
- \frac{296}{3}  \z3
+ \bigg(2\bigg)   \Lqf\bigg)  \Cf^2  \nf
+ \bigg(
- \frac{2617}{162}
+ 8  \z3
+ 24  \z2
+ \bigg(
- \frac{16}{3}\bigg)  \Lqf
+ \bigg(\frac{4}{3}\bigg)   \Lqf^2\bigg)  \Cf  \nf^2
+ \bigg(
- \frac{1487}{135}
- \frac{448}{45}  \z3
+ \frac{1019}{27}  \z2
+ \frac{304}{45}  \z2^2
+ \bigg(
- \frac{41}{9}\bigg)  \Lqf
+ \bigg(\frac{4}{3}\bigg)  \Lqf^2\bigg)  \Ca  \nf^2
+ \bigg(\frac{55546}{405}
+ \frac{2752}{9}  \z5
+ \frac{55229}{135}  \z3
- \frac{5369}{27}  \z2
- \frac{1508}{3}  \z2  \z3
- \frac{10393}{135}  \z2^2
+ \bigg(
- 14
- \frac{176}{3}  \z3
- \frac{160}{3}  \z2\bigg)  \Lqf^2
+ \bigg(\frac{523}{9}
+ \frac{760}{3}  \z3
- \frac{2368}{27}  \z2
+ \frac{72}{5}   \z2^2\bigg)  \Lqf
+ \bigg(\frac{64}{3}  \z2\bigg)  \Lqf^3\bigg)  \Ca^2  \nf
+ \bigg(\frac{65866}{405}
+ 80  \z5
- \frac{5552}{45}   \z3
- \frac{520}{9}  \z2
- \frac{128}{3}  \z2  \z3
- \frac{32}{45}  \z2^2
+ \bigg(
- \frac{22}{3}\bigg)  \Lqf^2
+ \bigg(\frac{55}{3}
- 16  \z2\bigg)  \Lqf\bigg)  \Ca  \Cf  \nf \bigg\}      
+ \D5    \bigg\{ \bigg(512\bigg)  \Ca^3 \bigg\}      
+ \D4    \bigg\{ \bigg(
- \frac{7040}{9}
+ \bigg(1280\bigg)  \Lqf\bigg)  \Ca^3
+ \bigg(\frac{1280}{9}\bigg)  \Ca^2  \nf \bigg\}      
+ \D3    \bigg\{ \bigg(
- \frac{18752}{27}
- 3584  \z2
+ \bigg(
- \frac{5632}{9}\bigg)  \Lqf
+ \bigg(1024\bigg)  \Lqf^2\bigg)  \Ca^3
+ \bigg(\frac{256}{27}\bigg)  \Ca  \nf^2
+ \bigg(\frac{2944}{27}
+ \bigg(\frac{1024}{9}\bigg)  \Lqf\bigg)  \Ca^2  \nf \bigg\}      
+ \D2    \bigg\{ \bigg(
- 1168
+ 11584  \z3
+ \frac{11968}{3}  \z2
+ \bigg(
- 1472
- 5376  \z2\bigg)  \Lqf
+ \bigg( 256\bigg)  \Lqf^3
+ \bigg(352\bigg)  \Lqf^2\bigg)  \Ca^3
+ \bigg(\frac{160}{9}\bigg)  \Ca  \nf^2
+ \bigg(\frac{656}{9}
- \frac{2176}{3}  \z2
+ \bigg(-64\bigg)  \Lqf^2
+ \bigg(320\bigg)  \Lqf\bigg)  \Ca^2  \nf
+ \bigg(32\bigg)  \Ca  \Cf  \nf \bigg\}      
+ \D1    \bigg\{ \bigg(
- \frac{6932}{9}
- \frac{20416}{3}  \z3
+ \frac{17120}{9}  \z2
- \frac{9856}{5}  \z2^2
+ \bigg(
- \frac{21536}{27}
+ 11520  \z3
+ \frac{5632}{3}  \z2\bigg)  \Lqf
+ \bigg(
- \frac{1472}{3}
- 2304  \z2\bigg)  \Lqf^2
+ \bigg(\frac{704}{3} \bigg)  \Lqf^3\bigg)  \Ca^3
+ \bigg(\frac{100}{9}
+ \frac{128}{9}  \z2\bigg)  \Ca  \nf^2
+ \bigg(\frac{1480}{9}
+ \frac{4096}{3}  \z3
- \frac{4288}{9}  \z2
+ \bigg(
- \frac{128}{3}\bigg)  \Lqf^3
+ \bigg(\frac{2720}{27}
- \frac{1024}{3}  \z2\bigg)  \Lqf
+ \bigg(\frac{320}{3}\bigg)   \Lqf^2\bigg)  \Ca^2  \nf
+ \bigg(\frac{536}{3}
- 128  \z3
+ \bigg(-32\bigg)  \Lqf\bigg)  \Ca  \Cf  \nf \bigg\}      
+ \D0    \bigg\{ \bigg(
- \frac{180844}{729}
+ \frac{3320}{9}  \z3
+ \frac{3200}{81}  \z2
- \frac{544}{15}  \z2^2
+ \bigg(
- \frac{352}{27}\bigg)  \Lqf^3
+ \bigg(\frac{140}{9}
+ 32  \z2\bigg)  \Lqf^2
+ \bigg(\frac{18052}{81}
+ \frac{800}{3}  \z3
- \frac{1184}{3}   \z2\bigg)  \Lqf\bigg)  \Ca^2  \nf
+ \bigg(\frac{19808}{729}
+ \frac{320}{27}  \z3
- \frac{160}{9}  \z2
+ \bigg(
- \frac{446}{81}
+ \frac{64}{3}  \z2\bigg)  \Lqf
+ \bigg(
- \frac{40}{9}\bigg)  \Lqf^2
+ \bigg(\frac{32}{27}\bigg)  \Lqf^3\bigg)  \Ca  \nf^2
+ \bigg(\frac{3422}{27}
- \frac{608}{9}  \z3
- 32  \z2
- \frac{64}{5}  \z2^2
+ \bigg(\frac{268}{3}
- 64  \z3\bigg)  \Lqf
+ \bigg(-24\bigg)  \Lqf^2\bigg)  \Ca   \Cf  \nf
+ \bigg(\frac{390086}{729}
+ 11904  \z5
- \frac{46952}{27}  \z3
+ \frac{29824}{81}  \z2
- \frac{23200}{3}   \z2  \z3
+ \frac{4048}{15}  \z2^2
+ \bigg(
- \frac{66746}{81}
- \frac{3344}{3}  \z3
+ \frac{4144}{3}  \z2
- \frac{4928}{5}   \z2^2\bigg)  \Lqf
+ \bigg(\frac{116}{9}
+ 2240  \z3
- 176  \z2\bigg)  \Lqf^2
+ \bigg(\frac{968}{27}
- 256  \z2\bigg)   \Lqf^3\bigg)  \Ca^3 \bigg\}
\Bigg]
\end{autobreak} 
\end{align}

%

\section{Resummed coefficients}
\subsection{Process dependent coefficients $g_0^I$}

Here we present the process-dependent coefficients used for N$^3$LL resummation for spin-2 production for both the quark and gluon initiated channels.

\begin{align} 
\begin{autobreak} 
\gGRQQ1 = 
  \Cf    \bigg\{ 
- 20
+ 16  \z2
+ \bigg(-6\bigg)  \Lfr
+ \bigg(6\bigg)  \Lqr \bigg\} ,   
\end{autobreak} 
\\ 
\begin{autobreak} 
\gGRQQ2 = 
  \Cf  \nf    \bigg\{ \frac{461}{18}
+ \frac{8}{9}  \z3
- 16  \z2
+ \bigg(
- 14
+ \frac{16}{3}  \z2\bigg)  \Lqr
+ \bigg(\frac{2}{3}
+ \frac{16}{3}  \z2\bigg)  \Lfr
+ \bigg(-2\bigg)  \Lfr^2
+ \bigg(2\bigg)  \Lqr^2 \bigg\}      
+ \Cf^2    \bigg\{ \frac{2293}{12}
- 124  \z3
- 230  \z2
+ \frac{552}{5}  \z2^2
+ \bigg(
- 117
+ 48  \z3
+ 72  \z2\bigg)  \Lqr
+ \bigg(117
- 48  \z3
- 72  \z2\bigg)  \Lfr
+ \bigg(-36\bigg)  \Lqrfr
+ \bigg(18\bigg)  \Lfr^2
+ \bigg( 18\bigg)  \Lqr^2 \bigg\}      
+ \Ca  \Cf    \bigg\{ 
- \frac{5941}{36}
+ \frac{1180}{9}  \z3
+ 96  \z2
- \frac{92}{5}  \z2^2
+ \bigg(
- \frac{17}{3}
+ 24   \z3
- \frac{88}{3}  \z2\bigg)  \Lfr
+ \bigg(79
- 24  \z3
- \frac{88}{3}  \z2\bigg)  \Lqr
+ \bigg(-11\bigg)  \Lqr^2
+ \bigg(11\bigg)   \Lfr^2 \bigg\} ,   
\end{autobreak} 
\\ 
\begin{autobreak} 
\gGRQQ3 = 
  \Cf  \nf^2    \bigg\{ 
- \frac{7345}{243}
- \frac{1136}{81}  \z3
+ \frac{848}{27}  \z2
+ \frac{448}{135}  \z2^2
+ \bigg(
- \frac{28}{3}
+ \frac{32}{9}  \z2\bigg)  \Lqr^2
+ \bigg(
- \frac{8}{9}\bigg)  \Lfr^3
+ \bigg(\frac{4}{9}
+ \frac{32}{9}  \z2\bigg)  \Lfr^2
+ \bigg( \frac{8}{9}\bigg)  \Lqr^3
+ \bigg(\frac{34}{9}
+ \frac{32}{9}  \z3
- \frac{160}{27}  \z2\bigg)  \Lfr
+ \bigg(\frac{820}{27}
- \frac{64}{27}  \z3
- \frac{416}{27}  \z2\bigg)  \Lqr \bigg\}      
+ \Cf^2  \nf    \bigg\{ 
- \frac{697}{3}
- \frac{608}{9}  \z5
+ \frac{18280}{27}  \z3
+ \frac{7096}{27}  \z2
- \frac{256}{3}  \z2   \z3
- \frac{27184}{135}  \z2^2
+ \bigg(
- 121
+ \frac{256}{3}  \z3
- \frac{40}{3}  \z2
+ \frac{272}{5}  \z2^2\bigg)  \Lfr
+ \bigg(
- 116
+ 32  \z3
+ 48  \z2\bigg)  \Lqr^2
+ \bigg(28
- 32  \z3
- 48  \z2\bigg)  \Lfr^2
+ \bigg( \frac{3022}{9}
- \frac{752}{3}  \z3
- \frac{784}{3}  \z2
+ \frac{464}{5}  \z2^2\bigg)  \Lqr
+ \bigg(-12\bigg)  \Lqrfrt
+ \bigg(-12\bigg)   \Lqrtfr
+ \bigg(12\bigg)  \Lfr^3
+ \bigg(12\bigg)  \Lqr^3
+ \bigg(88\bigg)  \Lqrfr \bigg\}      
+ \Cf^3    \bigg\{ 
- \frac{2807}{2}
+ \frac{3344}{3}  \z5
+ \frac{4300}{3}  \z3
+ 32  \z3^2
+ 1168  \z2
- 1424   \z2  \z3
- \frac{6388}{5}  \z2^2
+ \frac{169504}{315}  \z2^3
+ \bigg(
- \frac{2231}{2}
+ 480  \z5
+ 1568   \z3
+ 816  \z2
- 704  \z2  \z3
- \frac{1968}{5}  \z2^2\bigg)  \Lfr
+ \bigg(
- 342
+ 288  \z3
+ 144   \z2\bigg)  \Lfr^2
+ \bigg(
- 342
+ 288  \z3
+ 144  \z2\bigg)  \Lqr^2
+ \bigg(684
- 576  \z3
- 288  \z2 \bigg)  \Lqrfr
+ \bigg(\frac{2231}{2}
- 480  \z5
- 1568  \z3
- 816  \z2
+ 704  \z2  \z3
+ \frac{1968}{5}   \z2^2\bigg)  \Lqr
+ \bigg(-108\bigg)  \Lqrtfr
+ \bigg(-36\bigg)  \Lfr^3
+ \bigg(36\bigg)  \Lqr^3
+ \bigg(108\bigg)  \Lqrfrt \bigg\}      
+ \Ca  \Cf  \nf    \bigg\{ \frac{130871}{243}
+ \frac{136}{3}  \z5
- \frac{47984}{81}  \z3
- \frac{32756}{81}  \z2
+ \frac{928}{9}  \z2  \z3
- \frac{1076}{135}  \z2^2
+ \bigg(
- \frac{11732}{27}
+ \frac{5744}{27}  \z3
+ \frac{5392}{27}  \z2
- \frac{344}{15}  \z2^2\bigg)  \Lqr
+ \bigg(
- 40
- \frac{400}{9}  \z3
+ \frac{2672}{27}  \z2
- \frac{8}{5}  \z2^2\bigg)  \Lfr
+ \bigg(
- \frac{146}{9}
+ 16  \z3
- \frac{352}{9}  \z2\bigg)  \Lfr^2
+ \bigg(
- \frac{88}{9}\bigg)  \Lqr^3
+ \bigg(\frac{88}{9}\bigg)  \Lfr^3
+ \bigg(114
- 16  \z3
- \frac{352}{9}  \z2\bigg)  \Lqr^2 \bigg\}      
+ \Ca  \Cf^2    \bigg\{ \frac{328511}{108}
+ \frac{6968}{9}  \z5
- \frac{167428}{27}  \z3
+ \frac{592}{3}  \z3^2
- \frac{58912}{27}  \z2
+ \frac{6880}{3}  \z2  \z3
+ \frac{42128}{27}  \z2^2
- \frac{123632}{315}  \z2^3
+ \bigg(
- \frac{43727}{18}
+ 240  \z5
+ \frac{8216}{3}  \z3
+ \frac{4480}{3}  \z2
- 352  \z2  \z3
- \frac{2912}{5}  \z2^2\bigg)   \Lqr
+ \bigg(
- 508
+ 288  \z3\bigg)  \Lqrfr
+ \bigg(
- 175
+ 32  \z3
+ 264  \z2\bigg)  \Lfr^2
+ \bigg( 683
- 320  \z3
- 264  \z2\bigg)  \Lqr^2
+ \bigg(1028
- 240  \z5
- \frac{5488}{3}  \z3
+ \frac{580}{3}  \z2
+ 352  \z2  \z3
- \frac{1136}{5}  \z2^2\bigg)  \Lfr
+ \bigg(-66\bigg)  \Lfr^3
+ \bigg(-66\bigg)  \Lqr^3
+ \bigg(66\bigg)   \Lqrfrt
+ \bigg(66\bigg)  \Lqrtfr \bigg\}      
+ \Ca^2  \Cf    \bigg\{ 
- \frac{1970041}{972}
- \frac{4292}{3}  \z5
+ \frac{339325}{81}  \z3
- \frac{400}{3}  \z3^2
+ \frac{91067}{81}  \z2
- \frac{7660}{9}  \z2  \z3
- \frac{10673}{135}  \z2^2
+ \frac{7088}{63}  \z2^3
+ \bigg(
- \frac{971}{3}
+ 88  \z3
+ \frac{968}{9}  \z2\bigg)  \Lqr^2
+ \bigg(
- \frac{242}{9}\bigg)  \Lfr^3
+ \bigg(\frac{242}{9}\bigg)  \Lqr^3
+ \bigg(\frac{493}{9}
- 88  \z3
+ \frac{968}{9}  \z2\bigg)  \Lfr^2
+ \bigg(\frac{1657}{18}
- 80  \z5
+ \frac{3104}{9}  \z3
- \frac{8992}{27}   \z2
+ 4  \z2^2\bigg)  \Lfr
+ \bigg(\frac{36310}{27}
+ 80  \z5
- \frac{35272}{27}  \z3
- \frac{14912}{27}  \z2
+ \frac{1964}{15}  \z2^2\bigg)  \Lqr \bigg\} , 
\end{autobreak} 
\end{align}

\begin{align} 
\begin{autobreak} 
\gGRGG1 = 
  \nf    \bigg\{ \frac{35}{9}
+ \bigg(
- \frac{4}{3}\bigg)  \Lqr
+ \bigg(\frac{4}{3}\bigg)  \Lfr \bigg\}      
+ \Ca    \bigg\{ 
- \frac{203}{9}
+ 16  \z2
+ \bigg(
- \frac{22}{3}\bigg)  \Lfr
+ \bigg(\frac{22}{3}\bigg)  \Lqr \bigg\} ,   
\end{autobreak} 
\\ 
\begin{autobreak} 
\gGRGG2 = 
  \nf^2    \bigg\{ \frac{1225}{324}
+ \frac{8}{3}  \z2
+ \bigg(
- \frac{70}{27}\bigg)  \Lqr
+ \bigg(
- \frac{16}{9}\bigg)  \Lqrfr
+ \bigg(\frac{4}{9}\bigg)  \Lqr^2
+ \bigg(\frac{4}{3}\bigg)  \Lfr^2
+ \bigg(\frac{140}{27}\bigg)  \Lfr \bigg\}      
+ \Cf  \nf    \bigg\{ \frac{61}{3}
- 16  \z3
+ \bigg(-4\bigg)  \Lqr
+ \bigg(4\bigg)  \Lfr \bigg\}      
+ \Ca  \nf    \bigg\{ 
- \frac{2983}{162}
+ \frac{128}{9}  \z3
+ \frac{106}{9}  \z2
+ \bigg(
- \frac{1438}{27}
+ \frac{64}{3}  \z2\bigg)   \Lfr
+ \bigg(
- \frac{44}{3}\bigg)  \Lfr^2
+ \bigg(
- \frac{44}{9}\bigg)  \Lqr^2
+ \bigg(\frac{176}{9}\bigg)  \Lqrfr
+ \bigg(\frac{647}{27}
- \frac{32}{3}  \z2\bigg)  \Lqr \bigg\}      
+ \Ca^2    \bigg\{ \frac{7801}{324}
+ \frac{88}{9}  \z3
- \frac{1312}{9}  \z2
+ 92  \z2^2
+ \bigg(
- \frac{1657}{27}
+ 24   \z3
+ \frac{176}{3}  \z2\bigg)  \Lqr
+ \bigg(
- \frac{484}{9}\bigg)  \Lqrfr
+ \bigg(\frac{121}{9}\bigg)  \Lqr^2
+ \bigg(\frac{121}{3}\bigg)  \Lfr^2
+ \bigg(\frac{3890}{27}
- 24  \z3
- \frac{352}{3}  \z2\bigg)  \Lfr \bigg\} ,   
\end{autobreak} 
\\ 
\begin{autobreak} 
\gGRGG3 = 
  \nf^3    \bigg\{ \bigg(
- \frac{280}{81}\bigg)  \Lqrfr
+ \bigg(
- \frac{16}{9}\bigg)  \Lqrfrt
+ \bigg(\frac{16}{27}\bigg)  \Lqrtfr
+ \bigg( \frac{32}{27}\bigg)  \Lfr^3
+ \bigg(\frac{1225}{243}
+ \frac{32}{9}  \z2\bigg)  \Lfr
+ \bigg(\frac{140}{27}\bigg)  \Lfr^2 \bigg\}      
+ \Cf  \nf^2    \bigg\{ 
- \frac{2617}{162}
+ 8  \z3
+ 24  \z2
+ \bigg(
- \frac{32}{3}\bigg)  \Lqrfr
+ \bigg(
- \frac{16}{3}\bigg)   \Lqr
+ \bigg(\frac{4}{3}\bigg)  \Lqr^2
+ \bigg(\frac{28}{3}\bigg)  \Lfr^2
+ \bigg(\frac{362}{9}
- \frac{64}{3}  \z3\bigg)  \Lfr \bigg\}      
+ \Cf^2  \nf    \bigg\{ 
- \frac{241}{9}
+ 160  \z5
- \frac{296}{3}  \z3
+ \bigg(-2\bigg)  \Lfr
+ \bigg(2\bigg)  \Lqr \bigg\}      
+ \Ca  \nf^2    \bigg\{ 
- \frac{1487}{135}
- \frac{2944}{135}  \z3
+ \frac{1169}{27}  \z2
+ \frac{80}{3}  \z2^2
+ \bigg(
- \frac{668}{9}
+ \frac{64}{3}  \z2\bigg)  \Lfr^2
+ \bigg(
- \frac{1877}{54}
+ \frac{512}{27}  \z3
- \frac{104}{27}  \z2\bigg)  \Lfr
+ \bigg(
- \frac{176}{9}\bigg)  \Lfr^3
+ \bigg(
- \frac{88}{9}\bigg)  \Lqrtfr
+ \bigg(
- \frac{41}{9}\bigg)  \Lqr
+ \bigg(\frac{4}{3}\bigg)  \Lqr^2
+ \bigg( \frac{88}{3}\bigg)  \Lqrfrt
+ \bigg(\frac{1184}{27}
- \frac{128}{9}  \z2\bigg)  \Lqrfr \bigg\}      
+ \Ca  \Cf  \nf    \bigg\{ \frac{65866}{405}
+ 80  \z5
- \frac{6512}{45}  \z3
+ \frac{284}{9}  \z2
- \frac{320}{3}  \z2  \z3
- \frac{32}{45}  \z2^2
+ \bigg(
- \frac{1913}{9}
+ \frac{352}{3}  \z3
+ 64  \z2\bigg)  \Lfr
+ \bigg(
- \frac{154}{3}\bigg)  \Lfr^2
+ \bigg(
- \frac{22}{3}\bigg)  \Lqr^2
+ \bigg(\frac{55}{3}
- 32  \z2\bigg)  \Lqr
+ \bigg(\frac{176}{3}\bigg)  \Lqrfr \bigg\}      
+ \Ca^2  \nf    \bigg\{ \frac{55546}{405}
- \frac{3392}{9}  \z5
+ \frac{16223}{45}  \z3
- \frac{3149}{27}  \z2
+ \frac{284}{3}   \z2  \z3
- \frac{22681}{135}  \z2^2
+ \bigg(
- 190
+ 64  \z3
+ \frac{1408}{9}  \z2\bigg)  \Lqrfr
+ \bigg(
- \frac{484}{3}\bigg)  \Lqrfrt
+ \bigg(
- 14
- 16  \z3\bigg)  \Lqr^2
+ \bigg(
- \frac{829}{81}
- \frac{3544}{27}  \z3
- \frac{5132}{27}   \z2
+ \frac{376}{3}  \z2^2\bigg)  \Lfr
+ \bigg(\frac{484}{9}\bigg)  \Lqrtfr
+ \bigg(\frac{523}{9}
+ 40  \z3
- \frac{112}{3}  \z2
- \frac{8}{3}  \z2^2\bigg)  \Lqr
+ \bigg(\frac{968}{9}\bigg)  \Lfr^3
+ \bigg(\frac{3299}{9}
- 48  \z3
- \frac{704}{3}  \z2\bigg)  \Lfr^2 \bigg\}      
+ \Ca^3    \bigg\{ 
- \frac{303707}{810}
+ \frac{5156}{9}  \z5
- \frac{81074}{135}  \z3
+ 96  \z3^2
- \frac{697}{9}   \z2
+ 48  \z2  \z3
+ \frac{6248}{27}  \z2^2
+ \frac{3872}{15}  \z2^3
+ \bigg(
- \frac{17105}{27}
+ 264  \z3
+ \frac{1936}{3}  \z2\bigg)  \Lfr^2
+ \bigg(
- \frac{5324}{27}\bigg)  \Lfr^3
+ \bigg(
- \frac{1319}{9}
- 160  \z5
- 184   \z3
+ \frac{496}{3}  \z2
+ 352  \z2  \z3
+ \frac{44}{3}  \z2^2\bigg)  \Lqr
+ \bigg(
- \frac{2662}{27}\bigg)  \Lqrtfr
+ \bigg( \frac{110}{3}
+ 88  \z3\bigg)  \Lqr^2
+ \bigg(\frac{109651}{486}
+ 160  \z5
+ \frac{3032}{27}  \z3
+ \frac{19504}{27}   \z2
- 352  \z2  \z3
- \frac{2068}{3}  \z2^2\bigg)  \Lfr
+ \bigg(\frac{23782}{81}
- 352  \z3
- \frac{3872}{9}  \z2\bigg)   \Lqrfr
+ \bigg(\frac{2662}{9}\bigg)  \Lqrfrt \bigg\} . 
\end{autobreak} 
\end{align}
%

\subsection{Universal resummed exponents $G_N^I$} \label{app:B2}

Here we collect the universal resummed coefficients used for the N$^3$LL resummation. Taking $\omega = 2 \beta_0 a_s \ln \bar{N}, L_{qr} = \ln (Q^2/\mu_r^2), L_{fr} = \ln (\mu_f^2/\mu_r^2)$, we present the coefficients required up to N$^3$LL order.

\begin{align} 
\begin{autobreak} 
\gNB1 =
\bigg[ \AAo  ~  \bigg\{ 2
- 2 ~ \LogmW1
+ 2 ~ \LogmW1 ~ \iW \bigg\}
\bigg],   
\end{autobreak} 
\\ 
\begin{autobreak} 
\gNB2 =
\bigg[ \DDo  ~  \bigg\{ \frac{1}{2} ~ \LogmW1 \bigg\}      
+ \AAt  ~  \bigg\{ 
- \LogmW1
- \w \bigg\}      
+ \AAo  ~  \bigg\{ \bigg(\LogmW1
+ \frac{1}{2} ~ \LogmW1^2
+ \w\bigg) ~ \bigg(\frac{\beta_{1}}{\beta_0^{2}}\bigg)
+ \bigg(\w\bigg) ~ \Lfr
+ \bigg(\LogmW1\bigg) ~ \Lqr \bigg\}
\bigg],   
\end{autobreak} 
\\ 
\begin{autobreak} 
\gNB3 =
\bigg[ \btzAIII  ~  \bigg\{ 
- \WbimW
+ \w \bigg\}      
+ \btzAII  ~  \bigg\{ \bigg(2 ~ \WbimW\bigg) ~ \Lqr
+ \bigg(3 ~ \WbimW
+ 2 ~ \LogomWtIMW
- \w\bigg) ~ \bigg(\frac{\beta_{1}}{\beta_0^{2}}\bigg)
+ \bigg(
- 2 ~ \w\bigg) ~  \Lfr \bigg\}      
+ \btzAI  ~  \bigg\{ 
- 4 ~ \z2 ~ \WbimW
+ \bigg(
- \LogtmWtIMW
- \WbimW
- 2 ~ \LogomWtIMW
+ 2 ~  \LogmW1
+ \w\bigg) ~ \bigg(\frac{\beta_{1}}{\beta_0^{2}}\bigg)^2
+ \bigg(
- \WbimW\bigg) ~ \Lqr^2
+ \bigg(
- \WbimW
- 2 ~ \LogmW1
- \w\bigg) ~ \btoo
+ \bigg(\bigg(
- 2 ~ \WbimW
- 2 ~ \LogomWtIMW\bigg) ~ \bigg(\frac{\beta_{1}}{\beta_0^{2}}\bigg)\bigg) ~ \Lqr
+ \bigg(\w\bigg) ~ \Lfr^2 \bigg\}      
+ \btzDII  ~  \bigg\{ \WbimW \bigg\}      
+ \btzDI  ~  \bigg\{ \bigg(
- \WbimW\bigg) ~ \Lqr
+ \bigg(
- \WbimW
- \LogomWtIMW\bigg) ~ \bigg(\frac{\beta_{1}}{\beta_0^{2}}\bigg) \bigg\}
\bigg],   
\end{autobreak} 
\\ 
\begin{autobreak} 
\gNB4 =
\bigg[ \btztAIV  ~  \bigg\{ \frac{1}{6} ~ \WttmWtimWt
- \frac{1}{3} ~ \w \bigg\}      
+ \btztAIII  ~  \bigg\{ \bigg(
- \frac{1}{2} ~ \WttmWtimWt\bigg) ~ \Lqr
+ \bigg(
- \frac{5}{12} ~ \WttmWtimWt
- \frac{1}{2} ~  \LogomWtIMWt
+ \frac{1}{3} ~ \w\bigg) ~ \bigg(\frac{\beta_{1}}{\beta_0^{2}}\bigg)
+ \bigg(\w\bigg) ~ \Lfr \bigg\}      
+ \btztAII  ~  \bigg\{ 2 ~ \z2 ~ \WttmWtimWt
+ \bigg(\frac{1}{2} ~ \LogtmWtIMWt
- \frac{1}{12} ~ \WtbimWt
+ \frac{5}{6} ~  \WbimW
+ \frac{1}{2} ~ \LogomWtIMWt
- \frac{1}{3} ~ \w\bigg) ~ \bigg(\frac{\beta_{1}}{\beta_0^{2}}\bigg)^2
+ \bigg(\frac{1}{2} ~ \WttmWtimWt\bigg) ~ \Lqr^2
+ \bigg(\frac{1}{3}  ~ \WtbimWt
- \frac{1}{3} ~ \WbimW
+ \frac{1}{3} ~ \w\bigg) ~ \btoo
+ \bigg(
- \w\bigg) ~ \Lfr^2
+ \bigg(\bigg(\frac{1}{2} ~ \WttmWtimWt
+ \LogomWtIMWt\bigg) ~ \bigg(\frac{\beta_{1}}{\beta_0^{2}}\bigg)\bigg) ~ \Lqr \bigg\}      
+ \btztAI  ~  \bigg\{ \frac{8}{3} ~ \z3 ~ \WttmWtimWt
+ \bigg(
- \frac{1}{6} ~ \LogttmWtIMWt
+ \frac{1}{3} ~ \WtbimWt
- \frac{1}{3} ~ \WbimW
+ \frac{1}{2} ~ \LogomWtIMWt
- \LogomWtIMW
+ \frac{1}{2} ~ \LogmW1
+ \frac{1}{3} ~ \w\bigg) ~ \bigg(\frac{\beta_{1}}{\beta_0^{2}}\bigg)^3
+ \bigg(
- \frac{1}{6} ~ \WttmWtimWt\bigg) ~ \Lqr^3
+ \bigg(\frac{1}{12} ~ \WttmWtimWt
+ \frac{1}{2} ~ \LogmW1
+ \frac{1}{3} ~ \w\bigg) ~ \bthr
+ \bigg(
- \frac{5}{12} ~ \WtbimWt
+ \frac{1}{6} ~ \WbimW
- \frac{1}{2} ~ \LogomWtIMWt
+ \LogomWtIMW
- \LogmW1
- \frac{2}{3} ~ \w\bigg) ~ \bobt
+ \bigg(\frac{1}{3} ~ \w\bigg) ~ \Lfr^3
+ \bigg(
- 2 ~ \z2 ~ \WttmWtimWt
+ \bigg(
- \frac{1}{2} ~  \LogtmWtIMWt
+ \frac{1}{2} ~ \WtbimWt\bigg) ~ \bigg(\frac{\beta_{1}}{\beta_0^{2}}\bigg)^2
+ \bigg(
- \frac{1}{2} ~ \WtbimWt\bigg) ~ \btoo\bigg) ~ \Lqr
+ \bigg(
- 2 ~ \z2  ~ \LogomWtIMWt\bigg) ~ \bigg(\frac{\beta_{1}}{\beta_0^{2}}\bigg)
+ \bigg(\bigg(
- \frac{1}{2} ~ \LogomWtIMWt\bigg) ~ \bigg(\frac{\beta_{1}}{\beta_0^{2}}\bigg)\bigg) ~ \Lqr^2
+ \bigg(\bigg(
- \frac{1}{2} ~ \w\bigg) ~ \bigg(\frac{\beta_{1}}{\beta_0^{2}}\bigg)\bigg) ~  \Lfr^2 \bigg\}      
+ \btztDIII  ~  \bigg\{ 
- \frac{1}{4} ~ \WttmWtimWt \bigg\}      
+ \btztDII  ~  \bigg\{ \bigg(\frac{1}{4} ~ \WttmWtimWt
+ \frac{1}{2} ~ \LogomWtIMWt\bigg) ~ \bigg(\frac{\beta_{1}}{\beta_0^{2}}\bigg)
+ \bigg(\frac{1}{2} ~ \WttmWtimWt\bigg) ~  \Lqr \bigg\}      
+ \btztDI  ~  \bigg\{ 
- \z2 ~ \WttmWtimWt
+ \bigg(
- \frac{1}{4} ~ \LogtmWtIMWt
+ \frac{1}{4} ~ \WtbimWt\bigg) ~ \bigg(\frac{\beta_{1}}{\beta_0^{2}}\bigg)^2
+ \bigg(
- \frac{1}{4} ~ \WttmWtimWt\bigg) ~ \Lqr^2
+ \bigg(
- \frac{1}{4} ~ \WtbimWt\bigg) ~ \btoo
+ \bigg(\bigg(
- \frac{1}{2} ~  \LogomWtIMWt\bigg) ~ \bigg(\frac{\beta_{1}}{\beta_0^{2}}\bigg)\bigg) ~ \Lqr \bigg\}
\bigg].
\end{autobreak} 
\end{align}
The cusp anomalous dimensions $A_i$ are given as (with the recently known four-loops results \cite{Henn:2019swt}) ,

\begin{align} 
\begin{autobreak} 
\A1 = {\cal C}_{i}
\bigg\{ 4
\bigg\},   
\end{autobreak} 
\\ 
\begin{autobreak} 
\A2 = {\cal C}_{i}
\bigg\{ \nf    \bigg( 
- \frac{40}{9} \bigg)      
+ \Ca    \bigg( \frac{268}{9}
- 8  \z2 \bigg)
\bigg\},   
\end{autobreak} 
\\ 
\begin{autobreak} 
\A3 = {\cal C}_{i}
\bigg\{ \nf^2    \bigg( 
- \frac{16}{27} \bigg)      
+ \Cf  \nf    \bigg( 
- \frac{110}{3}
+ 32  \z3 \bigg)      
+ \Ca  \nf    \bigg( 
- \frac{836}{27}
- \frac{112}{3}  \z3
+ \frac{160}{9}  \z2 \bigg)      
+ \Ca^2    \bigg( \frac{490}{3}
+ \frac{88}{3}  \z3
- \frac{1072}{9}  \z2
+ \frac{176}{5}  \z2^2 \bigg)
\bigg\}, 
\end{autobreak} 
\\
\begin{autobreak} 
\A4 = {\cal C}_{i}
\bigg\{ 
 \nf^3    \bigg( 
- \frac{32}{81}
+ \frac{64}{27}  \z3 \bigg)      
+ \Cf  \nf^2    \bigg( \frac{2392}{81}
- \frac{640}{9}  \z3
+ \frac{64}{5}  \z2^2 \bigg)      
+ \Cf^2  \nf    \bigg( \frac{572}{9}
- 320  \z5
+ \frac{592}{3}  \z3 \bigg)      
+ \Ca  \nf^2    \bigg( \frac{923}{81}
+ \frac{2240}{27}  \z3
- \frac{608}{81}  \z2
- \frac{224}{15}  \z2^2 \bigg)      
+ \Ca  \Cf  \nf    \bigg( 
- \frac{34066}{81}
+ 160  \z5
+ \frac{3712}{9}  \z3
+ \frac{440}{3}  \z2
- 128  \z2  \z3
- \frac{352}{5}  \z2^2 \bigg)      
+ \Ca^2  \nf    \bigg( 
- \frac{24137}{81}
+ \frac{2096}{9}  \z5
- \frac{23104}{27}  \z3
+ \frac{20320}{81}  \z2
+ \frac{448}{3}  \z2  \z3
- \frac{352}{15}  \z2^2 \bigg)      
+ \Ca^3    \bigg( \frac{84278}{81}
- \frac{3608}{9}  \z5
+ \frac{20944}{27}  \z3
- 16  \z3^2
- \frac{88400}{81}  \z2
- \frac{352}{3}  \z2  \z3
+ \frac{3608}{5}  \z2^2
- \frac{20032}{105}  \z2^3 \bigg)
\bigg\}
+ \dRAoNR     \bigg( \frac{3520}{3}  \z5
+ \frac{128}{3}  \z3
- 384  \z3^2
- 128  \z2
- \frac{7936}{35}   \z2^3 \bigg)      
+ \nf  \dRFoNR    \bigg( 
- \frac{1280}{3}  \z5
- \frac{256}{3}  \z3
+ 256  \z2 \bigg)      \,. 
\end{autobreak} 
\end{align}
The quartic casimirs are given as
\begin{align}
\frac{d_A^{abcd}d_A^{abcd}}{N_A} &= \frac{n_c^2 (n_c^2 + 36)}{24}, 
\frac{d_A^{abcd}d_F^{abcd}}{N_A} = \frac{n_c (n_c^2 + 6)}{48}, \nn\\
\frac{d_F^{abcd}d_A^{abcd}}{N_F} &=  \frac{(n_c^2-1)(n_c^2+6)}{48}, 
\frac{d_F^{abcd}d_F^{abcd}}{N_F} = \frac{(n_c^2-1)(n_c^4 - 6 n_c^2 + 18)}{96n_c^3},
\end{align}
with $N_A = n_c^2 -1$ and $N_F = n_c$ where $n_c = 3$ for QCD.
%
The coefficients $D_i$ are given as,

\begin{align} 
\begin{autobreak} 
D_1 = {\cal C}_{i}
\bigg\{0
\bigg\},   
\end{autobreak} 
\\ 
\begin{autobreak} 
D_2 = {\cal C}_{i}
\bigg\{ \nf    \bigg( \frac{224}{27}
- \frac{32}{3}  \z2 \bigg)      
+ \Ca    \bigg( 
- \frac{1616}{27}
+ 56  \z3
+ \frac{176}{3}  \z2 \bigg)
\bigg\},   
\end{autobreak} 
\\ 
\begin{autobreak} 
D_3 = {\cal C}_{i}
\bigg\{ \nf^2    \bigg( 
- \frac{3712}{729}
+ \frac{320}{27}  \z3
+ \frac{640}{27}  \z2 \bigg)      
+ \Cf  \nf    \bigg( \frac{3422}{27}
- \frac{608}{9}  \z3
- 32  \z2
- \frac{64}{5}  \z2^2 \bigg)      
+ \Ca  \nf    \bigg( \frac{125252}{729}
- \frac{2480}{9}  \z3
- \frac{29392}{81}  \z2
+ \frac{736}{15}  \z2^2 \bigg)      
+ \Ca^2    \bigg( 
- \frac{594058}{729}
- 384  \z5
+ \frac{40144}{27}  \z3
+ \frac{98224}{81}  \z2
- \frac{352}{3}   \z2  \z3
- \frac{2992}{15}  \z2^2 \bigg)
\bigg\}, 
\end{autobreak} 
\end{align}
with ${\cal C}_i = C_A, C_F$ depending on the gluon or quark initiated process respectively.

\bibliographystyle{JHEP}
\bibliography{references}
\end{document}